\documentclass[twocolumn]{autart}    % Enable this line and disable the
 \usepackage[comma,numbers,square,sort&compress]{natbib}
 \usepackage{amsmath}
 \usepackage{amssymb}
 \usepackage{float}
 \usepackage{graphicx}
 \usepackage{epstopdf}
 \usepackage{color}
 \usepackage{verbatim}
 \usepackage{tikz,times}
 \usepackage{subfigure}
\usepackage{algorithmic,algorithm}
\DeclareMathOperator{\blockdiag}{blockdiag}
\DeclareMathOperator{\diag}{diag}

\DeclareMathOperator{\tr}{tr}

\theoremstyle{plain}
\newtheorem{theorem}{Theorem}[section]
\newtheorem{proposition}{\textbf{Proposition}}[section]
\newtheorem{lemma}{\textbf{Lemma}}[section]

\theoremstyle{definition}
\newtheorem{definition}{\textbf{Definition}}[section]
\newtheorem{assumption}{\textbf{Assumption}}[section]

\theoremstyle{remark}
\newtheorem{remark}{Remark}[section]

\begin{document}

\begin{frontmatter}
{
\title{Distributed Filtering for Uncertain  Systems  Under Switching Sensor Networks and Quantized Communications\thanksref{footnoteinfo}} % Title, preferably not more
}

\thanks[footnoteinfo]{ The material in this paper was not presented at any conference. }

\author[Beijing,kth]{Xingkang He}\ead{xingkang@kth.se},    % Add the
\author[Beijing,cor1]{ Wenchao Xue }\ead{wenchaoxue@amss.ac.cn},               % e-mail address
\corauth[cor1]{Corresponding author.}
\author[Beijing]{Xiaocheng Zhang}\ead{zhangxiaocheng16@mails.ucas.ac.cn},
\author[Beijing]{Haitao Fang}\ead{htfang@iss.ac.cn}  % (ead) as shown
\address[Beijing]{ LSC, NCMIS, Academy of Mathematics and Systems Science,
	Chinese Academy of Sciences, Beijing 100190, China}  % Please supply
%\address[UCAS]{School of Mathematical Sciences, University of Chinese Academy of Sciences, Beijing 100049, China}
\address[kth]{Division of Decision and Control Systems,  KTH Royal Institute of
	Technology, SE-100 44 Stockholm, Sweden}
%\address[Rome]{Senate House, Rome}             % full addresses
%\address[Baiae]{The White House, Baiae}        % here.

                             % keyword list or with the
                                          % help of the Automatica
                                          % keyword wizard

\begin{abstract}
	{ \color{black}
	This paper considers the distributed filtering problem for a class of stochastic uncertain systems under quantized data flowing over switching sensor networks.
Employing the biased noisy observations of the local sensor and interval-quantized messages from neighboring sensors successively,
 an extended state based distributed Kalman filter (DKF)  is proposed  for simultaneously estimating both system state and uncertain dynamics.
To alleviate the effect of observation biases, an event-triggered update  based DKF is presented with  a tighter mean square error bound   than  that of the time-driven one by designing a proper threshold.
Both the two DKFs are shown to provide the upper bounds of mean square errors online for each sensor.
Under mild conditions on systems and networks, the mean square error boundedness and asymptotic unbiasedness for the proposed two DKFs are proved.  Finally, the numerical simulations  demonstrate the effectiveness of the developed  filters.	
}
\end{abstract}

\end{frontmatter}

	\section{Introduction}
%This demo file is intended to serve as a ``starter file''
%for IEEE conference papers produced under \LaTeX\ using
%IEEEtran.cls version 1.8b and later.
%% You must have at least 2 lines in the paragraph with the drop letter
%% (should never be an issue)
%I wish you the best of success.

%State estimation problems for stochastic systems \cite{soderstrom2012discrete} exist in diverse areas, and they are widely studied in past decades.
In recent years,  sensor networks are broadly studied and applied to environment sensing, target tracking, smart grid, etc. {\color{black} As is well known,  state estimation problems over sensor networks are usually modeled as  distributed filtering studies.} Thus more and more  researchers and engineers around the world
are paying their attention  to the  methods and theories of distributed filtering.

%However, owing to  environmental complexity and network vulnerability, traditional centralized filtering methods confront many difficulties, such as link failure, transmission delay, channel restriction, etc.
%Because of the advantages in structure robustness, energy saving and parallel processing,

%increasing attention has been focused on the study of .
%Therefore, the  filtering (or state estimation) problems of multi-agent systems are becoming more and more attractive to the researchers and engineers around the world.
%Considering the existing state estimation methods on multi-agent systems, there are two main frameworks, namely, the centralized framework and the distributed framework. The main difference between the two frameworks lies in whether there exists a data center which is capable to collect the information messages from all agents.
%For the centralized framework, due to the existence of  data center, there are many practical problems, such as the robustness of system structure, the  burden of transmission channels, etc.
%As for the distributed framework, the communication strategy of agents follows a peer-to-peer protocol, where each agent can simply communicate with its neighbor agents.
%Because of the advantages in

%\subsection{Motivations}
{\color{black}In the existing literature on distributed filtering   over sensor networks}, many effective approaches and analysis tools have been provided. For linear time-invariant systems,    distributed filters with constant parameter gains were investigated in \cite{Khan2014Collaborative,Khan2010On}, which yet confined the scope of the system dynamics to be considered.
As is known, the optimal centralized Kalman filter for linear stochastic systems can provide the estimation error covariances online. However, in distributed
Kalman filters (DKFs)  \cite{Battistelli2016stability,Olfati2007Distributed,Cat2010Diffusion,Battistelli2014Kullback,He2017Consistent,wang2018convergence,liu2018distributed}, the covariances can not be obtained by each sensor due to the unknown correlation between state estimates of sensors, {\color{black}which essentially hinders the optimality of distributed filters}.  Then the distributed Kalman filters based on consensus or diffusion strategies were studied in \cite{Cat2010Diffusion,Olfati2007Distributed}, where the state estimates  were  fused by scalar weights. As a result, the neighboring information can not be well utilized.  
{\color{black}Owing to the important role of filtering consistency\footnote{The filtering consistency  means that an upper bound of estimation error covariance can be  calculated online.}  in real-time precision evaluation},  
 the consistent algorithms in  \cite{Battistelli2014Kullback,He2017Consistent,wang2018convergence,liu2017kalman} enabled their applications to design covariance weight based fusion schemes, though the system models were limited to be linear and the communications were required to be perfect. { In this paper, we will propose consistent filters for   general system models, communication schemes and network topologies.}

Stability is one of fundamental properties for filtering algorithms.
In the existing results on stability analysis,  local observability conditions of linear   systems  were assumed in \cite{Cat2010Diffusion,Yang2014Stochastic},  which confined the application scope of distributed filters.
%It is known that sensor bias prevalently exists in practice, resulted from many possible factors, such as calibration error,  sensor drift and fault, registration error, the range offset bias, and the positioning bias of moving sensor platforms, etc \cite{geng2018bias}. 
On the other hand,  the sensor observation bias, {\color{black}which  prevalently exists owing to factors like calibration error,  sensor drift, and registration error,}  can directly influence the consistency as well as the stability of filters.
{\color{black} This is attributed to the difficulty in dealing with biased observations of the local sensor and fusing the biased estimates from  neighboring sensors.}
%Hence, it is necessary to provide information metric for the quality of observations, so as to judge whether the observations corrupted by state-correlated bias can be utilized at the update stage or simply discarded.
The state estimation problems in the presence of observation bias were investigated in \cite{caglayan1983separated,Keller1997Optimal} by assuming the independence between the system state and the  random
bias, which yet is hard to be satisfied for feedback control systems with colored random bias processes.
More importantly, owing to the existence of outer disturbances or unmodeled dynamics, many practical systems contain  uncertain dynamics, which may be nonlinear.	To deal with the unknown dynamics, some robust estimation methods, such as $H_{\infty}$ filters and set valued filters, were studied by researchers
\cite{yang2008non,ding2012distributed}.  An  extended state based  Kalman filter was proposed  in \cite{Bai2017reliable} for a class of nonlinear uncertain systems.
{ However, the relation between the original system and the formulated system still needs further investigation.
	Compared with the  centralized filter \cite{Bai2017reliable}, more general system models and noise conditions will be studied in this work under a distributed framework.

{
Communication scheme between sensors is one of the essential features for decentralized algorithms. In the past years,  a considerable number of results have analyzed topology conditions in terms of network connectivity and graph types \cite{Battistelli2016stability,Olfati2007Distributed,Cat2010Diffusion,Battistelli2014Kullback,He2017Consistent,wang2018convergence,liu2018distributed,Yang2014Stochastic}. Most of these results assumed that the network is fixed over time. However, due to the  network vulnerability (e.g., link failure \cite{liu2017kalman}), the topologies of sensor networks may be changing with time. Another significant aspect is on the message transmission between neighboring sensors.
A majority of the existing literature on distributed filters required the accurate transmission. Nevertheless, due to limitations of energy and channels in practical networks, such as wireless sensor network, it is difficult to ensure perfect sensor communications. Thus, the filter design under quantized sensor communications seems to be an important issue of practice.}

 The main contributions of this paper are summarized in the following.
 {\color{black}
\begin{enumerate}
\item By utilizing  the techniques of interval quantization and state extension, we  propose a  quantized communication based distributed Kalman filter    for a class of stochastic systems suffering uncertain dynamics and observation biases. The filter enables the upper bounds of mean square estimation errors to be avaiable online.

\item  Under some mild  conditions including   uniformly collective observability of the system and jointly strong connectivity of the switching networks, we prove that the
mean square  estimation errors are uniformly upper bounded. Furthermore, it is shown that the estimation biases tend to zero under certain decaying conditions of uncertain dynamics and observation biases.

\item 	An event-triggered observation update   based DKF is presented with  a tighter mean square error bound   than  that of the time-driven one. Also, the mean square boundedness of the estimation error for the event-triggered filter is proved.
More importantly, we reveal that the estimation biases of the event-triggered filter  can tend to zero even if the observation biases of some sensors are not decaying over time.  	
\end{enumerate}

Compared with the existing literature \cite{Battistelli2016stability,Olfati2007Distributed,Cat2010Diffusion,Battistelli2014Kullback,He2017Consistent,wang2018convergence,liu2018distributed,Khan2014Collaborative,Khan2010On},
the studied systems  are more general by considering uncertain dynamics and observation biases.
 Furthermore, although there are some results on quantized distributed consensus \cite{kar2010distributed,zhu2011convergence},  the distributed filtering problems with quantized sensor communications have not been well investigated in the existing literature, especially for the scenario that the system is unstable and collectively observable. Moreover, the conditions of  noise in \cite{wang2018convergence}, the initial estimation error in \cite{Battistelli2016stability,reif1999stochastic}, and the non-singularity of time-varying system matrices in \cite{Battistelli2016stability,wang2018convergence} are all relaxed in this paper.

%For ensuring the uniformly observability of the reconstructed system, a sufficient condition was
%given in \cite{Bai2017reliable}, however, this paper provides a sufficient and necessary condition based on the
%original system (see Proposition 4.2).

}

%We note that this work gets some inspiration from \cite{Bai2017reliable} and \cite{He2017Consistent} on state extension and consistency, but the results developed in this work are not trivial or simply extensions of the two papers or other existing literature.

%We should notice that this work gets some inspiration from \cite{Bai2017reliable} and \cite{He2017Consistent}, but the results developed in this work are not trivial or simply extensions of the two papers or other existing literature.
%Compared with the  centralized filter \cite{Bai2017reliable}, the system model of this paper and noise conditions are more general by considering stochastic biased observations.
%Additionally, for ensuring the uniformly collective observability of the reconstructed system, unlike a sufficient condition given in \cite{Bai2017reliable}, this paper provides a sufficient and necessary condition. Besides, the filtering parameters are time-varying not constant \cite{Bai2017reliable}.
%Compared with distributed filters \cite{Battistelli2016stability,Olfati2007Distributed,Cat2010Diffusion,Battistelli2014Kullback,He2017Consistent,liu2018distributed,Khan2014Collaborative,Khan2010On}, this paper poses milder conditions on  system model, communication scheme and topology conditions.
}

The remainder of the paper is organized as follows. Section \uppercase\expandafter{2} is on the model description and preliminaries. Section \uppercase\expandafter{3}  studies the system reconstruction and filtering structure.
Section \uppercase\expandafter{4} analyzes the distributed filter with time-driven update scheme.
Section \uppercase\expandafter{5} studies the distributed filter with event-triggered update scheme.
Section \uppercase\expandafter{6}  shows the numerical simulations. The conclusion of this paper is given in Section \uppercase\expandafter{7}.
Some proofs are given in Appendix.

%\subsection{Notations}
%\begin{table}[htp]
%	\centering
%	\caption{Notations}\label{table_notations}
%	 \setlength{\tabcolsep}{0.1mm}{
%	\begin{tabular}{|c|c|}
%		\hline
%%		$\mathcal{G}_s=(\mathcal{V},\mathcal{E}_{\mathcal{G}_s},\mathcal{A}_{\mathcal{G}_s})$& the $s$th weighted digraph  \\\hline
%%		$\mathcal{V}=\{1,2,\cdots,N\}$ & sensor set  \\\hline
%%		$\mathcal{E}_{\mathcal{G}_s}$& edge set of digraph $\mathcal{G}_s$\\\hline
%%		$\mathcal{A}_{\mathcal{G}_s}=[a_{i,j}^{\mathcal{G}_s}]\in \mathbb{R}^{N\times N}$& weighted adjacency matrix of digraph $s$ \\\hline
%%		$\mathcal{N}_{i}(k)$& the neighbor set of sensor $i$ at time $k$ \\\hline
%%		$O, o$&   \\\hline
%		%&  \\\hline
%		%&  \\\hline
%	\end{tabular}}
%\end{table}

	\textbf{Notations}. The superscript ``T" represents the transpose. $I_{n}$ stands for the identity matrix with $n$ rows and $n$ columns. $E\{x\}$ denotes the mathematical expectation of the stochastic variable $x$, and  $\blockdiag\{\cdot\}$ means that the block elements are arranged in diagonals. $\diag\{\cdot\}$ represents the diagonalization of scalar elements. $\tr(P)$ is the trace of the matrix $P$. $\mathbb{N}^+$ denotes the set of positive natural numbers. $\mathbb{R}^n$ stands for the set of $n$-dimensional real vectors. $[a:b)$ stands for the set of integers $a,a+1,\cdots,b-1$. We denote $[a:b]=[a:b)\cup b$.
   And, $\mathrm{sat}(f,b)$   means $\max\{\min\{f,b\},-b\}$. We assume that $\lambda_{min}(A)$ and $\lambda_{max}(A)$ are the minimal eigenvalue and maximal eigenvalue of a real-valued square matrix $A$, respectively. $\sup_{k\in\mathbb{N}}A_{k}A_{k}^T<\infty$ means  $\sup_{k\in\mathbb{N}}\lambda_{max}\left(A_{k}A_{k}^T\right)<\infty$. {\color{black}Let $\|\cdot\|_2$ be the standard Euclidean norm.}

\section{Model Description and Preliminaries}
%In this section, some graph preliminaries and useful definitions will be provided.
\subsection{Network topology and definitions}
We model the communication topologies of sensor networks by switching weighted digraphs
{\color{black}$\{\mathcal{G}_{s}=(\mathcal{V},\mathcal{E}_{s},\mathcal{A}_{s})\}$,  where $\mathcal{V}$, $\mathcal{E}_{s}$ and $\mathcal{A}_{s}$ stand for the node set,  the edge set  and  the weighted adjacency matrix, respectively.}
%
% $\mathcal{G}=(\mathcal{V},\mathcal{E}_\mathcal{G},\mathcal{A}_\mathcal{G})$.
% $\mathcal{V}=\{1,2,\cdots,N\}$ stands for the node set,  $\mathcal{E}_\mathcal{G}\subseteq \mathcal{V}\times \mathcal{V}$ stands for  the edge set  and $\mathcal{A}_\mathcal{G}=[a_{i,j}]\in \mathbb{R}^{N\times N}$ represents the weighted adjacency matrix.
We assume that {\color{black}$\mathcal{A}_{s}$ is row stochastic  with nonnegative off-diagonal elements and positive diagonal elements,} i.e., $a_{i,i}^{\mathcal{G}_s}> 0,a_{i,j}^{\mathcal{G}_s}\geq 0,\sum_{j\in \mathcal{V}}a_{i,j}^{\mathcal{G}_s}=1$. 
As node $i$ can receive  information from its neighboring sensors, the neighbor set of node $i$ is denoted by $\mathcal{N}_{i}^{\mathcal{G}_s}\triangleq \{j\in\mathcal{V}|a_{i,j}^{\mathcal{G}_s}>0\}$, which  includes node $i$.
%If $a_{i,j}^{\mathcal{G}_s}>0,j\neq i$, there is a link $(i,j)\in \mathcal{E}_{s}$, which means that node $i$ can directly receive the information of node $j$ through the communication channel. In this situation, node $j$ is called the neighbor of node $i$ and
% all the neighbors of node $i$ including itself can be represented by the set $\mathcal{N}_{i}^{\mathcal{G}_s}\triangleq \{j\in\mathcal{V}|(i,j)\in \mathcal{E}_{s}\}\bigcup\{i\}$.
 We denote $a_{i,j}^{\mathcal{G}_s}=a_{i,j}(k)$ and $\mathcal{N}_{i}^{\mathcal{G}_s}=\mathcal{N}_{i}(k)$, if the graph is $\mathcal{G}_s$ at time $k$.
$\mathcal{G}_s$ is called strongly connected if for any pair nodes $(i_{1},i_{l})$, there exists a direct path from $i_{1}$ to $i_{l}$ consisting of edges $(i_{1},i_{2}),(i_{2},i_{3}),\cdots,(i_{l-1},i_{l})$. We say that $\{\mathcal{G}_{s}\}_{s=1}^K$ is jointly strongly connected if the union graph $\bigcup_{s=1}^K\mathcal{G}_s$ is strongly connected, where $K\in\mathbb{N}^+.$

%For the linear time-varying systems with known exact noise statistics, by Kalman filter, the estimation error covariances  can be recursively calculated. However, for the distributed Kalman filters \cite{He2017Consistent,Battistelli2014Kullback}, due to the unknown correlation between state estimates of sensors, the error covariances are usually unaccessible. In order to evaluate the estimation performance, the following definition of consistency is introduced.
The following definitions are needed in this paper.
\begin{definition}(\cite{Julier1997A})\label{def_consistency}
	Suppose that $x_{k}$ is a random vector and $\hat x_{k}$ is the estimate of $x_{k}$. Then the pair ($\hat x_{k},\varPi_{k}$) is said to be \textbf{consistent}  if $E\{(\hat x_{k}-x_{k})(\hat x_{k}- x_{k})^T\}\leq \varPi_{k}.$ An algorithm is of \textbf{consistency}  if it provides consistent pairs ($\hat x_{k},\varPi_{k}$) for all $k\in\mathbb{N}$.
\end{definition}
%We note that the consistency introduced here is different from the one in parameter identification \cite{S1989System}.
%To study the estimation properties of filtering algorithms, we provide the definition in the following .
\begin{definition}
	Let $e_{k,i}$ be the state estimation error of sensor $i$ at time $k$, then the sequence of estimation error covariances $E\{e_{k,i}e_{k,i}^T\}, k\in\mathbb{N}, $ is said to be \textbf{stable} if  $\sup_{k\in\mathbb{N}} E\{e_{k,i}e_{k,i}^T\}<\infty.$ And, the sequence of estimates is said to be \textbf{asymptotically unbiased} if $\lim\limits_{k\rightarrow \infty}E\{e_{k,i}\}=0$.
\end{definition}
%\begin{definition} (Asymptotically unbiased)
%We say the  estimates $\{\hat x_{k}\}$ of $\{x_{k}\}$ are asymptotically unbiased if $\lim\limits_{k\rightarrow \infty}E\{ \hat x_{k}-x_{k}\}=0$.
%\end{definition}

Let $(\Omega,\mathcal{F},P)$ be the basic probability space.  $\mathcal{F}_k$ stands for a filtration of $\sigma$-algebra $\mathcal{F}$, {  i.e., for $\mathcal{F}_k\subset \mathcal{F}$,  $\mathcal{F}_i\subset \mathcal{F}_j$ if $i< j$. Here, the $\sigma$-algebra is a collection of subsets of $\Omega$ and satisfies certain algebraic structure.}    A discrete-time  sequence $\{\xi_k,\mathcal{F}_k\}$ is said to be adapted if $\xi_k$ is measurable to $\mathcal{F}_k$.
{ In principle, $\xi_k$ is a function of past events within $\mathcal{F}_k$.}  We refer the readers to the formal definitions of `filtration', `$\sigma$-algebra' and `measurable'   in \cite{chow2012probability}.

\begin{definition}
	A discrete-time adapted sequence $\{\xi_k,\mathcal{F}_k\}$ is called a \textbf{martingale difference sequence (MDS)}, if $E\{\|\xi_k\|_2\}<\infty$ and $E\{\xi_k|\mathcal{F}_{k-1}\}=0$, almost surely.
\end{definition}
%
%Since this paper studies a class of time-varying  systems, a useful definition is provided.	
{\color{black}
	\begin{definition}	
A time sequence $\{T_{l},l\in\mathbb{N}\}$ is called an \textbf{$L$-step supporting sequence (L-SS)} of a matrix sequence $\{M_{k},k\in\mathbb{N}\}$, if
there exists a scalar $\beta>0$, such that the sequence  $\{T_{l},l\in\mathbb{N}\}$ is well defined in the following manner
		\begin{align}\label{notation_T}
		\begin{cases}
		T_{0}
		=\inf\bigg\{k\geq 0\bigg|\lambda_{min}(M_{k+s}M_{k+s}^T)\geq \beta,\forall s\in[0:L)\bigg\}\\
			T_{l+1}
			=\inf\bigg\{k\geq T_l+L\bigg|\\
			 \qquad\qquad\lambda_{min}(M_{k+s}M_{k+s}^T)\geq \beta,\forall s\in[0:L)\bigg\}\\
			\mathop{\sup}\limits_{l\in \mathbb{N}}\{T_{l+1}-T_{l}\}<\infty.
			%		\inf\left\lbrace k| k\in [T_l+L:T_{l}+T], \lambda_{min}(M_{k+s}M_{k+s}^T)\geq \beta,s\in[0:L)\right\rbrace,\nonumber
		\end{cases}
		\end{align}
\end{definition}
}
%	\begin{definition}	
%		%	 $T_{l+1}=\inf\{k>T_l|\lambda_{min}(A_{k_{l}+s}A_{k_{l}+s}^T)\geq \beta,s=0,\cdots,L-1\}$
%		NEEDS MORE CONCISE:
%Given a positive integer $L$, a matrix sequence  $\{M_{k},k\in\mathbb{N}\}$,
%		if there is a scalar $\beta>0$, such that  the following time sequence $\{T_{l},l\in\mathbb{N}\}$ is well defined
%		\begin{align}
%		&T_{l+1}\label{notation_T}\\
%		=&\inf\left\lbrace k| k\geq T_l+L, \lambda_{min}(M_{k+s}M_{k+s}^T)\geq \beta,s\in[0:L)\right\rbrace,\nonumber
%		\end{align} and there is an integer $T>0$ such that  $T_{l+1}-T_{l}\leq T,$
%		then $\{T_{l},l\in\mathbb{N}\}$ is called \textbf{$L$-step supporting sequence (L-SS)} of $\{M_{k},k\in\mathbb{N}\}$.	
%		\end{definition}}	
\begin{remark}
	The definition of L-SS is introduced  to study the nonsingularity of the time-varying transition matrices $\{\bar A_{k,i}\}$ given in the sequel. In many existing results \cite{Battistelli2016stability,wang2018convergence}, $\bar A_{k,i}$
	is assumed to be  nonsingular for any $  k\in\mathbb{N}$, which is removed in this paper.
\end{remark}
%\section{Model Description and Problem setup}
\subsection{Model description and assumptions}
Consider the following  model for a class of stochastic  systems with uncertain dynamics and biased observations
\begin{equation}\label{system}
\begin{cases}
x_{k+1}=\bar A_{k}x_{k}+\bar G_{k}f(x_k, k)+\bar \omega_{k},\\
y_{k,i}=\bar H_{k,i}x_{k}+b_{k,i}+v_{k,i},\quad i\in\mathcal{V},
\end{cases}
\end{equation}
where $x_{k}\in \mathbb{R}^n$ is the unknown  system state, $\bar A_{k}\in \mathbb{R}^{n\times n}$ is the known state transition matrix and $\bar\omega_{k}\in \mathbb{R}^n$ is the unknown zero-mean white process  noise.
$f(x_k, k)\in \mathbb{R}^p$ is the  uncertain  dynamics (e.g., some unknown disturbance). $\bar G_{k}\in \mathbb{R}^{n\times p}$ is the known matrix subject to $\sup_{k\in\mathbb{N}}\{\bar G_{k}\bar G_{k}^T\}<\infty$.
Here, $y_{k,i}\in \mathbb{R}^{m_i}$ is the observation vector obtained via sensor $i$, $\bar H_{k,i}\in \mathbb{R}^{m_i\times n}$ is the known observation matrix, subject to $\sup_{k\in\mathbb{N}}\{\bar H_{k,i}^T\bar H_{k,i}\}<\infty$, $b_{k,i}\in \mathbb{R}^{m_i}$ is the unknown state-correlated stochastic observation bias of sensor $i$,  and $v_{k,i}\in \mathbb{R}^{m_i}$ is the stochastic zero-mean observation noise.
$N$ is the number of sensors over the system, thus $\mathcal{V}=\{1,2,\dots,N\}$.
{\color{black}Note that  $\bar A_{k}$, $\bar G_{k}$ are known to all the local sensors, while $\bar H_{k,i}$  and $y_{k,i}$ are only known to the local sensor $i$.}
%The above matrices and vectors have compatible dimensions.
%\subsection{Assumptions on system and communication scheme}
%Let $\mathcal{F}_k=\sigma\{F_j, \bar\omega_{j},y_{j,i},i=1,\cdots,N,j\leq k\}$.
%Let $\mathcal{F}_k=\sigma\{F_{j}, \bar\omega_{j},b_{j,i},v_{j,i},i=1,\cdots,N,j\leq k\}$.
{ Suppose that $\sigma\{x\}$ stands for the minimal sigma algebra generated by the random vector $x$.}
Let $\mathcal{F}_k\triangleq \sigma\{x_{0},b_{0,i}, \bar\omega_{j},v_{j,i}, i\in\mathcal{V},0\leq j\leq k\}$ and $f_k\triangleq f(x_k, k)$ for simplicity.
%We denote $\mathcal{S}_k=\mathcal{F}_k\bigcup\sigma\{b_{k+1,i},v_{k+1,i},i=1,\cdots,N,\}$.
In the following, we will provide several assumptions on the system structure and communication scheme.
%On the conditions of noise statistics and stochastic biases, the following assumption is needed.
%\begin{assumption}
%	 On the stochastic biases, it holds that $E\{b_{k,i}b_{k,i}^T\}\leq B_{k,i},i=1,2,\cdots,N$. $\{v_{k,i},\mathcal{F}_k\},i=1,2,\cdots,N,$ are  $N$ martingale difference sequences such that $E\{v_{k,i}v_{k,i}^T\}\leq R_{k,i}$, where $R_{k,i}>0$. $\{\omega_{k},\mathcal{F}_k\}$ is a martingale difference sequence such that $E\{\omega_{k}\omega_{k}^T\}\leq Q_{k}$, where sup$_{k}$$Q_{k}<\infty$.
%%Let $\{b_{k,i},\mathcal{F}_k\},i=1,\cdots,N,$ be $N$ adapted sequences, subject to
%%	, i.e.,
%%	\begin{align*}
%%	E\{v_{k,i}|\mathcal{F}_k\}=0, i=1,2,\cdots,N.
%%	\end{align*}
%\end{assumption}	

{\color{black}
\begin{assumption}\label{ass_noise}
The following conditions hold:
	\begin{itemize}
		\item 1)	The sequence $\{\bar \omega_{k}\}_{k=0}^{\infty}$ is independent of $x_{0}$ and $\{v_{k,i}\}_{k=0}^{\infty}$, $i\in\mathcal{V}$,  with $E\{\bar\omega_{k}\bar\omega_{k}^T\}\leq Q_{k}$, where
		$\inf_{k}Q_{k}>0$ and $\sup Q_{k}<\infty$, $\forall k\in\mathbb{N}$.
%		 $\inf_{k\in\mathbb{N}} Q_{k}\geq \underline Q_0>0$ and sup$_{k\in\mathbb{N}}$$Q_{k}\leq \bar Q_0<\infty$.
		%			\item $\{\bar\omega_{k},\mathcal{S}_k\}$ is a martingale difference sequence subject to $E\{\bar\omega_{k}\bar\omega_{k}^T\}\leq Q_{k}$, where sup$_{k}$$Q_{k}<\infty$.
		\item 2) The sequences $\{b_{k,i}\}_{k=0}^{\infty}$, $i\in\mathcal{V}$, are measurable to $\mathcal{F}_{k-1}$, $k\geq 1,$ and $E\{b_{k,i}b_{k,i}^T\}\leq B_{k,i}$.
	\item 3) The sequence $\{v_{k,i},\mathcal{F}_k\}_{k=0}^{\infty}$ is MDS and $E\{v_{k,i}v_{k,i}^T\}\leq R_{k,i}$  as well as $\sup_{k\in\mathbb{N}} R_{k,i}>0$ holds $\forall i\in\mathcal{V}$.
        \item 4) It holds that $	E\{(X_{0}-\hat X_{0,i})(X_{0}-\hat X_{0,i})^T\}\leq P_{0,i},$	
	where $\hat X_{0,i}$ is the estimate of   $X_{0}\triangleq [x_0^T, f_0^T]^T$, $i\in\mathcal{V}$.
	\end{itemize}

	%	On the stochastic biases, it holds that
	%Let $\{b_{k,i},\mathcal{F}_k\},i=1,\cdots,N,$ be $N$ adapted sequences, subject to
	%	, i.e.,
	%	\begin{align*}
	%	E\{v_{k,i}|\mathcal{F}_k\}=0, i=1,2,\cdots,N.
	%	\end{align*}
\end{assumption}
}
%\begin{remark}
	Compared with \cite{wang2018convergence} where the observation noises of sensors are independent,  	{\color{black}   the MDS	assumption on $v_{k,i}$ is milder.}	The condition 4) of Assumption \ref{ass_noise} is more general than that in \cite{Battistelli2016stability,reif1999stochastic} which  the initial estimation error is required to be sufficiently small.

%\end{remark}
{\color{black}
\begin{assumption}\label{ass_A}
	There is an $L\in\mathbb{N}^+$, such that the sequence $\{\bar A_{k},k\in\mathbb{N}\}$ has an L-SS and $\sup\limits_{k\in\mathbb{N}}\{\bar A_{k}\bar A_{k}^T\}<\infty$.
\end{assumption}
}
%\begin{remark}
	Assumption \ref{ass_A} poses no requirement on the stability of the original system (\ref{system}). Besides, within the scope of distributed filtering for time-varying systems, Assumption \ref{ass_A} is milder than that in \cite{Battistelli2016stability,wang2018convergence}, where the non-singularity of the system state transition matrix is needed for each time.
%\end{remark}		
{\color{black}
\begin{assumption}\label{ass_F}	
	The following conditions hold.
	\begin{itemize}		
		\item 1) $f_k$ is measurable to  $\mathcal{\bar F}_{k} $, where  $\mathcal{\bar F}_{k}=\sigma\{x_{0},b_{0,i},$ $\bar\omega_{j-1},v_{j,i},i\in\mathcal{V},1\leq j\leq k\}$ for $k\geq 1$, and $\mathcal{\bar F}_{0}=\sigma\{x_{0},b_{0,i}, v_{0,i},i\in\mathcal{V}\}$.
		
		\item 2) $E\{u_{k}u_{k}^T\}\leq \hat Q_{k},$ where $u_{k}\triangleq f_{k+1}-f_k$,
		subject to $\inf_{k\in\mathbb{N}} \lambda_{min}(\hat Q_{k})>0$ and $\sup_{k\in\mathbb{N}}\lambda_{max}(\hat Q_{k})< \infty$.
	\end{itemize}
\end{assumption}}

%\begin{remark}

%\end{remark}
%\begin{remark}
The first condition of Assumption \ref{ass_F} permits $f_k$ to be  implicitly related with $\{x_{j},y_{j,i}\}_{j=0}^{k},i\in\mathcal{V}$.  Under this setting, the  model built in (\ref{system}) also considers the distributed  output feedback control systems, such as the system  of coupled tanks \cite{orihuela2016suboptimal}.
The condition in 2) of Assumption \ref{ass_F} relies on the boundedness of the dynamic increment, which   is milder than the boundedness of uncertain dynamics required  by \cite{Cai2006Robust}.
%	It is a mild condition for the systems with nonlinear uncertain dynamics. 	
%\end{remark}
%			\begin{assumption}\label{ass_sys}
%				The system (\ref{system3}) satisfies the following conditions
%				\begin{equation}
%				\begin{split}
%				& E\{\omega_{k}\omega_{k}^T\}\leq Q_{k},\\
%				&E\{v_{k,i}v_{k,i}^T\}\leq R_{k,i},i=1,2,\cdots,N,\\\
%				& E\{u_{k}^2(j)\}\leq q_{k}(j), j=1,2,\cdots,p,
%				\end{split}
%				\end{equation}
%				where $R_{k,i}>0$ and $q_{k}(j)>0$. Also, $Q_{k}$ and $q_{k}(j)$ are uniformly upper bounded.
%			\end{assumption}

%On Assumption \ref{ass_F}, we have the following remark.

{
	\begin{definition}\label{def_obser}
		The system \eqref{system} is said to be \textbf{uniformly collectively observable} if there exist two positive integers $\bar N$, $\bar M$, and a constant $\alpha>0$ such that for any $k\geq \bar M$, there is $\sum_{i=1}^{N}\left[\sum_{j=k}^{k+\bar N}\bar\Phi^T_{j,k} \bar H_{j,i}^T(R_{j,i}+B_{j,i})^{-1} \bar H_{j,i}\bar\Phi_{j,k}\right]\geq \alpha I_n,$
%		\begin{equation}\label{eq_obser}
%		\sum_{i=1}^{N}\left[\sum_{j=k}^{k+\bar N}\bar\Phi^T_{j,k} \bar H_{j,i}^T(R_{j,i}+B_{j,i})^{-1} \bar H_{j,i}\bar\Phi_{j,k}\right]\geq \alpha I_n,
%		\end{equation}	
		where $\bar\Phi_{k,k}=I_{n},\bar\Phi_{k+1,k}=\bar A_{k}$, $\bar\Phi_{j,k}=\bar\Phi_{j,j-1}\times\cdots \times\bar\Phi_{k+1,k},$ if $j>k.$
	\end{definition}
	
\begin{assumption}\label{ass_observable}
The system \eqref{system} is uniformly collectively observable.
\end{assumption}
}
	Assumption \ref{ass_observable} is a mild collective observability condition for time-varying stochastic systems. If the system is time-invariant, then Assumption \ref{ass_observable} degenerates to $(\bar A,\bar H)$ being observable \cite{Battistelli2014Kullback,He2017On}, where $\bar H=[\bar H_1^T,\dots,\bar H_N^T]^T$. Besides, if the local observability conditions are satisfied \cite{Cat2010Diffusion,Yang2014Stochastic}, then Assumption \ref{ass_observable}  holds, but not vice versa.

The topologies of the networks are assumed to be switching digraphs $\{\mathcal{G}_{\sigma_{k}},k\in\mathbb{N}\}$.
$\sigma_{k}$ is the switching signal defined by $\sigma_{k}:\mathbb{N}\rightarrow \Omega$, where $\Omega$ is the set of the underlying network topology numbers. For convenience, the weighted adjacency matrix of the digraph $\mathcal{G}_{\sigma_{k}}$ is denoted by $\mathcal{A}_{\sigma_{k}}=[a_{i,j}(k)]\in \mathbb{R}^{N\times N}$.
%In this paper, the topologies of the networks are assumed to be switching digraphs $\{\mathcal{G}_{l},l\in[1:L]\}$, where  $\mathcal{G}_{l}=(\mathcal{V},\mathcal{E}_{l},\mathcal{A}_{l})$, which consists of  the set of nodes $\mathcal{V}=\{1,2,\cdots,N\}$, the set of edges $\mathcal{E}_{l}\subseteq \mathcal{V}\times \mathcal{V}$ and the weighted adjacency matrix $\mathcal{A}_{l}=[a_{i,j}(l)]$.
%Since the interconnected topology of the network may vary with time, it is necessary to consider all possible digraphs $\{\mathcal{G}_{l},l\in[1:L]\}$ on the node set $\mathcal{V}$. Define the graph switching signal $\sigma_{k}:[0,\infty)\rightarrow[1:L]$, which is piecewise constant variable. Thus, one can obtain the underling digraph  $\mathcal{G}_{\sigma_{k}}$ at $k$th moment.
To analyze the switching topologies, we consider the infinite interval sequence consisting of  non-overlapping and contiguous time intervals $[k_{l},k_{l+1}),l=0,1,\dots,$ with $k_{0}=0$. 	It is assumed that there exists an integer $k_0$, such that $k_{l+1}-k_{l}\leq k^0$.  On the switching topologies, the following assumption is needed.
{\color{black}\begin{assumption}\label{ass_primitive}
		The union graph { $\bigcup_{k=k_{l}}^{k_{l+1}}\mathcal{G}_{\sigma_{k}}$ is strongly connected,} and the elements of $\mathcal{G}_{\sigma_{k}}$, i.e., $a_{i,j}(k)$, belong to  a set consisting of finite  nonnegative real numbers.
		%	, i.e., the union of the digraphs in the interval $[k_{l},k_{l+1}-1]$ is strongly connected.\\
		%	2)The elements of the weighted adjacency matrix $a_{i,j}(s)\in\bar\varPsi$? where $\bar\varPsi$ is a finite set of arbitrary nonnegative numbers.
\end{assumption}}
%\begin{remark}
Since the joint connectivity of the switching digraphs admits that the network is unconnected at each moment, Assumption \ref{ass_primitive} is  milder for the networks confronting  link failures. If the network remains connected at each moment or fixed \cite{Battistelli2014Kullback,He2017Consistent}, then Assumption \ref{ass_primitive} holds as well.
%\end{remark}
%On Assumptions \ref{ass_noise}-\ref{ass_primitive} given above, we provide the following remarks.

\subsection{Quantized communications}

{
%	\subsection{Quantized Communications over Sensors}
	In sensor networks, such as wireless sensor networks, the  signal transmission between two sensors may confront the problems of channel limitation and  energy restriction. Thus, without losing too much accuracy, some quantization operators can be considered to reduce the package size in the encoding process with respect to the transmitted messages.
	In this paper, we study the case  that the messages to be transmitted are quantized element by element through a given quantizer before transmission.	
	Let $z_{k,i}$  be a scalar element remaining to be sent out by sensor $i$, then we consider the following interval quantizer with quantizing step $\Delta_i>0$ and quantizing function $g(\cdot):\mathbb{R}\longrightarrow \mathcal{Q}_i$
	\begin{align}\label{eq_quantization}
g(z_{k,i})=m\Delta_i, \text{ if $(m-\frac{1}{2})\Delta_i\leq z_{k,i}< (m+\frac{1}{2})\Delta_i$}
	\end{align}
	where $\mathcal{Q}_i=\{m\Delta_i|m\in\mathbb{Z}\}$ is the quantization alphabet with countably infinite elements.
Then we write $g(z_{k,i})=z_{k,i}+\vartheta(z_{k,i}),$
%\begin{align}\label{eq_quantization_error}
%
%\end{align}
where $\vartheta(z_{k,i})$ is the quantization error, which is deterministic conditioned on the input $z_{k,i}$. In addition, we have $\vartheta(z_{k,i})\in[-\frac{\Delta_i}{2},\frac{\Delta_i}{2})$.
A technique to deal with the correlation between $\vartheta(z_{k,i})$ and $z_{k,i}$ is to consider a dither operator by adding a random sequence $\{\xi_{k,i}\}$, which can randomize the quantization error and make it independent of the input data.
Write
\begin{align}\label{eq_dither}
\vartheta(k,i)=g(z_{k,i}+\xi_{k,i})-(z_{k,i}+\xi_{k,i}),
\end{align}
 then the quantization error sequence  $\{\vartheta(k,i)\}$ is  independent and identically distributed (i.i.d.), uniformly distributed on $[-\frac{\Delta_i}{2},\frac{\Delta_i}{2})$ and independent of $z_{k,i}$, if the following assumption holds.
\begin{assumption}\label{ass_dither}
The sequence $\{\xi_{k}\}_{k\geq 0}$ satisfies Schuchman conditions \cite{kar2010distributed} and it is independent of $\mathcal{\bar F}_{k}$ defined in Assumption \ref{ass_F}.
\end{assumption}
A sufficient condition such that   $\{\xi_{k}\}_{k\geq 0}$ satisfies Schuchman conditions is  that
$\{\xi_{k,i}\}_{k\geq 0}$ is an i.i.d. sequence uniformly distributed on $[-\frac{\Delta_i}{2},\frac{\Delta_i}{2})$ and independent of  input sequence $\{z_{k,i}\}$.
%Let $\{\xi_{k,i}\}_{k\geq 0}$ be  .

%In a distributed filtering problem, each sensor may transmit its state estimate vector, some parameter vectors or matrices. They all can be quantized element by element through the above defined quantizing function and dither  before transmission. 	We also notice that different sensors may  have different quantizing steps, which enables a distributed algorithm design.	
%	The dither operator considered in this paper can remove the correltion between $z_{k,i}$ and the quantization error $\vartheta(k,i)$, however, more uncertainties should  be considered due to the addition of the random variable sequences $\{\xi_{k,i}\}_{k\geq 0}$.
%\begin{remark}
%	In the existing literature related with quantized sensor communications, there are some results on distributed consensus \cite{kar2010distributed,zhu2011convergence}. However,  the distributed filtering problems with quantized sensor communications have not been well investigated, especially for the scenario that the system is unstable and collectively observable.
%\end{remark}

}

\section{System Reconstruction and Filtering Structure}
\subsection{System reconstruction}
By constructing   a new state vector, consisting of the original state $x_{k}$ and the  uncertain dynamics $f_{k}$,  a modified system model  is given as follows.
\begin{equation}\label{system3}
\begin{cases}
X_{k+1}= A_{k}X_{k}+Du_{k}+\omega_{k}\\
y_{k,i}= H_{k,i}X_{k}+b_{k,i}+v_{k,i}
\end{cases}
\end{equation}
where
\begin{align*}
%		\begin{cases}
&X_{k}\triangleq \begin{pmatrix}
x_{k} \\
f_{k}
\end{pmatrix}\in\mathbb{R}^{n+p},
\quad A_{k}\triangleq \begin{pmatrix}
\bar A_{k} & \bar G_{k} \\
0 & I_p
\end{pmatrix}\in\mathbb{R}^{(n+p)\times (n+p)},\\
&\omega_{k}\triangleq\begin{pmatrix}
\bar \omega_{k} \\
0
\end{pmatrix}\in\mathbb{R}^{n+p},\quad D\triangleq \begin{pmatrix}
0 \\
I_p
\end{pmatrix}\in\mathbb{R}^{(n+p)\times p},\\
& u_{k}\triangleq f_{k+1}-f_k\in\mathbb{R}^{ p},\quad H_{k,i}\triangleq \begin{pmatrix}
\bar H_{k,i} & 0
\end{pmatrix}\in\mathbb{R}^{m_i\times (n+p)}.
%		\end{cases}
\end{align*}
Considering the system (\ref{system}) and the reformulated system (\ref{system3}), Propositions \ref{lem_iff} and \ref{lem_observable}, {\color{black}which are proved in Appendices \ref{app_iff}-\ref{pf_observ}, show the equivalence between the properties of the two systems}.
\begin{proposition} \label{lem_iff}
	On the relation between  $\bar A_{k}$ in (\ref{system}) and $ A_{k}$ in (\ref{system3}), the following conclusions hold.	
	\begin{itemize}
		\item \text{1)} $\sup_{k\in\mathbb{N}}\{ A_{k} A_{k}^T\}<\infty$ if and only if $\sup_{k\in\mathbb{N}}\{\bar A_{k}\bar A_{k}^T\}<\infty$.	
		\item\text{2)} $\{A_{k}|A_{k}\in\mathbb{R}^{(n+p)\times(n+p)},k\in\mathbb{N}\}$ has an L-SS if and only if $\{\bar A_{k}|\bar A_{k}\in\mathbb{R}^{n\times n},k\in\mathbb{N}\}$ has an L-SS.
	\end{itemize}
	%2) $\sup_k\{\bar A_{k}\bar A_{k}^T\}<\infty$ if and only if $\sup_k\{ A_{k} A_{k}^T\}<\infty$.
	%
	%
	%1) $\{A_{k}|A_{k}\in\mathbb{R}^{n+p},k\in\mathbb{N}\}$ has an L-SS if and only if $\{\bar A_{k}|\bar A_{k}\in\mathbb{R}^{n},,k\in\mathbb{N}\}$ has an L-SS.
\end{proposition}
%\begin{pf}
%	See Appendix \ref{app_iff}.
%\end{pf}
{
\begin{proposition}\label{lem_observable}
	The reformulated system (\ref{system3}) is  uniformly collectively observable if and only if there exist  $ M, \bar N\in\mathbb{N}^+$ and $\alpha>0$, such that for any $k\geq M$, 	
	\begin{align}\label{eq_iff}
\begin{split}
&\bar\Theta_{k,\bar N}^{1,1}>\alpha I_{n}\\
&\bar\Theta_{k,\bar N}^{2,2}-( \bar\Theta_{k,\bar N}^{1,2})^T(\bar\Theta_{k,\bar N}^{1,1}-\alpha I_{n})^{-1} \bar\Theta_{k,\bar N}^{1,2}>\alpha I_{p}
\end{split}
	\end{align}
where $\bar\Theta_{k,\bar N}^{1,1},\bar\Theta_{k,\bar N}^{1,2},$ and $\bar\Theta_{k,\bar N}^{2,2}$ are given in \eqref{eq_notation}.
% , i.e.,
%	there exist two positive integers $\bar N$, $\bar M$, and a constant $\alpha>0$ such that for any $k\geq \bar M$, there is
%	$\sum_{i=1}^{N}\left[\sum_{j=k}^{k+\bar N}\Phi^T_{j,k} H_{j,i}^T(R_{j,i}+B_{j,i})^{-1} H_{j,i}\Phi_{j,k}\right]\geq \alpha I_{n+p},$
%%	\begin{equation*}
%%	\sum_{i=1}^{N}\left[\sum_{j=k}^{k+\bar N}\Phi^T_{j,k} H_{j,i}^T(R_{j,i}+B_{j,i})^{-1} H_{j,i}\Phi_{j,k}\right]\geq \alpha I_{n+p},
%%	\end{equation*}	
%	where $\Phi_{j,k}=\Phi_{j,j-1}\cdots \Phi_{k+1,k}(j>k),$ 	$\Phi_{k,k}=I_{n+p}$, and $\Phi_{k+1,k}=A_{k}$,
\end{proposition}
According to the proof of Proposition \ref{lem_observable} and Definition \ref{def_obser}, we can obtain the following lemma.
\begin{lemma}\label{lem_obser}
	Let  Assumption \ref{ass_observable} holds, then  there exist  $ M, \bar N\in\mathbb{N}^+$ and $\alpha>0$, such that for any $k\geq M$, 	$\bar\Theta_{k,\bar N}^{1,1}>\alpha I_{n}$, where $\bar\Theta_{k,\bar N}^{1,1}$ is given in \eqref{eq_notation}.
\end{lemma}
{\color{black}\begin{remark}\label{rem_rank}
		By Proposition \ref{lem_observable} and Lemma \ref{lem_obser}, the observability gap between the reformulated system (\ref{system3}) and the system \eqref{system} is the second equation in \eqref{eq_iff}, which will reduce to $\mathrm{rank}\left(\begin{smallmatrix}
		I_n-\bar A&\,-\bar G\\
		\bar H&\,0
		\end{smallmatrix}\right)=n+p$ if the system is time-invariant and observable, where $\bar H=[\bar H_1^T,\dots,\bar H_N^T]^T$, $\bar H_{i}\in\mathbb{R}^{m_i\times n}$, $\bar A\in\mathbb{R}^{n\times n}$ and $\bar G\in\mathbb{R}^{n\times p}$.
		\end{remark}}

\textbf{Example}\\
To show the feasibility of \eqref{eq_iff}, we consider a connected network with three sensors.
{\color{black}Suppose that $b_{k,i}=\mathrm{sat}\left(\sin\left(x_{k}^2(1)+x_{k}^2(2)\right),1\right).$}
 For the system \eqref{system}, we let $\bar A=\left(\begin{smallmatrix}
1 & 1 \\
0 & 2
\end{smallmatrix}\right)$, $\bar G=[1,1]^T$, $\bar H_1=[1, 0],$ $\bar H_2=[0, 1],$ $\bar H_3=[1, 1],$ $R_i=1,B_i=1,i=1,2,3.$  On one hand, according to \eqref{eq_iff} and the notations in the proof of Proposition \ref{lem_observable}, by choosing $\alpha=2$ and $\bar N=10$, we have $\lambda_{min}(\bar\Theta_{k,\bar N}^{1,1}-\alpha I_{n})>5.77$, and $\lambda_{min}\left(\bar\Theta_{k,\bar N}^{2,2}-( \bar\Theta_{k,\bar N}^{1,2})^T(\bar\Theta_{k,\bar N}^{1,1}-\alpha I_{n})^{-1} \bar\Theta_{k,\bar N}^{1,2}-\alpha I_{p}\right)> 1.2\times 10^{6}$. On the other hand,  by Remark \ref{rem_rank}, we have $\mathrm{rank}\left(\begin{smallmatrix}
I_n-A&\,-G\\
\bar H&\,0
\end{smallmatrix}\right)=3$. Then due to $\mathrm{rank}\left(\begin{smallmatrix}
\bar H\\
\bar H\bar A
\end{smallmatrix}\right)=2$, the condition \eqref{eq_iff} holds.
%\begin{remark}
%	Note that (\ref{eq_obser}) is required to be met not
%\end{remark}

In the sequel,  we require the following assumption.
{\color{black}\begin{assumption}\label{ass_observable2}
		There exist  $ M, \bar N\in\mathbb{N}^+$ and $\alpha>0$, such that for any $k\geq M$, 	the second inequality in \eqref{eq_iff} holds.
\end{assumption}
It is straightforward to prove that if Assumptions \ref{ass_observable} and \ref{ass_observable2} are both satisfied, there is a common $\alpha>0$, such that the two inequalities in \eqref{eq_iff}  holds simultaneously. By Proposition \ref{lem_iff}, Assumptions \ref{ass_observable} and \ref{ass_observable2} can lead to the uniformly  collective observability of the reformulated system (\ref{system3}).}
}

%\begin{assumption}\label{ass_A}
%	There exist a sequence set $\mathcal{K}=\{k_{l}\}$, a integer $L\geq N+\bar N$ and two scalars $\beta_{1}\geq\beta_{2}>0$, such that
%	\begin{itemize}
%		\item $\sup_{l\geq1} (k_{l+1}-k_{l})<\infty$
%		\item $\inf_{l\geq1} (k_{l+1}-k_{l})>0$
%		\item $\lambda_{max}(A_{k}A_{k}^T)\leq\beta_{1}$, $\forall k\geq0$  and $ \lambda_{min}(A_{k_{l}+s}A_{k_{l}+s}^T)\geq\beta_{2},\forall k_{l}\in \mathcal{K},s=0,\cdots,L-1$.
%	\end{itemize}
%\end{assumption}
%For the time-varying systems, Assumption \ref{ass_A} first proposed in our previous work \cite{He2017Consistent}  is  milder than existing many results which requires the system matrices are non-singular at each moment \cite{Battistelli2015Consensus,Battistelli2016stability,He16con,wang2018convergence}.

%\begin{assumption}\label{ass_primitive}
%	(Topology) The communication topology of the network is  jointly connected.
%\end{assumption}

\subsection{Filtering structure}
In this paper, we consider two observation update schemes, namely, time-driven update and event-triggered update, whose difference lies in whether the biased noisy observation $y_{k,i}$ is utilized at the update stage or not.
We propose the following distributed filter structure of the system (\ref{system3}) for sensor $i$, $\forall i\in \mathcal{V}$,
{
\begin{equation}\label{filter_stru}
%	\text{\scriptsize{Filter Structure}}
\begin{cases}
\bar X_{k,i}=A_{k-1}\hat X_{k-1,i}\\
\text{if $y_{k,i}$ is utilized:}\\
\quad\tilde X_{k,i}=\bar X_{k,i}+K_{k,i}(y_{k,i}- H_{k,i}\bar X_{k,i})\\
\text{if $y_{k,i}$ is discarded:}\\
\quad\tilde X_{k,i}=\bar X_{k,i}\\
\hat X_{k,i}=\sum_{j\in \mathcal{N}_{i}(k)}W_{k,i,j}\mathcal{\tilde X}_{k,j}
\end{cases}
\end{equation}}
{where $\mathcal{\tilde X}_{k,j}=\tilde X_{k,i},$ if $i=j$, otherwise $\mathcal{\tilde X}_{k,j}=\mathbf{G}_1(\tilde X_{k,j}), j\in\mathcal{N}_{i}$,  and
	$\mathbf{G}_1(\tilde X_{k,j})=:[g(\tilde X_{k,j}(1)),\dots,$\\$g(\tilde X_{k,j}(n))]^T$ is the  quantized message sent out by sensor $j$ with element-wise quantization  by \eqref{eq_quantization}.  }
Also, $\bar X_{k,i}$,  $\tilde X_{k,i}$,  and $\hat X_{k,i}$ are the extended state's prediction,  update, and  estimate by  sensor $i$ at the $k$th moment, respectively.
$K_{k,i}$ and $W_{k,i,j},j\in \mathcal{N}_{i}(k)$, are the filtering gain matrix and the local fusion matrices, respectively.
They remain to be designed.
%Additionally,
%\begin{equation}\label{eq_u}
%\begin{split}
%&\hat u_{k-1,i}=(\hat u_{k-1,i}(1),\dots,\hat u_{k-1,i}(p))^T\\
%&\hat u_{k-1,i}(j)=sat\big(\hat {\bar u}_{k-1,i}(j),\sqrt{ q_{k-1}(j)}\big)\\
%&\hat {\bar u}_{k-1,i}(j)=\hat{ f}_{k-1,i}(j)-\hat{f}_{k-2,i}(j),j=1,2,\dots,p
%\end{split}
%\end{equation}
%where { $\hat u_{k-1,i}(s)$ is $s$th element of  $\hat u_{k-1,i}$,  $\hat{f}_{k,i}$ is the estimate of uncertain dynamics $f_{k,i}$.  Note that $q_{k-1}$ is given in Assumption \ref{ass_F}.} It is noted that the saturation function $sat(\cdot,\cdot)$ is utilized to guarantee the boundedness of $\hat u_{k-1,i}$.

{
\begin{remark}
	We utilize the dither quantization of state estimates in \eqref{filter_stru} to relax the conservativeness in handling correlation between the quantization error and  state estimates.
	 For example, if random variables $x$ and $y$ are correlated, a common technique is to use $E\{(x+y)^2\}\leq (1+\frac{1}{\alpha})E\{x^2\}+(1+\alpha)E\{y^2\}$ for $\alpha>0.$ However, this operator is usually conservative. Thus, we use the dithered quantization to remove the correlation.
\end{remark}
}

{\color{black}
In the sequel, we will study the consistency, stability and asymptotic unbiasedness of the proposed distributed filters under Assumptions 2.1-3.1. Table \ref{table_con} is to provide the connection between the main results and assumptions.
\begin{table}[htbp]%%%%%if use table*, it will cover two colums
%	\centering  % 显示位置为中间
	{\color{black}
		\caption{Connection between results and assumptions}  % 表格标题
		\label{table_con}  % 用于索引表格的标签
		\scalebox{1}{
		{\renewcommand{\arraystretch}{1.0} %<- modify value to suit your needs
			\begin{tabular}{|c|c|}% 通过添加 | 来表示是否需要绘制竖线
				\hline  % 在表格最上方绘制横线
				Assumptions  & Results\\
				\hline
                Assumptions 2.1, 2.3, 2.6 & Lemma 4.1/5.2 (Consistency) \\
				\hline
				Assumptions 2.1-2.3 & Lemma 4.3 \\
					\hline
				Assumptions 2.1-2.6, 3.1 & Theorem  4.1/5.1(Stability) \\
									\hline
				Assumptions 2.1-2.6, 3.1 & Theorem  4.2/5.2(Unbiasedness) \\
				%在第一行和第二行之间绘制横线
				\hline % 在表格最下方绘制横线
			\end{tabular}
		}
	}
}
\end{table}
}

%	The objective of this paper is to 1) construct scalable distributed recursive filters under a typical observation update protocol with guaranteed consistency to estimate the extended state $X_{k}$, and 2) analyze the boundedness of the estimation covariance for each sensor under the above mild conditions, and 3) study the distributed filtering problem based a new observation update method.

%Denote the estimation error of sensor $i$ as $e_{k,i}\triangleq \hat x_{k,i}-x_{k}$.

%Since $F_{k}$  can be related to $x_{k}$,  the correlation between them should not be ignored. Under the weak observability condition, i.e., the local sensor with its own observations can not achieve the effective estimation, it is more difficult to deal with the distributed estimation problem over time-varying topologies.

%\section{Filter Design}
\section{Distributed filter: time-driven update}
In this section, for the filtering structure (\ref{filter_stru}) with $y_{k,i}$ employed at each time, we will study the design methods of $K_{k,i}$ and $W_{k,i,j}$. Then we will find the conditions to guarantee the stability of the estimation error covariances for the proposed filter with the designed $K_{k,i}$ and $W_{k,i,j}$.

%	we will propose a distributed filtering structure and study the design methods for the structure, so as to raise the extended state distributed Kalman filter of this paper. Besides, we will study the performance of this filter in terms of the boundedness of estimation covariance.

\subsection{Filter design}
The next lemma, {\color{black} proved in Appendix \ref{pf_1},} provides a design method of fusion matrices $\{W_{k,i,j}\}_{j\in\mathcal{N}_i}$, which can lead to the consistent estimate of each sensor.
\begin{lemma}\label{thm_W}
	%	For the distributed filter structure (\ref{filter_stru}) with initial setting $\bar x_{0|0,i}=0$, $P_{0,i}=P_0$,
{\color{black}	Consider the filtering structure (\ref{filter_stru}) with   a  non-stochastic filtering gain $K_{k,i}$}.
	Under Assumptions \ref{ass_noise}, \ref{ass_F} and \ref{ass_dither},  for any $i\in\mathcal{V}$, positive scalars $\theta_{k,i}$ and $\mu_{k,i}$, the pairs ($\bar X_{k,i},\bar P_{k,i}$),($\tilde X_{k,i},\tilde P_{k,i}$), ($\mathcal{\tilde X}_{k,i},\mathcal{\tilde P}_{k,i}$) and ($\hat X_{k,i},P_{k,i}$) are all consistent, %i.e.,
	%		\begin{equation}\label{thm_consis}
	%		\begin{cases}
	%		E\{(\bar X_{k,i}- X_{k})(\bar X_{k,i}- X_{k})^T\}\leq \bar P_{k,i}\\
	%		E\{(\tilde X_{k,i}-X_{k})(\tilde X_{k,i}-X_{k})^T\}\leq \tilde P_{k,i}\\
	%		E\{(\hat X_{k,i}-X_{k})(\hat X_{k,i}-X_{k})^T\}\leq P_{k,i},
	%		\end{cases}
	%		\end{equation}	
	provided by 	{  $	W_{k,i,j}=a_{i,j}(k)P_{k,i}\mathcal{\tilde P}_{k,j}^{-1} $,
		where
	\begin{equation*}
	\begin{split}
\mathcal{\tilde P}_{k,j}=\begin{cases}
	\tilde P_{k,j}, \qquad\qquad\qquad\qquad\text{ if $i=j$}, j\in\mathcal{N}_i(k)\\
\check P_{k,j}+\frac{\bar n\Delta_j(2\Delta_j+1)}{2}I_{\bar n},  \quad \text{ if $i\neq j$}, j\in\mathcal{N}_i(k),
	\end{cases}
	\end{split}
	\end{equation*}
and $\bar P_{k,i},\tilde P_{k,i}$, $\check P_{k,i}$ and $P_{k,i}$ are recursively calculated through
	\begin{equation*}
	\begin{cases}
	\bar P_{k,i}=(1+\theta_{k,i}) A_{k-1}P_{k-1,i} A_{k-1}^T+ \frac{1+\theta_{k,i}}{\theta_{k,i}} \bar Q_{k-1}+\tilde Q_{k-1},\\
	\tilde P_{k,i}=(1+\mu_{k,i})(I-K_{k,i} H_{k,i})\bar P_{k,i}(I-K_{k,i} H_{k,i})^T\\
	\qquad\quad+	K_{k,i}\bigg(R_{k,i}+\frac{1+\mu_{k,i}}{\mu_{k,i}}B_{k,i}\bigg)K_{k,i}^T,	\\
	\check P_{k,i}=\mathbf{G_2}(\tilde P_{k,i}),\\
	P_{k,i}=\bigg(\sum_{j\in \mathcal{N}_{i}(k)}a_{i,j}(k) \mathcal{\tilde P}_{k,j}^{-1}\bigg)^{-1},
	\end{cases}	
	\end{equation*}
}
	with $\bar  n=n+p$,
	$\tilde Q_{k-1}=\blockdiag\{Q_{k-1},0^{p\times p}\}, \bar Q_{k-1}=D\hat Q_{k-1}D^T.$  Here, $\mathbf{G_2}(\tilde P_{k,i})$ is the element-wise quantization operator by \eqref{eq_quantization} without dither.
\end{lemma}
{
\begin{remark}
	In some distributed filtering algorithms \cite{Battistelli2014Kullback,He2017Consistent}, besides state estimates,  parameter matrices are also transmitted  between neighboring sensors. In this work, the quantized messages  $\{\check X_{k,j},\check P_{k,j}\},$ will be received by sensor $i$ from its neighbor sensor $j$, $ j\neq i$.
%	we also require the transmission of  parameter matrices (i.e., $\tilde  P_{k,j}$), whose  quantizations are conducted by \eqref{eq_quantization} to obtain $\check P_{k,j}$. 
	To keep the consistency in Definition \ref{def_consistency}, the quantization error of the parameter matrix is compensated by  $\mathcal{\tilde P}_{k,j}=\check P_{k,j}+\frac{\bar n\Delta_j(2\Delta_j+1)}{2}I_{\bar n}$.
\end{remark}
}

{\color{black}The next lemma, proved in Appendix \ref{pf_K}, considers the design of  the filtering gain matrix $K_{k,i}$. }
\begin{lemma}\label{thm_K}
	Solving the optimization problem $K_{k,i}^*=\arg\min_{K_{k,i}}\tr(\tilde P_{k,i})$
	yields
	\begin{align*}
K_{k,i}^*=\bar P_{k,i} H_{k,i}^T\bigg( H_{k,i}\bar P_{k,i}H_{k,i}^T+\frac{R_{k,i}}{1+\mu_{k,i}}+\frac{B_{k,i}}{\mu_{k,i}}\bigg)^{-1}.
	\end{align*}
%	\begin{equation*}\label{eq_design_para2}
%	\begin{split}
%	&K_{k,i}^*=\bar P_{k,i} H_{k,i}^T\bigg( H_{k,i}\bar P_{k,i}H_{k,i}^T+\frac{R_{k,i}}{1+\mu_{k,i}}+\frac{B_{k,i}}{\mu_{k,i}}\bigg)^{-1}.
%	\end{split}
%	\end{equation*}
\end{lemma}
Summing up the results of Lemmas \ref{thm_W} and  \ref{thm_K}, the extended state based distributed Kalman filter (ESDKF) with quantized communications is provided in Algorithm \ref{ODKF2}.

\begin{algorithm}[htp]
	\caption{Extended State Based Distributed Kalman Filter (ESDKF):}
	\label{ODKF2}
%	\begin{algorithmic}
		{\textbf{Prediction:} Each sensor carries out a prediction operation}\\
		$\bar X_{k,i}=A_{k-1}\hat X_{k-1,i},$\\  % \hline ?????????
		$\bar P_{k,i}=(1+\theta_{k,i}) A_{k-1}P_{k-1,i} A_{k-1}^T+ \frac{1+\theta_{k,i}}{\theta_{k,i}} \bar Q_{k-1}+\tilde Q_{k-1},\forall \theta_{k,i}>0.$   \\      % \\ ????????
		where $\bar Q_{k-1}$ and $\tilde Q_{k-1}$ are given in  Lemma \ref{thm_W}.
				\vspace{0.1cm}\\
	 {\textbf{Update:} Each sensor uses its own observations to update the estimation}\\
		$\tilde X_{k,i}=\bar X_{k,i}+K_{k,i}(y_{k,i}-H_{k,i}\bar X_{k,i})$\\        % & ???????
		$K_{k,i}=\bar P_{k,i}H_{k,i}^T\bigg(H_{k,i}\bar P_{k,i}H_{k,i}^T+\frac{R_{k,i}}{1+\mu_{k,i}}+\frac{B_{k,i}}{\mu_{k,i}}\bigg)^{-1}$\\
		$\tilde P_{k,i}=(1+\mu_{k,i})(I-K_{k,i}H_{k,i})\bar P_{k,i}$.\vspace{0.1cm}\\
			{
	 {\textbf{Quantization:} Each sensor uses element-wise dither quantization to $\tilde X_{k,i}$ and interval quantization to $\tilde P_{k,i}$}, i.e.,
	$\check X_{k,i}=\mathbf{G}_1(\tilde X_{k,i})$, $	\check P_{k,i}=\mathbf{G_2}(\tilde P_{k,i})$,
	where $\mathbf{G_1}(\cdot)$ and $\mathbf{G_2}(\cdot)$ are the element-wise quantization operators with and without dither, respectively. 	\vspace{0.1cm} \\
		{\textbf{Local Fusion:} Each sensor receives ($\check X_{k,i}$, $\check P_{k,i}, \Delta_j$ )  from its neighbors, denoting $\bar n=n+p,$ for $j\in\mathcal{N}_i(k),$}\\
					\begin{equation*}
		\begin{split}
		(\mathcal{\tilde X}_{k,j},\mathcal{\tilde P}_{k,j})=\begin{cases}
		(\tilde X_{k,j},\tilde P_{k,j}), \qquad\qquad\text{ if $i=j$}, \\
		(\check X_{k,i},\check P_{k,j}+\frac{\bar n\Delta_j(2\Delta_j+1)}{2}I_{\bar n}),   \text{otherwise}
		\end{cases}
		\end{split}
		\end{equation*}
		$\hat X_{k,i}=P_{k,i}\sum\limits_{j\in \mathcal{N}_{i}(k)}a_{i,j}(k) \mathcal{\tilde P}_{k,j}^{-1}\mathcal{\tilde X}_{k,j}$, \\	
		$P_{k,i}=\bigg(\sum\limits_{j\in \mathcal{N}_{i}(k)}a_{i,j}(k) \mathcal{\tilde P}_{k,j}^{-1}\bigg)^{-1}$.	
%		$\mathcal{\tilde X}_{k,j}=\tilde X_{k,i},$ if $i=j$, otherwise $\mathcal{\tilde X}_{k,j}=\mathbf{G}_1(\tilde X_{k,j}), j\in\mathcal{N}_{i}$.
	}
%\left(\check P_{k,j}+\frac{ \bar n\Delta_j(\bar n\Delta_j+1)}{2}I_n\right)^{-1}
%
		%\REPEAT
		%\STATE set $h(t)=r(t)$
		%\REPEAT
		%\STATE set $h(t)=r(t)$
		%\UNTIL{B}
		%\UNTIL{B}
%	\end{algorithmic}
\end{algorithm}	
\subsection{Stability}
%In this subsection, we will prove the  boundedness of the estimation error covariances of Algorithm \ref{ODKF2} under mild assumptions. Before that, we provide the following lemma for proof convenience.
The next lemma, proved in Appendix \ref{app_inver}, is useful for further analysis.
\begin{lemma}\label{lem_inver}
	Under Assumptions \ref{ass_noise}-\ref{ass_F}, if there are positive constants $\{\theta_1,\theta_2,\mu_1,\mu_2\}$ such that $\theta_{k,i}\in(\theta_1,\theta_2)$ and $\mu_{k,i}\in(\mu_1,\mu_2)$, the following two conclusions hold.
	\begin{itemize}
		\item 	1) It holds that $\tilde P_{k,i}^{-1}=\frac{\bar P_{k,i}^{-1}}{1+\mu_{k,i}}+H_{k,i}^T\Delta R_{k,i}^{-1}H_{k,i}, k\in\mathbb{N},i\in\mathcal{V},$
		where $\Delta R_{k,i}=R_{k,i}+\frac{1+\mu_{k,i}}{\mu_{k,i}}B_{k,i}$.	
		\item 	2) There exists a positive scalar $\eta $ such that $\bar P_{T_l+s+1,i}^{-1}\geq \eta  A_{T_l+s}^{-T}P_{T_l+s,i}^{-1}A_{T_l+s}^{-1}, s\in[0:L),$	
		where $\{T_{l},l\in\mathbb{N}\}$ is an L-SS of $\{A_{k},k\in\mathbb{N}\}$.
	\end{itemize}		
\end{lemma}
%\begin{pf}
%	See the proof in Appendix \ref{app_inver}.	
%\end{pf}
\begin{theorem}\label{thm_stability}
	Consider the  system (\ref{system3}) with Algorithm \ref{ODKF2}. Under Assumptions \ref{ass_noise}-\ref{ass_dither} and \ref{ass_observable2}, if the parameter $L$ given in \eqref{notation_T} satisfies $L>\max\{k^0,N\}+\bar N$  and  there are positive constants $\{\theta_1,\theta_2,\mu_1,\mu_2\}$ such that $\theta_{k,i}\in(\theta_1,\theta_2)$ and $\mu_{k,i}\in(\mu_1,\mu_2)$,
	then  the sequence of estimation error covariances is stable, i.e., $\sup_{k\in\mathbb{N}}\left\lbrace E\{(\hat X_{k,i}-X_k)(\hat X_{k,i}-X_k)^T\} \right\rbrace <+\infty, \forall i\in \mathcal{V}.$	
%	\begin{equation*}
%	\sup_{k\in\mathbb{N}}P_{k,i}<+\infty, \forall i\in \mathcal{V}.
%	\end{equation*}
\end{theorem}
\begin{pf}
{\color{black}
	Due to the consistency in Lemma \ref{thm_W}, we turn to prove $\sup_{k\in\mathbb{N}}P_{k,i}<\infty$.
	Under Assumption \ref{ass_A}, $\{A_{k}|A_{k}\in\mathbb{R}^{(n+p)\times (n+p)},k\in\mathbb{N}\}$ has an L-SS, which is supposed to be $\{T_{l},l\in\mathbb{N}\}$ subject to $ L\leq T_{l+1}-T_{l}\leq T<\infty$, $\forall l\geq 0$, where $L>\max\{k^0,N\}+\bar N$.
Without loss of generality, we assume $T_0\geq \bar M$, where $\bar M$ is given in Assumption \ref{ass_observable}. Otherwise, a subsequence of $\{T_{l},l\in\mathbb{N}\}$ can always be obtained to satisfy the requirement.
	We divide the sequence set $\{T_{l},l\geq 0\}$ into two  non-overlapping time set:   $\{T_{l}+L,l\geq 0\}$ and $\bigcup_{l\geq 0} [T_{l}+L+1:T_{l+1}+L-1]$. 
	
	1) First, we consider the case of $k=T_{l}+L$, $l\geq1$. 	
	For convenience, let $\bar k\triangleq T_{l}+L$. It can be easily shown that $\left(\check P_{k,i}+\frac{\bar n\Delta_i(2\Delta_i+1)}{2}I_{\bar n}\right)^{-1}\geq \varrho\tilde P_{\bar k,i}^{-1}$, where $\varrho\in(0,1]$ and $\bar n=n+p$.	
	According to Lemma \ref{lem_inver}, 
	we obtain
	\begin{align}\label{proof_stability13}
	P_{\bar k,i}^{-1}
	\geq&\varrho\sum_{j\in \mathcal{N}_{i}(\bar k)}a_{i,j}(\bar k)  \tilde P_{\bar k,j}^{-1}\nonumber\\
%	\geq&\varrho\sum_{j\in \mathcal{N}_{i}(\bar k)}a_{i,j}(\bar k) \left( \frac{\bar P_{\bar k,j}^{-1}}{1+\mu_{\bar k,j}}+ H_{\bar k,j}^T  \Delta R_{\bar k,j}^{-1}H_{\bar k,j}\right),\nonumber\\
	%			\geq&  \frac{\eta}{(1+\mu_{\bar k,i}) }\sum_{j\in \mathcal{N}_{i}(\bar k)}a_{i,j}(\bar k) A_{\bar k}^{-T}P_{ \bar k-1,j}^{-1} A_{\bar k}^{-1}\nonumber\\
	%			&+\sum_{j\in \mathcal{N}_{i}(\bar k)} a_{i,j}(\bar k)H_{\bar k,j}^T  \Delta R_{\bar k,j}^{-1}H_{\bar k,j}\nonumber\\
	\geq&  \frac{\varrho\eta}{(1+\mu_2) }\sum_{j\in \mathcal{N}_{i}(\bar k)}a_{i,j}(\bar k) A_{\bar k-1}^{-T}P_{ \bar k-1,j}^{-1} A_{\bar k-1}^{-1}\nonumber\\
	&+\sum_{j\in \mathcal{N}_{i}(\bar k)} a_{i,j}(\bar k)H_{\bar k,j}^T  \Delta R_{\bar k,j}^{-1}H_{\bar k,j}.
	\end{align}

	By recursively applying  (\ref{proof_stability13}) for $L$ times, denoting  $\delta=\frac{\varrho\eta}{(1+\mu_2) }$, one has
	\begin{align}\label{proof_stability3}
	P_{ \bar k,i}^{-1} 		\geq&\frac{a_{min}\delta^{ L-1}\mu_1}{\mu_1+1}\sum_{s= k^0}^{L-1}\Bigg[\Phi_{\bar k,\bar k-s}^{-T}\\
	&\cdot
	\sum_{j\in \mathcal{V}} H_{\bar k-s,j}^T (R_{\bar k-s,j}+B_{\bar k-s,j})^{-1}H_{\bar k-s,j}\Bigg] \Phi_{\bar k,\bar k-s}^{-1},\nonumber
	\end{align}
	where  $a_{min}=\min_{i,j\in \mathcal{V}}{a_{i,j}^{\bar k,\bar k-s}>0,s\in [\max\{k^0,N\}: L)}$, 	and $a_{i,j}^{\bar k,\bar k-s}$ is the $(i,j)$th element of $\Pi_{l=\bar k-s}^{\bar k}\mathcal{A}_{\sigma_{l}}$. Note that $a_{min}$ can be obtained under Assumption \ref{ass_primitive}, since the elements of $\mathcal{A}_{\sigma_{ k}}$ belong to a finite set and  the jointly strongly connected network can lead to $a_{i,j}^{\bar k,\bar k-s}>0,s\geq \max\{k^0,N\}.$	
	 Under Assumptions \ref{ass_A} and \ref{ass_observable},
	 there exists a constant positive definite matrix $\breve{P}$, such that $P_{k,i}\leq \breve{P}, k=T_{l}+L.$}
	
	2) Second, we consider the time set $\bigcup_{l\geq 0} [T_{l}+L+1:T_{l+1}+L-1]$. 		
	Considering (\ref{proof_stability3}), we have $P_{T_{l}+L,i}\leq \breve{P}, k\geq L.$
	Since the  length of the interval $k\in[T_{l}+L+1:T_{l+1}+L-1]$ is bounded by $T<\infty$, we can just consider the prediction stage to study the boundedness of $P_{k,i}$, for $k\in[T_{l}+L+1:T_{l+1}+L-1]$.
	
	Under Assumption \ref{ass_A} and Proposition \ref{lem_iff}, there is a scalar $\beta_1>0$ such that $A_{k}A_{k}^T\leq \beta_{1}I_{n}$.
	Due to sup$_{k}$$(\bar Q_{k}+\tilde Q_{k})<\infty$,  it is safe to conclude that there exists an constant matrix $P^{mid}$, such that
	\begin{align}\label{P2}
	P_{k,i}\leq P^{mid}, k\in[T_{l}+L+1:T_{l+1}+L-1]
	\end{align}	
	3) Finally, for the time interval $[0:T_{0}+L]$, there exists a constant matrix $\hat P$, such that
	\begin{align}\label{P3}
	P_{k,i}\leq \hat P, k\in[0:T_{0}+L].
	\end{align}	
	According to (\ref{P2}) and (\ref{P3}),  we have $\sup_{k\in\mathbb{N}}P_{k,i}<\infty$.
\end{pf} 	
%Based on the conclusions of Lemma \ref{thm_W} and Theorem \ref{thm_stability}, the following corollary is obtained.
%\begin{corollary}\label{coro_1}
%	Consider the multi-sensor system (\ref{system}) with Algorithm \ref{ODKF2}. Under the same conditions as Theorem \ref{thm_stability}, the filtering error covariance $E\{(\hat x_{k,i}-x_k)(\hat x_{k,i}-x_k)^T\}$ is mean square stable, i.e.,
%	\begin{align*}
%	\sup_{k\in\mathbb{N}}\left\lbrace E\{(\hat x_{k,i}-x_k)(\hat x_{k,i}-x_k)^T\} \right\rbrace <+\infty, \forall i\in \mathcal{V}.
%	\end{align*}	
%\end{corollary}
 { An upper bound of error covariance can be derived by the proof of  Theorem \ref{thm_stability}, from which the influence of quantization interval size $\Delta_{i}$ can be seen through the scalar $\varrho$. With the increase of $\Delta_{i}$, the upper bound  will become larger.}
\subsection{Design of parameters}
In this subsection, {\color{black} the  design methods for the parameters $\theta_{k,i}$ and $\mu_{k,i}$  are considered by solving two optimization problems. } 
%First of all, the objective functions ought to be given.
%Due to the unknown correlation between estimates of sensors, the estimation covariance of DKFs is usually not attainable. By the consistency of Algorithm \ref{ODKF2}, we can use $\bar P_{k,i}$ and $\tilde P_{k,i}$  to take the roles.

 \textbf{a) Design of $\theta_{k,i}$}

At the prediction stage, the design of the parameter $\theta_{k,i}$ is aimed to minimize the trace of $\bar P_{k,i}$, which is an upper bound of mean square error by Lemma \ref{thm_W}.
Mathematically, the  optimization problem on $\theta_{k,i}$  is given as
\begin{align}\label{eq_opt_1}
\theta_{k,i}^*=\arg\min\limits_{\theta_{k,i}}\tr(\bar P_{k,i}),
\end{align}
where $\bar P_{k,i}=(1+\theta_{k,i}) A_{k-1}P_{k-1,i} A_{k-1}^T+ \frac{1+\theta_{k,i}}{\theta_{k,i}} \bar Q_{k-1}+\tilde Q_{k-1}$.

Since \eqref{eq_opt_1} is  convex, which can be numerically solved by many existing convex optimization methods.
The next  proposition \ref{prop_theta}, proved in Appendix \ref{app_propo_theta},  provides the closed-form solution of \eqref{eq_opt_1}.
\begin{proposition}\label{prop_theta}
	Solving   \eqref{eq_opt_1} yields the closed-form solution
	\begin{align*}
	\theta_{k,i}^*=\sqrt{\frac{\tr(\bar Q_{k-1})}{\tr(A_{k-1}P_{k-1,i} A_{k-1}^T)}}, i\in\mathcal{V}.
	\end{align*}
	 Furthermore, it holds that $\theta_{k,i}^*>0.$
\end{proposition}
%\begin{pf}
%		See Appendix \ref{app_propo_theta}.
%		\end{pf}
%\begin{pf}
%%	See the proof in Appendix \ref{app_prop_theta}.
%Consider $tr(\bar P_{k,i})$, then we have $tr(\bar P_{k,i})=(1+\theta_{k,i}) tr(A_{k-1}P_{k-1,i} A_{k-1}^T)
%+ \frac{1+\theta_{k,i}}{\theta_{k,i}} tr(\bar Q_{k-1})+tr(\tilde Q_{k-1}).$
%Then $\theta_{k,i}^*=\arg\min\limits_{\theta_{k,i}}\mathop{tr}(\bar P_{k,i})= \arg\min\limits_{\theta_{k,i}}f_k(\theta_{k,i}),$
%where
%$f_k(\theta_{k,i})=\theta_{k,i}tr(A_{k-1}P_{k-1,i} A_{k-1}^T)+\frac{tr(\bar Q_{k-1}) }{\theta_{k,i}},$
%which is minimized if $\theta_{k,i}^*tr(A_{k-1}P_{k-1,i} A_{k-1}^T)=\frac{tr(\bar Q_{k-1}) }{\theta_{k,i}^*}.$
%As a result, $\theta_{k,i}^*=\sqrt{\frac{tr(\bar Q_{k-1})}{tr(A_{k-1}P_{k-1,i} A_{k-1}^T)}}$.		
%Due to $P_{k-1,i}>0$, $\bar Q_{k-1}>0$ and $A_{k-1}\neq 0$, we have $\theta_{k,i}^*>0$.	 \textbf{Q.E.D.}
%\end{pf}

%From Proposition \ref{prop_theta}, it can be seen that the time-varying $\theta_{k,i}^*$ can be calculated in a step-wise manner based on the current information of sensor $i$.

\textbf{b) Design of $\mu_{k,i}$ }

 Considering the consistency, we  cast  the design of $\mu_{k,i}$ into the following optimization problem:
\begin{align}\label{eq_opt_2}
\mu_{k,i}^*=\arg\min\limits_{\mu_{k,i}}\tr(\tilde P_{k,i}), \text{ s.t. }  \mu_{k,i}>0.
\end{align}
%where
%\begin{align*}
%\tilde P_{k,i}&=(1+\mu_{k,i})(I-K_{k,i}H_{k,i})\bar P_{k,i},\\
%K_{k,i}&=\bar P_{k,i}H_{k,i}^T\bigg(H_{k,i}\bar P_{k,i}H_{k,i}^T+\frac{R_{k,i}}{1+\mu_{k,i}}+\frac{B_{k,i}}{\mu_{k,i}}\bigg)^{-1}.
%\end{align*}
\begin{remark}
{\color{black} Although the optimization in \eqref{eq_opt_2} is not convex, it can be solved by some existing methods}, such as the  quasi-Newton methods or the nonlinear least square  \cite{bartholomew2008nonlinear}.
\end{remark}

{
\subsection{Asymptotically unbiased}
%In this section, we study the conditions ensuring that the state estimate by Algorithm \ref{ODKF2} is asymptotically unbiased in presence of uncertainty and observation biases of sensors.
 We write $f_k=O(g_k)$ (or $f_k=o(g_k)$) if and only if $f_k\leq Mg_k$ for all $k\geq 0$ (or $\lim\limits_{k\rightarrow \infty}\frac{f_k}{g_k}=0$). The following two lemmas, proved in \cite{He2018tac} and Appendix \ref{app_conver} respectively, are to find conditions ensuring the asymptotic unbiasedness of Algorithm \ref{ODKF2}.
\begin{lemma}\label{lem_ineq}
	Suppose that $\varPi_{0}$ and $\varPi_{1}$ satisfy  $0\leq\varPi_{0}\leq \varPi_{1}$ and $\varPi_{1}>0$, then  $\varPi_{0}\varPi_{1}^{-1}\varPi_{0}\leq\varPi_{0}$.
\end{lemma}
\begin{lemma}\label{lem_conver}
Let $\{d_{k}\}\in\mathbb{R}$ be generated by $d_{k+1}=\rho d_{k}+m_k$ with $\rho\in(0,1)$ and $d_0<\infty$, then
\begin{enumerate}
	\item if $m_{k}=o(1)$, then $d_{k}=o(1)$, i.e., $d_{k}\rightarrow 0$ as $k\rightarrow\infty$;
	\item if $m_{k}=o(\delta^k)$ and $\rho<\delta<1$, then $d_{k}=o(\delta^k)$;
	\item if $m_{k}=o(\frac{1}{k^{M}})$ with $M\in\mathbb{N}^+$, then $d_{k}=o(\frac{1}{k^{M-1}})$.
\end{enumerate}
\end{lemma}
%\begin{pf}
%	See Appendix \ref{app_conver}.
%%	By Theorem 1 of \cite{Cat2010Diffusion}, (1) can be proved. 	
%%	For (2), we have $\frac{x_{k}}{\delta^k}=\left(\frac{\rho}{\delta}\right)^kx_0+\sum_{i=0}^{k}\left(\frac{\rho}{\delta}\right)^{k-i}\frac{m_i}{\delta^i}$.
%%	Due to $\rho<\delta<1$, $\left(\frac{\rho}{\delta}\right)^kx_0\rightarrow 0.$ Denote $\bar \rho=\frac{\rho}{\delta}\in(0,1)$ and $\bar m_{i}=\frac{m_i}{\delta^i}=o(1)$, then we construct a sequence $\{\bar x_{k}\}$ satisfying $\bar x_{k+1}=\bar \rho \bar  x_{k}+\bar m_k$ with $\bar  x_{0}=0$. 	By (1), we have $\bar  x_{k}=o(1)$. In light of  $\bar  x_{k}=\sum_{i=0}^{k}\left(\bar \rho \right)^{k-i}\bar m_i$, we have $\sum_{i=0}^{k}\left(\frac{\rho}{\delta}\right)^{k-i}\frac{m_i}{\delta^i}=\sum_{i=0}^{k}\left(\bar \rho \right)^{k-i}\bar m_i \rightarrow 0$  as $k\rightarrow\infty$. Hence,  $x_{k}=o(\delta^k)$.
%%	For (3), we consider $x_{k}k^{M-1}=\rho^kk^{M-1}x_0+k^{M-1}\sum_{i=0}^{k}\rho ^{k-i} m_i$. Notice $\rho^kk^{M-1}=o(1)$, then we consider the convergence of the second term, namely, $k^{M-1}\sum_{i=0}^{k}\rho ^{k-i} m_i$. Due to $m_{k}=o(\frac{1}{k^{M}})$, we have
%%	\begin{align*}
%%k^{M-1}\sum_{i=0}^{k}\rho ^{k-i} m_i
%%=&\sum_{i=0}^{k}o(\rho ^{k-i} \frac{k^{M-1}}{i^{M}})\\
%%= &\sum_{i=0}^{k}o(\frac{k^{M-1}}{i^{M}(k-i)^M })\\
%%= &\sum_{i=0}^{k} o(\frac{1}{k})\\
%%=&o(1),
%%	\end{align*}
%%where the second equality is obtained by $\rho ^{k-i}=o(\frac{1}{(k-i)^M})$ and the third equality is obtained by  $i(k-i)\leq \frac{k}{2}$. Thus, $x_{k}k^{M-1}=o(1)$, which means $x_{k}=o(\frac{1}{k^{M-1}})$.
%\end{pf}

{\color{black}
\begin{theorem}\label{thm_bias}
Consider the system (\ref{system3}) satisfying the same conditions as Theorem \ref{thm_stability} and $\lambda_{min}\{A_{k}A_{k}^T\}\geq \beta_0>0$. Denoting $\mathcal{M}_{k}:=\max_{j\in\mathcal{V}}\|E\{b_{k,j}\}\|_2^2+\|E\{u_{k}\}\|_2^2$, 
\begin{itemize}
		\item if $\mathcal{M}_{k}=o(1)$, then $\|E\{e_{k,i}\}\|_2=o(1)$;
	\item if $\mathcal{M}_{k}=o(k^{-M})$ with $M\in\mathbb{N}^+$, then $\|E\{e_{k,i}\}\|_2=o(k^{\frac{1-M}{2}})$;
	\item if $\mathcal{M}_{k}=o(\delta^k)$ and $\tilde\varrho <\delta<1$, then $\|E\{e_{k,i}\}\|_2=o(\delta^{\frac{k}{2}})$, where  $\tilde\varrho \in(0,1)$ is defined in \eqref{eq_rho}.
\end{itemize}
%the uncertain dynamics $f_{k}$  and the observation bias   tend to a constant vector and zero in expectation, respectively, then the sequence of estimates is asymptotically unbiased, i.e., $\|E\{e_{k,i}\}\|_2=o(1)$.
%Furthermore, 
\end{theorem}
}

\begin{pf}
Since $\bar P_{k,i}$ is positive definite, we define the following function $V_{k,i}(E\{\bar e_{k,i}\})=E\{\bar e_{k,i}\}^T\bar P_{k,i}^{-1}E\{\bar e_{k,i}\}.$
From the fact (\expandafter{\romannumeral3}) of Lemma 1 in \cite{Battistelli2014Kullback} and the non-singularity of $A_{k}$, we have $V_{k+1,i}(E\{\bar e_{k+1,i}\})\leq \varrho $ $E\{\bar e_{k+1,i}\}^TA_{k}^{-T}P_{k,i}^{-1}A_{k}^{-1}E\{\bar e_{k+1,i}\},$
where $0<\varrho<1$.
Since $w_{k} $ is zero-mean,  $E\{\bar e_{k+1,i}\}=A_{k}E\{e_{k,i}\}-DE\{u_{k}\}$. Then, there exists a scalar $\alpha_0>0$, such that  $\bar\varrho_1 =\varrho (1+\alpha_0)\in (0,1)$, and
\begin{align}\label{V_ineq}
&V_{k+1,i}(E\{\bar e_{k+1,i}\})\nonumber\\
\leq& \bar\varrho_1 E\{e_{k,i}\}^TP_{k,i}^{-1}E\{e_{k,i}\}+O(\|E\{u_{k}\}\|_2^2).
\end{align}
As the quantization error is uniformly  distributed in $[-\frac{\Delta_i}{2},\frac{\Delta_i}{2}]$, the quantization error is zero-mean.
%Then  the estimation error $e_{k,i}$ satisfies $E\{e_{k,i}\}=P_{k,i}$$\sum\limits_{j\in \mathcal{N}_{i}(k)}a_{i,j}(k)\bar{\check P}_{k,j}^{-1}E\{\tilde e_{k,j}\},$
%%\begin{equation}\label{ek}
%%\begin{split}
%%
%%%=&P_{k,i}\sum_{j\in \mathcal{N}_{i}}a_{i,j}\tilde P_{k,j}^{-1}E\{\tilde e_{k,j}\}\\
%%%=&P_{k,i}\sum_{j\in \mathcal{N}_{i}}a_{k,i,j}\tilde P_{k,j}^{-1}(\bar x_{k,j}+K_{k,j}(y_{k,j}-H_{k,j}\bar x_{k,j})-x_{k})\\
%%%=&P_{k,i}\sum_{j\in \mathcal{N}_{i}}a_{i,j}\tilde P_{k,j}^{-1}(I_n-\tau_{k,j}K_{k,j}C_{k,j})E\{\bar e_{k,j}\}.
%%\end{split}
%%\end{equation}
%where $\bar{\check P}_{k,j}=\check P_{k,j}+\frac{ n\Delta_j(n\Delta_j+1)}{2}I_n$.
By \eqref{V_ineq} and Lemma 2 in \cite{Battistelli2014Kullback}, there exists a scalar $\bar\varrho_2 \in(0,1)$ such that
\begin{align}\label{V_ineq2}
V_{k+1,i}(E\{\bar e_{k+1,i}\})\nonumber
\leq& \bar\varrho_2 \sum\limits_{j\in \mathcal{N}_{i}(k)}a_{i,j}(k) E\{\tilde e_{k,j}\}^T\mathcal{\tilde P}_{k,j}^{-1}E\{\tilde e_{k,j}\}\nonumber\\
&+O(\|E\{u_{k}\}\|_2^2).
\end{align}
Notice that  $\tilde P_{k,j}=(1+\mu_{k,j})(I-K_{k,j}H_{k,j})\bar P_{k,j}$ and $E\{e_{k,j}\}=(I-K_{k,j}H_{k,j})E\{\bar e_{k,j}\}+K_{k,j}E\{b_{k,j}\}$, then we have
$\frac{\bar P_{k,j}^{-1}}{1+\mu_{k,j}}=\tilde P_{k,j}^{-1}(I-K_{k,j}H_{k,j})$ and $\mathcal{\tilde P}_{k,j}^{-1}\leq \tilde P_{k,j}^{-1}.$
%where $\alpha_{\bar\Delta}$ goes to zero as $\bar\Delta$ goes to zero.
There exists a sufficiently small scalar $\alpha_1>0$ such that $\bar\varrho_2(1+\alpha_1)<1$. Denote
\begin{align}\label{eq_rho}
\tilde \varrho=\bar\varrho_2(1+\alpha_1),
\end{align}
For this $\alpha_1$, we have
\begin{align*}
&E\{\tilde e_{k,j}\}^T\mathcal{\tilde P}_{k,j}^{-1}E\{\tilde e_{k,j}\}\\
\leq & (1+\alpha_1)E\{\bar e_{k,j}\}^T(I-K_{k,j}H_{k,j})^T\\
&\quad\times\tilde P_{k,j}^{-1}(I-K_{k,j}H_{k,j})E\{\bar e_{k,j}\}+O(\|E\{b_{k,j}\}\|_2^2)\\
%\leq & (1+\alpha_1)E\{\bar e_{k,j}\}^T(I-K_{k,j}H_{k,j})^T\frac{\bar P_{k,j}^{-1}}{1+\mu_{k,j}}E\{\bar e_{k,j}\}\\
%&+O(\|E\{b_{k,j}\|_2^2)\\
\leq &(1+\alpha_1)E\{\bar e_{k,j}\}^T\frac{\bar P_{k,j}^{-1}}{1+\mu_{k,j}} \tilde P_{k,j}\frac{\bar P_{k,j}^{-1}}{1+\mu_{k,j}}E\{\bar e_{k,j}\}\\
&+O(\|E\{b_{k,j}\|_2^2)\\
\leq &(1+\alpha_1)E\{\bar e_{k,j}\}^T\bar P_{k,j}^{-1}E\{\bar e_{k,j}\}+O(\|E\{b_{k,j}\|_2^2)
\end{align*}
where the last inequality is obtained by Lemma \ref{lem_ineq} and 1) of Lemma \ref{lem_inver}.
By (\ref{V_ineq2}) and (\ref{eq_rho}), we have
\begin{align}\label{V_ineq3}
&V_{k+1,i}(E\{\bar e_{k+1,i}\})\nonumber\\
%\leq& \bar\varrho\sum\limits_{j\in \mathcal{N}_{i}(k)}\frac{a_{i,j}(k) }{1+\mu_{k,j}}E\{\bar e_{k,j}\}^T\bar P_{k,j}^{-1}E\{\bar e_{k,j}\}\nonumber\\
%&+O(\max_{j\in\mathcal{V}}\|B_{k,j}\|_2)+O(\|q_{k}\|_2)\\
\leq&\tilde\varrho\sum\limits_{j\in \mathcal{N}_{i}(k)}a_{i,j}(k)V_{k,j}(E\{\bar e_{k,j}\})\nonumber\\
&+O(\max_{j\in\mathcal{V}}\|b_{k,j}\|_2^2)+O(\|u_{k}\|_2^2)
\end{align}
Due to $\mathcal{A}_k=[a_{i,j}(k)],i,j=1,2,\cdots,N$, summing up  (\ref{V_ineq3}) for $i=1,2,\cdots,N$, then
\begin{align}\label{V_ineq_final4}
&V_{k+1}(E\{\bar e_{k+1}\})\\
\leq&\tilde\varrho\mathcal{A}_kV_{k}(E\{\bar e_{k}\})+\mathbf{1}_N\otimes (O(\max_{j\in\mathcal{V}}\|b_{k,j}\|_2^2)+O(\|u_{k}\|_2^2)), \nonumber
\end{align}
where $V_{k}(E\{\bar e_{k}\})=[V_{k,1}^T(E\{\bar e_{k,1}\}),\dots,V_{k,N}^T(E\{\bar e_{k,N}\})]^T.$
Taking 2-norm operator on both sides of \eqref{V_ineq_final4} and considering $\|\mathcal{A}_k\|_2=1$ yields
$\|V_{k+1}(E\{\bar e_{k+1}\})\|_2\leq \tilde\varrho \|V_{k}(E\{\bar e_{k}\})\|_2+O\left(\max_{j\in\mathcal{V}}\|b_{k,j}\|_2^2+\|u_{k}\|_2^2\right).$
% Recall  $\|E\{f_{k+1}-f_{k}\|_2^2\}=O(\|q_{k}\|_2)$ and $\|b_{k,i}\|_2^2=O(\|B_{k,j}\|_2)$.
Due to $\mathcal{M}_{k}=\max_{j\in\mathcal{V}}\|b_{k,j}\|_2^2+\|u_{k}\|_2^2=o(1)$ and (1) of Lemma \ref{lem_conver}, the estimate sequence of each sensor by Algorithm \ref{ODKF2} is asymptotically unbiased. Furthermore, if $\mathcal{M}_{k}$ satisfies  certain convergence rates (i.e., $\mathcal{M}_{k}=o(k^{-M})$ or $\mathcal{M}_{k}=o(\delta^k)$), by (2) and (3) of Lemma \ref{lem_conver}, the estimation bias is convergent to zero with certain rates (i.e., $o(k^{\frac{1-M}{2}})$ or $o(\delta^{\frac{k}{2}})$), respectively.

	\end{pf}
%\begin{remark}
%	The design in Proposition \ref{prop_theta} ensures that the sequence of $\{\theta_{k,i}\}_{k=1}^{\infty
%	}$ is uniformly upper bounded, since $\bar Q_{k-1}$ and $A_{k-1}P_{k-1,i} A_{k-1}^T$ are uniformly upper and lower bounded by constant positive definite matrices, respectively.
%\end{remark}
\begin{remark}
	Theorem \ref{thm_bias} shows the polynomial and exponential convergence rates  of estimation bias by Algorithm \ref{ODKF2} in presence of decaying observation biases and uncertain dynamics.
	In practical applications, if the observation biases of sensors do not converge to zero or the uncertainty does not converge to a constant vector, one can analyze an upper bound (i.e., $b_1$) of estimation bias with the similar procedure as the proof of Theorem \ref{thm_bias}. Additionally, due to $E\{\hat X_{k,i}-X_k\}E\{\hat X_{k,i}-X_k\}^T+Cov\{\hat X_{k,i}-X_k\}=E\{(\hat X_{k,i}-X_k)(\hat X_{k,i}-X_k)^T\}\leq P_{k,i}$, it holds that $b_2:=\|E\{(\hat X_{k,i}-X_k)\}\|_2\leq \sqrt{\|P_{k,i}\|_2 }$. Then one can utilize the value $\min\{b_1,b_2\}$ to evaluate the estimation bias in real time.
\end{remark}

}

\section{Distributed filter: event-triggered update}
{\color{black}In this section, we will study an even-triggered update based DKF and analyze the conditions to ensure the mean square boundedness  and asymptotic unbiasedness.}
\subsection{Event-triggered update scheme}
Due to the influence of random noise and observation bias over the system (\ref{system}), some corrupted observations may lead to the performance degradation of filters.
 Thus, we aim to provide a scheme to decide when the observation is utilized or discarded.
We introduce the  information metric $S_{k,i}$, defined as $S_{k,i}\triangleq H_{k,i}^T\left( R_{k,i}+\frac{(1+\mu_{k,i})}{\mu_{k,i}}B_{k,i}\right) ^{-1}H_{k,i}.$
%Note that $S_{k}$ is defined based on the known local rough statistics, which admits that individual sensor can calculate the metric in real time to  statistically evaluate the quantity of current observation information.
%%If $H_{k,i}=0$, then $S_{k,i}=0$, which makes sense since the observation $y_{k,i}$ only contains the bias and noise information in this scenario.
%If the eigenvalues of  $R_{k,i}$ and $B_{k,i}$ are pretty larger than the observation matrix $H_{k,i}$, then the eigenvalues of $S_{k,i}$ will be quite small, which means that the quantity of the observation information is not sufficient.
In the following, we define the update event and the event-triggered  scheme.

%\textbf{Update Event $\mathbb{E}$:}
\begin{definition}
	We say that an update event $\mathbb{E}$ of sensor $i$ is triggered at time $k$, if  sensor $i$ utilizes the observation $y_{k,i}$ to update the estimate  as Algorithm \ref{ODKF2}.
\end{definition}
%If the  update event $\mathbb{E}$ of sensor $i$ is not triggered at some time $k$, then the observation vector $y_{k,i}$ will be discarded.
%{Event-triggered Scheme:}
The event $\mathbb{E}$ is triggered at time $k$ (i.e., $y_{k,i}$ is utilized) if
\begin{align}\label{eq_trigger2}
\lambda_{max}\left( S_{k,i}-\frac{\mu_{k,i}}{1+\mu_{k,i}}\bar P_{k,i}^{-1}\right) >\tau,
\end{align}		
where $\tau\geq0$ is the preset triggering threshold of the observation update.
Otherwise, $y_{k,i}$ will be discarded.
\begin{remark}
	The triggering scheme in (\ref{eq_trigger2}) shows that if the current information is more sufficient in at least  one   channel than the prediction information,
	{\color{black}then it is worth using the available observation in the update stage}.
	The triggering threshold $\tau$ is used as a measure of information increment.
%	It can be seen that if $\tau$ goes to infinity, then the event $\mathbb{E}$ will not be triggered, which explicitly means that the observations are not utilized in the whole process. As a result of this case, the estimation performance may be deteriorated.
\end{remark}
The following lemma, proved in Appendix \ref{app_final}, provides an equivalent form of the triggering scheme.
\begin{lemma}\label{lem_trigger2}
	The event-triggered scheme (\ref{eq_trigger2}) is satisfied if and only if
	\begin{align}\label{eq_trigger}
	\lambda_{max}(\tilde P_{k,i}^{-1}-\bar P_{k,i}^{-1})> \tau,
	\end{align}
	where $\tilde  P_{k,i}=(1+\mu_{k,i})(I-K_{k,i}H_{k,i})\bar P_{k,i}$ and $\tau\geq 0.$
%	\begin{align*}
%	&\check  P_{k,i}\triangleq(1+\mu_{k,i})(I-K_{k,i}H_{k,i})\bar P_{k,i}.
%%	&K_{k,i}=\bar P_{k,i}H_{k,i}^T\bigg(H_{k,i}\bar P_{k,i}H_{k,i}^T+\frac{R_{k,i}}{1+\mu_{k,i}}+\frac{B_{k,i}}{\mu_{k,i}}\bigg)^{-1}.
%	\end{align*}
\end{lemma}
%\begin{pf}
%	See the proof in Appendix \ref{app_lem_trigger2}.
%%	Employing the matrix inverse formula on $\check  P_{k,i}$ yields $\check  P_{k,i}^{-1}=\frac{\bar P_{k,i}^{-1}}{1+\mu_{k,i}}+H_{k,i}^T\Delta R_{k,i}^{-1}H_{k,i}$.
%%	where $\Delta R_{k,i}=R_{k,i}+\frac{1+\mu_{k,i}}{\mu_{k,i}}B_{k,i}$.	
%%	Substituting $\check  P_{k,i}^{-1}$ into (\ref{eq_trigger}), the conclusion of this lemma holds. \textbf{Q.E.D.}
%\end{pf}
%\begin{remark}
	Since  Algorithm \ref{ODKF2} is consistent, $\tilde P_{k,i}^{-1}$ and $\bar P_{k,i}^{-1}$ stand for the lower bounds of  information matrices at the update stage and the prediction stage, respectively. Then $\tilde P_{k,i}^{-1}-\bar P_{k,i}^{-1}$ reflects the variation of statistical information resulted from a new observation.
	In light of Lemma \ref{lem_trigger2}, the event $\mathbb{E}$ is triggered if
	sufficiently new information is accumulated at the update stage.
	%	
	%	The triggering scheme (\ref{eq_trigger}) shows that the observation should contain sufficiently new information so as to be utilized in the update stage.  For example, if $H_{k,i}=0$, the observation $y_{k,i}$ contain no  information on the system state $X_{k}$. Thus, $y_{k,i}$ ought to be discarded in the update stage, otherwise the estimation performance may be deteriorated.
%\end{remark}
%where $\Delta R_{k,i}=R_{k,i}+\frac{(1+\mu_{k,i})}{\mu_{k,i}}B_{k,i}$.
%The triggering scheme (\ref{eq_trigger2}) decides whether the degraded observations are employed or not
%at  time instant $k$.
%If the condition (\ref{eq_trigger}) is not satisfied, then $\bar P_{k,i}^{-1}+\tau I_n \geq\check P_{k,i}^{-1}$.
\subsection{ESKDF with event-triggered update}
Based on the event-triggered scheme (\ref{eq_trigger2}) and Algorithm \ref{ODKF2}, we can obtain the extended state based DKF with event-triggered update scheme in Algorithm \ref{ODKF3}.
\begin{algorithm}
	\caption{ESDKF based on event-triggered update:}
	\label{ODKF3}
%	\begin{algorithmic}
		 {\textbf{Prediction:} the same as the one of Algorithm \ref{ODKF2}}\\
%		\vspace{0.1cm}\\
	 {\textbf{Event-triggered update:} }\\
		If (\ref{eq_trigger2}) is satisfied, then \\
			$\qquad K_{k,i}=\bar P_{k,i}H_{k,i}^T\bigg(H_{k,i}\bar P_{k,i}H_{k,i}^T+\frac{R_{k,i}}{1+\mu_{k,i}}+\frac{B_{k,i}}{\mu_{k,i}}\bigg)^{-1}$\\
		$\qquad \tilde P_{k,i}=(1+\mu_{k,i})(I-K_{k,i}H_{k,i})\bar P_{k,i}$.\\
		%				\textbf{$\qquad$observation Update:}\\
		$\qquad\tilde X_{k,i}=\bar X_{k,i}+K_{k,i}(y_{k,i}-H_{k,i}\bar X_{k,i})$\\        % & ???????
%		$\qquad\tilde P_{k,i}=\check P_{k,i}$\\
		Otherwise, \\
		$\qquad\tilde X_{k,i}=\bar X_{k,i}$, $\tilde P_{k,i}=\bar P_{k,i}$. \vspace{0.1cm}\\
			{\textbf{Quantization:} the same as the one of Algorithm \ref{ODKF2}}\\
%			\vspace{0.1cm}\\
			{\textbf{Local Fusion:} the same as the one of Algorithm \ref{ODKF2}}
%			\vspace{0.1cm}\\	
		%\REPEAT
		%\STATE set $h(t)=r(t)$
		%\REPEAT
		%\STATE set $h(t)=r(t)$
		%\UNTIL{B}
		%\UNTIL{B}
%	\end{algorithmic}
\end{algorithm}	
According to Algorithms \ref{ODKF2} and \ref{ODKF3}, we can easily obtain  Lemma \ref{thm_filter2_consistency}.
\begin{lemma}\label{thm_filter2_consistency}
	%	For the distributed filter structure (\ref{filter_stru}) with initial setting $\bar x_{0|0,i}=0$, $P_{0,i}=P_0$,
	Consider the system (\ref{system3}) with Algorithm \ref{ODKF3}. Under Assumptions \ref{ass_noise},  \ref{ass_F} and \ref{ass_dither}, the pairs ($\bar X_{k,i},\bar P_{k,i}$), ($\tilde X_{k,i},\tilde P_{k,i}$), ($\mathcal{\tilde X}_{k,i},\mathcal{\tilde P}_{k,i}$)  and ($\hat X_{k,i},P_{k,i}$) are all consistent.
	%	\begin{equation}\label{thm_consis2}
	%%	\begin{cases}
	%	E\{(\hat X_{k,i}-X_{k})(\hat X_{k,i}-X_{k})^T\}\leq P_{k,i}.
	%%	\end{cases}
	%	\end{equation}	
%	\begin{equation*}
%	\begin{cases}
%	E\{(\bar X_{k,i}- X_{k})(\bar X_{k,i}- X_{k})^T\}\leq \bar P_{k,i}\\
%	E\{(\tilde X_{k,i}-X_{k})(\tilde X_{k,i}-X_{k})^T\}\leq \tilde P_{k,i}\\
%	E\{(\hat X_{k,i}-X_{k})(\hat X_{k,i}-X_{k})^T\}\leq P_{k,i}?
%	\end{cases}
%	\end{equation*}
\end{lemma}	
%\begin{pf}
%%	See the proof in Appendix \ref{app_consistent2}.
%Since the difference between Algorithm \ref{ODKF2} and Algorithm \ref{ODKF3} lies in the observation update stage, here for convenience, we simply consider the observation update of Algorithm \ref{ODKF3}.
%If (\ref{eq_trigger2}) is satisfied, then the estimate $\bar X_{k,i}$ will be updated with the current observation $y_{k,i}$. In this case, the proof  is the same as that of Lemma \ref{thm_K}. 	If (\ref{eq_trigger2}) is not satisfied, then $\tilde X_{k,i}=\bar X_{k,i}$, $\tilde P_{k,i}=\bar P_{k,i}$, which guarantees the consistency of $(\tilde X_{k,i},\tilde P_{k,i})$. \textbf{Q.E.D.}
%\end{pf}
%Lemma \ref{thm_filter2_consistency} shows that, although the observation update scheme has been modified, Algorithm \ref{ODKF3} inherits the consistency of Algorithm \ref{ODKF2}.
The next lemma, proved in Appendix \ref{app_final}, is on Algorithm \ref{ODKF3}.
\begin{lemma}\label{lem_trigger}
	Under the event-triggered update scheme, for Algorithm \ref{ODKF3}, it holds that $	\tilde P_{k,i}^{-1}\geq \frac{\bar P_{k,i}^{-1}}{1+\mu_{k,i}}+H_{k,i}^T\Delta R_{k,i}^{-1}H_{k,i}-\tau I_{n+p},$
%	\begin{align*}
%
%	\end{align*}
	where $\Delta R_{k,i}=R_{k,i}+\frac{1+\mu_{k,i}}{\mu_{k,i}}B_{k,i}$.
\end{lemma}
%\begin{pf}
%		See the proof in Appendix \ref{app_trigger}.
%%	%Employing the matrix inverse formula on $\check  P_{k,i}$ yields $\check  P_{k,i}^{-1}=\frac{\bar P_{k,i}^{-1}}{1+\mu_{k,i}}+H_{k,i}^T\Delta R_{k,i}^{-1}H_{k,i}$.
%%	%where $\Delta R_{k,i}=R_{k,i}+\frac{(1+\mu_{k,i})}{\mu_{k,i}}B_{k,i}$.	
%%	%Substituting $\check  P_{k,i}^{-1}$ into (\ref{eq_trigger}) and considering (\ref{eq_S}), the conclusion of this lemma holds.
%%	If the update event in (\ref{eq_trigger}) is triggered, then $\tilde P_{k,i}=\check P_{k,i}$. According to 2) of Lemma \ref{lem_inver}, we have $\tilde P_{k,i}^{-1}= \frac{\bar P_{k,i}^{-1}}{1+\mu_{k,i}}+H_{k,i}^T\Delta R_{k,i}^{-1}H_{k,i}.$
%%	%\begin{align*}
%%	%
%%	%\end{align*}
%%	Thus, conclusion of Lemma \ref{lem_trigger} holds in this case.
%%	If the update event in (\ref{eq_trigger}) is not triggered, then $\tilde P_{k,i}=\bar P_{k,i}$. Besides, according to the scheme, it follows that $\bar P_{k,i}^{-1} \geq\check P_{k,i}^{-1}-\tau I_n.$
%%	%		$\bar P_{k,i}^{-1}+\tau I_n \geq\check P_{k,i}^{-1}$.
%%	%Thus, the conclusion holds.
%%	\textbf{Q.E.D.}
%\end{pf}
The next proposition, proved in Appendix \ref{app_final}, studies the relation between Algorithm \ref{ODKF2} and Algorithm \ref{ODKF3}, which shows that the event-triggered observation update scheme can lead to a tighter bound of error covariance than the typical time-driven observation update.
\begin{proposition}\label{prop_comp}
	Let $P_{k,i}^{As}$, $s=1,2$, be the $P_{k,i}$ matrix of Algorithm \ref{ODKF2} and Algorithm \ref{ODKF3}, respectively. If the two algorithms share the same initial setting and $\tau=0$, then $P_{k,i}^{A2}\leq P_{k,i}^{A1}$.
\end{proposition}
%\begin{pf}
%			See the proof in Appendix \ref{app_prop_comp}.
%%	In light of Lemma \ref{lem_trigger}, for $\tau=0$, $\tilde P_{k,i}^{-1}\geq \frac{\bar P_{k,i}^{-1}}{1+\mu_{k,i}}+H_{k,i}^T\Delta R_{k,i}^{-1}H_{k,i}=\check  P_{k,i}^{-1},$
%%	%			\begin{align*}
%%	%			
%%	%			\end{align*}
%%	which means $\tilde P_{k,i}\leq \check  P_{k,i}$, where $\check  P_{k,i}$ corresponds to the observation update of  Algorithm \ref{ODKF2}. By using the mathematical induction method, the proof of this proposition can be finished. \textbf{Q.E.D.}
%\end{pf}
\begin{remark}
	Compared with Algorithm \ref{ODKF2}, Algorithm \ref{ODKF3} is able to obtain better estimation performance since it discards corrupted observations that may deteriorate the estimation performance. Meanwhile, {\color{black}in the scenarios as  \cite{han2015optimal,weerakkody2016multi}  where the estimator and sensor are distributed 	}
	 at different geographical locations with energy-constrained communication channels, it is suggested to judge which observations contain novel information and to decide when the observations are transmitted from the sensor to the remote estimator. These tasks can be achieved by  the proposed event-triggered update scheme.
\end{remark}
\subsection{Stability and asymptotic unbiasedness}
%In this subsection, considering the system (\ref{system3}) with Algorithm \ref{ODKF3}, we will provide the stability analysis in terms of the boundedness of the sequence of estimation error covariances.
%The following lemma serves for the proof of the subsequent theorem.
%%	 and potentially contributes to the enhancement of the estimation performance.
%Since the event-triggered scheme can abandon redundant observations, it is meaningful to  applications associated with observation scheduling \cite{han2015optimal,weerakkody2016multi}.
{\color{black} 
%MORE CONCISE! TOO MANY ``then'' and ``if''. Why not directly say ``Consider the  system (\ref{system3}) with Algorithm \ref{ODKF3} under ....... If..., then... ''
%
\begin{theorem}\label{thm_stability2}
	Consider the  system (\ref{system3}) and Algorithm \ref{ODKF3} under  Assumptions \ref{ass_noise}-\ref{ass_dither} and \ref{ass_observable2}. If the parameter $L$ given in \eqref{notation_T} satisfies $L>\max\{k^0,N\}+\bar N$ and  $\theta_{k,i}\in(\theta_1,\theta_2)$, $\mu_{k,i}\in(\mu_1,\mu_2)$ for positive constants $\{\theta_1,\theta_2,\mu_1,\mu_2\}$,  then
	there exists a scalar $\vartheta>0$, such that for $0\leq\tau<\vartheta$, 
	\begin{align*}
\sup_{k\in\mathbb{N}}\left\lbrace E\{e_{k,i}e_{k,i}^T\} \right\rbrace <+\infty, \forall i\in \mathcal{V}.
	\end{align*}
	 { Furthermore, if $\inf_{k}\lambda_{min}\left(A_{k}A_{k}^T\right)>0,$ then 
	 	\begin{align*}
\max_{i\in\mathcal{V}}\sup_{k\geq L}\lambda_{max}\left(E\{e_{k,i}e_{k,i}^T\} \right) \leq \frac{1}{a_0-\tau a_1}, 
	 	\end{align*}
	where $e_{k,i}=\hat X_{k,i}-X_{k}$, $a_0$ and $a_1$ are given in \eqref{eq_a}.}
\end{theorem}
}
\begin{pf}
	According to Lemma \ref{thm_filter2_consistency}, we turn to prove $\sup_{k\in\mathbb{N}}P_{k,i}<\infty.$
	Similar to the proof of Theorem \ref{thm_stability}, with the same notations, we consider the time set $\bigcup_{l\geq 0} [T_{l}+L+1:T_{l+1}+L-1]$. The rest part can be similarly proved by taking the   method as the proof of Theorem \ref{thm_stability}.
We have
	$	P_{ \bar k,i}^{-1} 		\geq\delta^{ L}\Phi_{\bar k,\bar k-L}^{-T}\left[\sum_{j\in \mathcal{V}} a_{i,j}^{\bar k,\bar k-L}P_{\bar k-L,j}^{-1} \right] \Phi_{\bar k,\bar k-L}^{-1}+
	\breve{P}_{\bar k,i}^{-1},$
%	\begin{align*}
%	P_{ \bar k,i}^{-1} 		\geq&\delta^{ L}\Phi_{\bar k,\bar k-L}^{-T}\left[\sum_{j\in \mathcal{V}} a_{i,j}^{\bar k,\bar k-L}P_{\bar k-L,j}^{-1} \right] \Phi_{\bar k,\bar k-L}^{-1}+
%	\breve{P}_{\bar k,i}^{-1},
%	\end{align*}
	where $\delta=\frac{\varrho\eta}{(1+\mu_2) }$ and
	\begin{align}\label{proof_stability402}
	\breve{P}_{\bar k,i}^{-1}			=&\varrho
	\sum_{s=0}^{L-1}\delta^{s}\Bigg[\Phi_{\bar k,\bar k-s}^{-T}\sum_{j\in \mathcal{V}} a_{i,j}^{\bar k,\bar k-s}\\
	&\cdot
	\left( H_{\bar k-s,j}^T\Delta R_{\bar k-s,j}^{-1}H_{\bar k-s,j}-\tau I_{n+p} \right)\Phi_{\bar k,\bar k-s}^{-1}\Bigg].	\nonumber	
	\end{align}	
	%			and $a_{i,j}^{\bar k,\bar k-L}$ is the $(i,j)$th element of $\Pi_{s=\bar k-L}^{\bar k}\mathcal{A}_{\sigma_{s}}$.	
	We have $P_{ \bar k,i}^{-1}\geq \breve{P}_{\bar k,i}^{-1}$. To prove the conclusion, we turn to prove there is a constant matrix $S>0$, such that  $\breve{P}_{\bar k,i}^{-1}\geq S$.
	Under the conditions of this theorem, it follows from the proof of Theorem \ref{thm_stability} that there is a  scalar $\pi>0$, such that
	$\sum_{s=0}^{L-1}\delta^{s}\Phi_{\bar k,\bar k-s}^{-T}
	\Big[
	\sum_{j\in \mathcal{V}} a_{i,j}^{\bar k,\bar k-s}H_{\bar k,j}^T\Delta R_{\bar k,j}^{-1}H_{\bar k,j}\Big]\Phi_{\bar k,\bar k-s}^{-1}
	\geq	 \pi I_{n+p}.$
%	\begin{align*}
%	&\sum_{s=0}^{L-1}\delta^{s}\Phi_{\bar k,\bar k-s}^{-T}
%	\Bigg[
%	\sum_{j\in \mathcal{V}} a_{i,j}^{\bar k,\bar k-s}H_{\bar k,j}^T\Delta R_{\bar k,j}^{-1}H_{\bar k,j}\Bigg]\Phi_{\bar k,\bar k-s}^{-1}
%	\geq	 \pi I_{n+p}.\nonumber
%	\end{align*}		
	Denote $\Xi_{\bar k}=\sum_{s=0}^{L-1}\delta^{s}\Phi_{\bar k,\bar k-s}^{-T}\Phi_{\bar k,\bar k-s}^{-1}.$
	To guarantee $\inf_{\bar k}\breve{P}_{\bar k,i}^{-1}>0$, it is sufficient to prove that there exists a constant matrix $S$, such that $\pi I_{n+p}-\tau\Xi_{\bar k}\geq S>0.$
	Under Assumption \ref{ass_A}, there exists a constant scalar $\bar\mu>0$, such that $	\text{sup}_{\bar k}\Xi_{\bar k}\leq \bar\mu I_{n+p}.$	
	Let $S=(\pi-\tau\bar\mu)I_{n+p} $,
	then a sufficient condition is  $	0\leq\tau<\frac{\pi }{\bar\mu}.$
	Choosing $\vartheta=\frac{\pi }{\bar\mu}>0$, then the boundedness is obtained.
	{
	Furthermore, if $\inf_k\lambda_{min}\{A_{k}A_{k}^T\}>0,$ then the above analysis holds for any $k\geq L$ with  no prediction step in the proof of Theorem \ref{thm_stability}. Thus, for any $k\geq 0$, it holds that $P_{k,i}^{-1}\geq \breve{P}_{\bar k,i}^{-1}\geq \varrho(\pi-\tau\bar\mu)I_{n+p}$.
	Let
	\begin{align}\label{eq_a}
	a_0=\varrho\pi, a_1=\varrho\bar\mu,
	\end{align}
	then we have $\max_{i\in\mathcal{V}}\sup_{k\geq L}\lambda_{max}(P_{k,i})\leq \frac{1}{a_0-\tau a_1}$. In light of the consistency in Lemma \ref{thm_W}, the conclusion holds.}
\end{pf}
%
%Based on the conclusions of Lemma \ref{thm_filter2_consistency} and Theorem \ref{thm_stability2}, the following corollary is obtained.
%\begin{corollary}\label{coro_filter2}
%	Consider the multi-sensor system (\ref{system}) with Algorithm \ref{ODKF2}. Under the same conditions as Theorem \ref{thm_stability2}, there exists a scalar $\vartheta>0$ such that for $0\leq\tau<\vartheta$,  the estimation error covariances of each sensor are stable, i.e.,
%	\begin{align*}
%	\sup_{k\in\mathbb{N}}\left\lbrace E\{(\hat x_{k,i}-x_k)(\hat x_{k,i}-x_k)^T\} \right\rbrace <+\infty, \forall i\in \mathcal{V}.
%	\end{align*}	
%\end{corollary}	
\begin{remark}
%	According to Theorem \ref{thm_stability2}, the estimation error covariances of Algorithm \ref{ODKF3} remain stable by designing a proper event-triggered update parameter $\tau.$
{\color{black}
If better performance of Algorithm \ref{ODKF3} is pursued, one could let $\tau$ be zero or sufficiently small, which can ensure a smaller bound of mean square error  by Proposition \ref{prop_comp}. If one aims to reduce the update frequency with stable estimation error, a relatively large $\tau$ can be set by satisfying the requirement in Theorem \ref{thm_stability2}. }Note that	if  a too large $\tau$ is given such that the triggering condition (\ref{eq_trigger2})  is hardly satisfied,
	then most of observation information will be discarded. As a result, the collective observability condition in Assumption \ref{ass_observable} may not hold, which means that the boundedness of estimation error covariances is not guaranteed.
%	
%	It makes sense, since if we set a quite large $\tau$, the observation update scheme will be not frequently triggered. Thus, many useful observations will be discarded, and then no sufficient information can be employed to guarantee the estimation boundedness.
\end{remark}
%	\eqref{eq_opt_2}:
%\begin{align}
%\min\limits_{\theta_{k,i}}\mathop{tr}(\bar P_{k,i})
%\end{align}
%where $\bar P_{k,i}=(1+\theta_{k,i}) A_{k-1}P_{k-1,i} A_{k-1}^T+ \frac{1+\theta_{k,i}}{\theta_{k,i}} \bar Q_{k-1}+\tilde Q_{k-1}$.
%
%\begin{proposition}
%	Solving the \eqref{eq_opt_2} yields the solution
%	\begin{align*}
%	\theta_{k,i}=\sqrt{\frac{tr(\bar Q_{k-1})}{tr(A_{k-1}P_{k-1,i} A_{k-1}^T)}}
%	\end{align*}
%\end{proposition}
%
%	Note that under the observation update scheme (\ref{eq_trigger2}) the update rate is monotonically decreasing with respect to the threshold $\tau$. Thus, we design the following optimization problem to lower the update rate while the desired covariance bound $P_f$ is met.
%	
%		Problem 3:
%	\begin{align}
%	\max \tau, \qquad\text{s.t.}\quad \breve{P}_{k,i}^{-1}(\tau)\geq P_f^{-1}
%	\end{align}
%%	s.t. $\breve{P}_{k,i}^{-1}(\tau)\geq P_f^{-1}$.
{
	To analyze the asymptotic unbiasedness of Algorithm \ref{ODKF3}, for convenience, we denote
		\begin{align}\label{eq_notations}
		\begin{split}
	\mathcal{I}_{k,i}&=H_{k,i}^T\left(R_{k,i}+\frac{1+\mu_{2}}{\mu_{2}}B_{k,i}\right)^{-1}H_{k,i}\\
\bar a&=\sup_{k\in\mathbb{N}}\{ \lambda_{max}(A_{k} A_{k}^T)\},
b_w=\sup\{\lambda_{max}(\bar Q_{k})\}\\
r_{0}&=(1+\theta_2)(a_0-\tau a_1)\bar a+\frac{1+\theta_1}{\theta_1}b_w+b_q\\
	r_1&=\frac{\mu_1}{(\mu_1+1)r_0}+\tau,b_q=\sup\{\lambda_{max}(\tilde Q_{k})\},
		\end{split}
	\end{align}
	where $a_0$ and $a_1$ are given in \eqref{eq_a}, $\{\theta_1,\theta_2,\mu_1,\mu_2\}$ are given in Theorem \ref{thm_stability2}.

{\color{black} 
\begin{theorem}\label{thm_bias2}
	Consider the system (\ref{system3}) with Algorithm \ref{ODKF3} under the same conditions as Theorem \ref{thm_stability2}.
	 If 	there is an integer $\tilde M>L$ such that the set $	\mathcal{S}=\{i\in\mathcal{V}|\sup_{k\geq\tilde M}\lambda_{max}\left(\mathcal{I}_{k+1,i}\right)\leq r_1\}$ is non-empty, and 
\begin{itemize}
		\item if $\mathcal{\bar M}_{k}=o(1)$, then $\|E\{e_{k,i}\}\|_2=o(1)$;
	\item if $\mathcal{\bar M}_{k}=o(k^{-M})$ with $M\in\mathbb{N}^+$, then $\|E\{e_{k,i}\}\|_2=o(k^{\frac{1-M}{2}})$;
	\item if $\mathcal{\bar M}_{k}=o(\delta^k)$ and $\tilde\varrho <\delta<1$, then $\|E\{e_{k,i}\}\|_2=o(\delta^{\frac{k}{2}})$, where  $\tilde\varrho \in(0,1)$ is defined in \eqref{eq_rho},
\end{itemize}
where $\mathcal{\bar M}_{k}:=\max\limits_{j\in\mathcal{V}-\mathcal{S}}\|E\{b_{k,j}\}\|_2^2+\|E\{u_{k}\}\|_2^2$.
%%	Under Assumptions \ref{ass_noise}-\ref{ass_observable2}, let $L>\max\{k^0,N\}+\bar N$ and  there be positive constants $\{\theta_1,\theta_2,\mu_1,\mu_2\}$ such that $\theta_{k,i}\in(\theta_1,\theta_2)$ and $\mu_{k,i}\in(\mu_1,\mu_2)$,
%	 and  and $\tau>0$ , such that $\inf_{k\geq L}\lambda_{\min}(P_{k,i}^{-1})\geq a_0-\tau a_1$ and the set $\mathcal{S}\subset\mathcal{V}$ is non-empty, where
%%	\begin{align}
%%	\mathcal{S}=\{i\in\mathcal{V}|\sup_{k\geq\tilde M}\lambda_{max}\left(\mathcal{I}_{k+1,i}\right)\leq r_{0}\}.
%%%	\end{align}
%	Then, if  the observation biases  $\{b_{k,i},i\in\mathcal{V}-\mathcal{S}\}$ tend to  zero in expectation, then   the sequence of estimates is asymptotically unbiased, i.e., $\|E\{e_{k,i}\}\|_2=o(1)$.
%	Furthermore,
	%Assume that   the uncertain dynamics asymptotically goes to a constant vector $f_0$, and the observation bias of each sensor  asymptotically goes to zero,   i.e., $\|f_{k}-f_0\|_2\rightarrow 0$ and $\|b_{k,i}\|_2\rightarrow 0$, as $k\rightarrow\infty$, respectively.  If $\lambda_{min}\{A_{k}A_{k}^T\}\geq \beta_0>0,$  the  estimation bias is asymptotically convergent to zero with the speed $O(\min\{\|q_{k}\|_2,\|B_{k,i}\|_2,\rho_0^k\})$,
	%then there exists  $\bar \Delta>0$, such that for $\Delta_{i}\leq\bar \Delta$,
\end{theorem}
}
\begin{remark}
	Theorem \ref{thm_bias2} shows that the estimation biases of Algorithm \ref{ODKF3} tend to zero even if the observation biases of some sensors do not decay. The proof of Theorem \ref{thm_bias2} is similar to that of Theorem \ref{thm_bias}, by noting that $\tilde P_{k,i}=\bar P_{k,i}$, for $i\in\mathcal{V}-\mathcal{S}.$
\end{remark}
In the following, we show the feasibility of the condition that there is a positive integer $\tilde M>L$ such that the set $	\mathcal{S}=\{i\in\mathcal{V}|\sup_{k\geq\tilde M}\lambda_{max}\left(\mathcal{I}_{k+1,i}\right)\leq r_1\}$ is non-empty.

\textbf{Feasibility for a non-empty $\mathcal{S}$}\\
To find the condition under which the set $\mathcal{S}$ is non-empty, we prove that the condition $\sup_{k\geq\tilde M}\lambda_{max}\left(\mathcal{I}_{k+1,i}\right)\leq r_1$ can be satisfied for some sensors. Note that under the conditions of Theorem \ref{thm_bias2}, then the conclusions of Theorem \ref{thm_stability2} holds as well. Thus, recalling the notations in  \eqref{eq_notations}, we have  $a_0-\tau a_1>0$, which leads to $r_0>0$ and then $r_1>\tau>0$. Meanwhile, whatever how $\tau$ is small, $r_1$ has  a uniformly lower bound by noting $0<r_0\leq (1+\theta_2)a_0\bar a+\frac{1+\theta_1}{\theta_1}b_w+b_q $. If there is  $M\geq L$, such that $\sup_{k\geq\tilde M}\lambda_{max}\left(\mathcal{I}_{k+1,i}\right)\leq r_1$, then $\mathcal{S}$ is non-empty. Simple examples ensuring this condition include that  for $k\geq \tilde M\geq L$, $\|H_{k,i}\|_2$ being very small,  $\lambda_{max}\left(R_{k,i}+\frac{1+\mu_{2}}{\mu_{2}}B_{k,i}\right)$ being very large, and so on.

The event-triggered scheme and the time-driven scheme have some similar properties in estimation consistency, boundedness of error covariances, and asymptotic unbiasedness of estimates. The differences between the two schemes are explicitly shown in Table \ref{table_comp}.}
\begin{table*}[htp]
		\centering  % 显示位置为中间
		{
	\caption{Differences of time-driven  and event-triggered update schemes}  % 表格标题
	\label{table_comp}  % 用于索引表格的标签
		{\renewcommand{\arraystretch}{1.0} %<- modify value to suit your needs
\begin{tabular}{|c|c|c|c|c|}% 通过添加 | 来表示是否需要绘制竖线
	\hline  % 在表格最上方绘制横线
			update scheme  &observation&threshold& stability condition& asymptotic unbiasedness\\
						\hline
		time-driven  &always&no& system+topology& decaying observation biases of all sensors \\
			\hline
	event-triggered  &selective&yes&system+topology+threshold& non-decaying  observation biases of some sensors\\
  %在第一行和第二行之间绘制横线
	\hline % 在表格最下方绘制横线
\end{tabular}
}
}
\end{table*}
\section{Numerical Simulations}
In this section, numerical simulations are carried out to demonstrate the aforementioned theoretical results and show the effectiveness of the proposed algorithms.

\subsection{Performance Evaluation}\label{subsec:Per}
In this and next subsections, let us consider an object whose motion is described by the kinematic model \cite{simon2010kalman} with uncertain dynamics:
\begin{equation}\label{eq:sim_mdl}
\begin{aligned}
x_{k+1}=&\begin{pmatrix}
1&0&T&0\\
0&1&0&T\\
0&0&1&0\\
0&0&0&1
\end{pmatrix}x_k+
\begin{pmatrix}
0&0\\0&0\\T&0\\0&T
\end{pmatrix}
f\left(x_k,k\right)+\omega_k,
\end{aligned}
\end{equation}
where $T=0.1$ is the sampling step, $x_{k}$ is the unknown state vector consisting of four-dimensional components along the coordinate axes and $f\left(x_k,k\right)$ is the uncertain dynamics. The covariance of process noise $\omega_k$ is $Q_k=\diag\left(\left[4,4,1,1\right]\right)$.
The kinematic state of the object is observed by means of four sensors modeled as
\begin{equation*}
y_{k,i}=\bar{H}_{k,i}x_k+b_{k,i}+v_{k,i},i=1,2,3,4.
\end{equation*}
The observation noise $[v_{k,1},\dots,v_{k,N}]^T$ is i.i.d. Gaussian with covariance $R_{k}={ 4}\times I_4$.
Additionally, the sensor network's communication topology is assumed to be directed and switching, whose adjacency matrix is selected from
$\mathcal{A}_1=\left(\begin{smallmatrix}
1&0&0&0\\
0.5&0.5&0&0\\
0&0.5&0.5&0\\
0&0&0.5&0.5
\end{smallmatrix}\right)$, $\mathcal{A}_2=\left(\begin{smallmatrix}
0.5&0.5&0&0\\
0&1&0&0\\
0&0.3&0.4&0.3\\
0&0.5&0&0.5
\end{smallmatrix}\right)$ and $\mathcal{A}_3=\left(\begin{smallmatrix}
0.5&0.5&0&0\\
0&0.5&0.5&0\\
0&0&1&0\\
0.25&0.25&0.25&0.25
\end{smallmatrix}\right).$
%
%
%\begin{equation*}
%\left\{\begin{aligned}
%\mathcal{A}_1&=\begin{pmatrix}
%1&0&0&0\\
%0.5&0.5&0&0\\
%0&0.5&0.5&0\\
%0&0&0.5&0.5
%\end{pmatrix},\mathcal{A}_2=\begin{pmatrix}
%0.5&0.5&0&0\\
%0&1&0&0\\
%0&0.3&0.4&0.3\\
%0&0.5&0&0.5
%\end{pmatrix}\\
%\mathcal{A}_3&=\begin{pmatrix}
%0.5&0.5&0&0\\
%0&0.5&0.5&0\\
%0&0&1&0\\
%0.25&0.25&0.25&0.25
%\end{pmatrix}\\
%\end{aligned}\right ..
%\end{equation*}
%The corresponding topologies of the sensors are given in Fig. \ref{topology}.
And the topology switching signal $\sigma_k=\mathrm{mod}\left(\lfloor \frac{k}{5}\rfloor,3\right)+1$, where mod$(a,b)$  stands for the modulo operation of $a$ by $b$. 
In the following, we conduct the numerical simulations through Monte Carlo experiment, in which 500 runs for the considered algorithms are implemented, respectively. The Root Mean Square Error (RMSE) averaged over all the sensors is defined as $RMSE_k=\sqrt{\frac14\sum_{i=1}^4 {MSE_{k,i}}}$, where $MSE_{k,i}=\frac1{500}\sum_{j=1}^{500}{\left(x_{k,i}^j-\hat{x}_{k,i}^j\right)^T\left(x_{k,i}^j-\hat{x}_{k,i}^j\right)},$
%\begin{equation*}
%
%\end{equation*}
and $\hat{x}_{k,i}^j$ is the state { (position or velocity)} estimate of the $j$th run of sensor $i$ at the $k$th time instant. Besides, we denote $P_k=\frac14\sum_{i=1}^4 P_{k,i}$. { The mean estimation error (ME) averaged over all  sensors is defined by $ME_k=\frac14\sum_{i=1}^4{\frac1{500}\sum_{j=1}^{500}{\frac14\sum_{l=1}^4{\left(x_{k,i}^j(l)-\hat{x}_{k,i}^j(l)\right)}}}$.} It is assumed that the initial state is a zero-mean random vector with covariance matrix $P_0=\diag\left(\left[10,10,1,1\right]\right)$.

\begin{figure}
	\centering
	\includegraphics[scale=0.5]{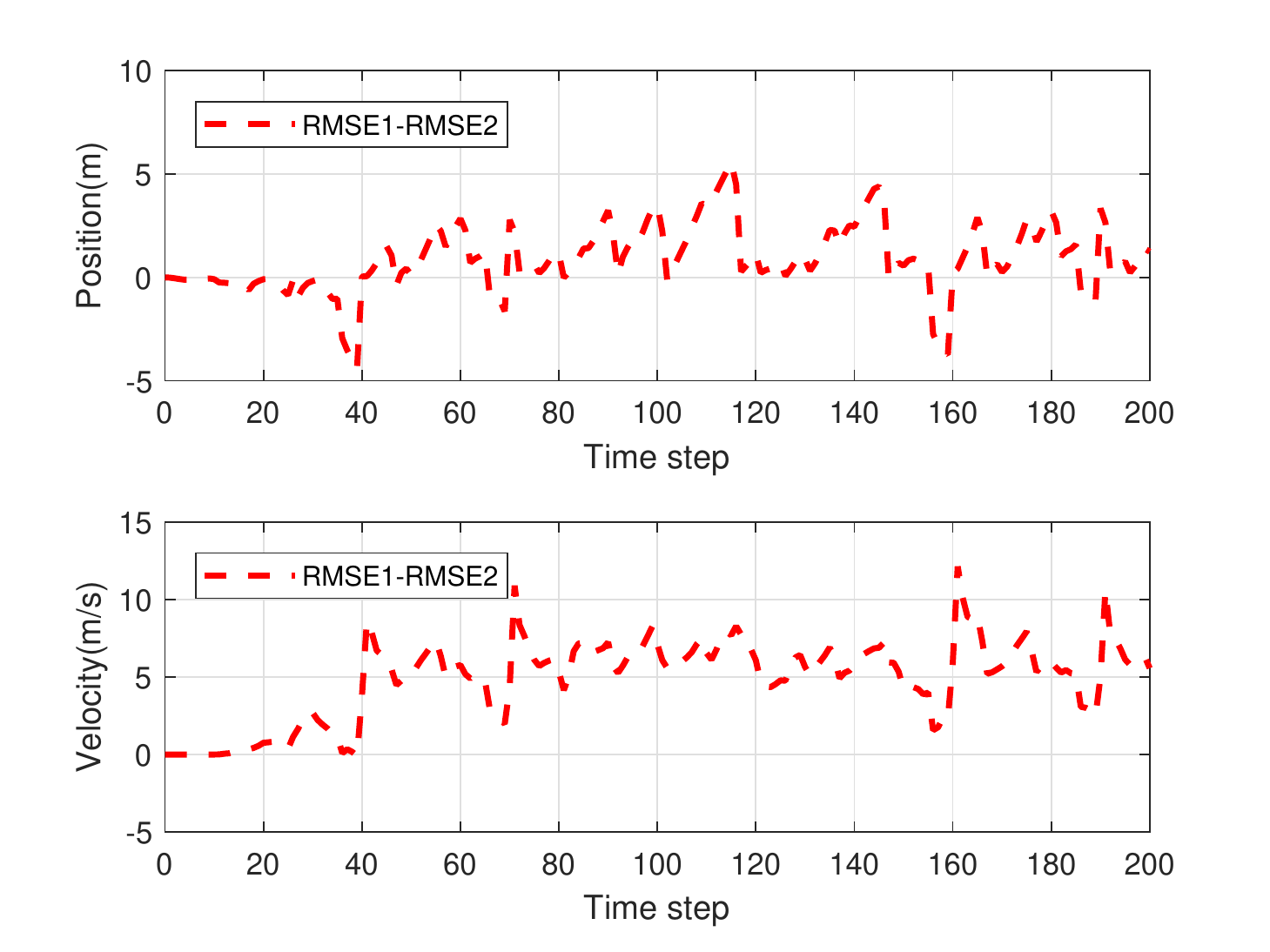}
	\caption {RMSE difference between Algorithm \ref{ODKF2} and Algorithm \ref{ODKF3}}
	\label{fig:cmp_trigger}
\end{figure}

\begin{figure}
	\centering
	\includegraphics[scale=0.5]{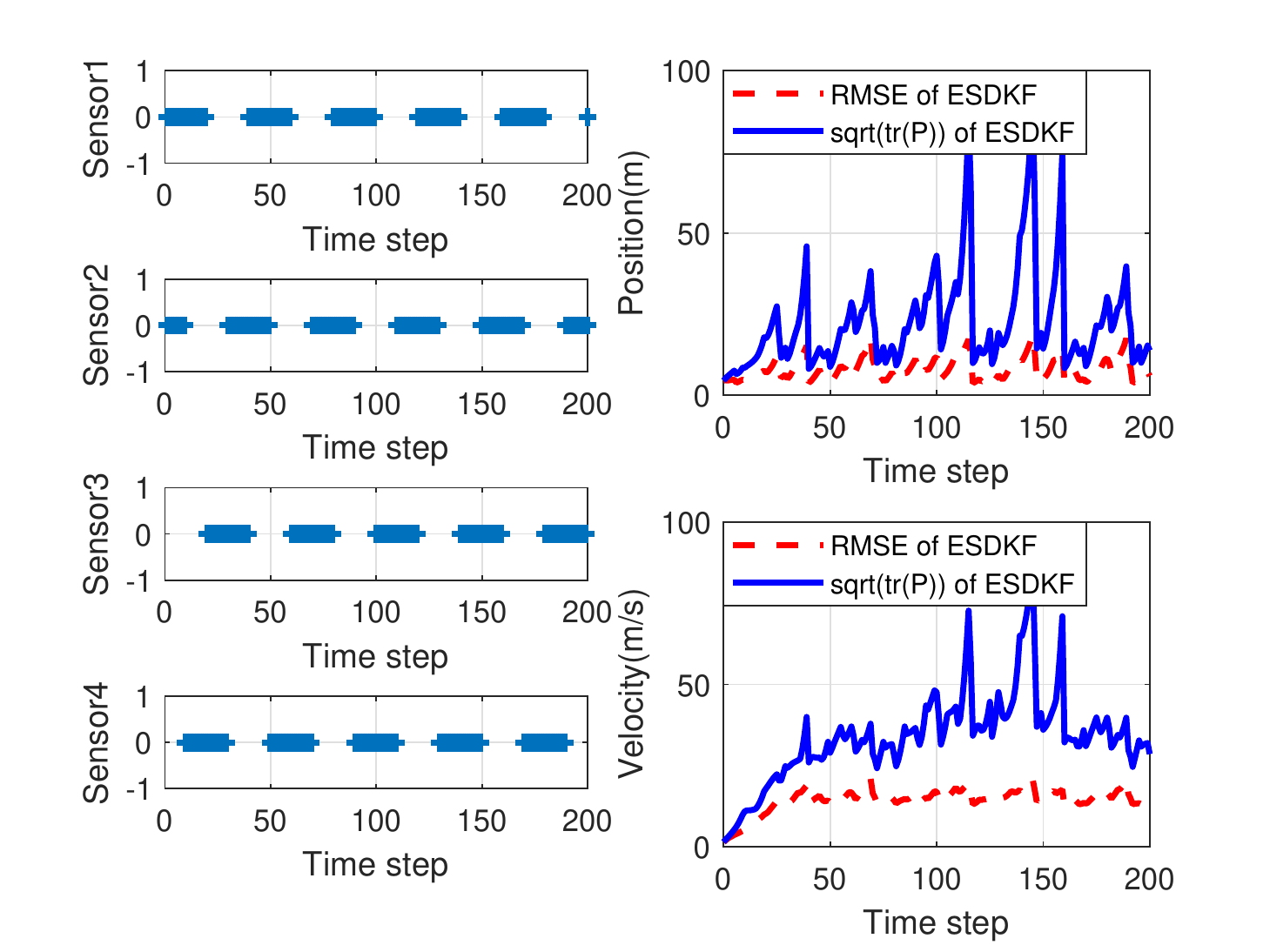}
	\caption{ Triggering times, RMSE and $\sqrt{\tr(P)}$ of  Algorithm \ref{ODKF3}}
	\label{fig:Consistence_Trigger}
\end{figure}

{The uncertain dynamics and the state-correlated bias are assumed to be $	f\left(x_k,k\right)=\frac13\left(\begin{smallmatrix}
	\sin\left(x_k\left(3\right)\right)+k\\\sin\left(x_k\left(4\right)\right)+k
	\end{smallmatrix}\right)$ and 
	 $	b_{k,i}=\mathrm{sat}\left(2\sin\left(x_{k}^2(1)+x_{k}^2(2)\right)+b_{0,i},2\right),$ respectively,
	where the initial bias $b_{0,i}$ is generated uniformly within [-2,2]. {\color{black}We assume $B_{k,i}=4$.} The observation matrices are supposed to be switching with time, following $\bar{H}_{k,i}=\bar{H}_{\mathrm{mod}\left(i+\lfloor \frac{k}{10}\rfloor,4\right)+1}$, and $
	\bar{H}_1=\left(\begin{smallmatrix}
	1&0&0&0
	\end{smallmatrix}\right),\bar{H}_2=\left(\begin{smallmatrix}
	0&1&0&0
	\end{smallmatrix}\right),\bar{H}_3=\bar{H}_4=\left(\begin{smallmatrix}
	0&0&0&0
	\end{smallmatrix}\right).$
	And the parameters of Algorithm 1 and Algorithm 2 are set  $X_{i,0}=0_{6\times1}$, $P_{i,0}=\left(\begin{smallmatrix}
	P_0&0_{4\times2}\\
	0_{2\times4}&I_2
	\end{smallmatrix}\right)$, { $\hat{Q}_k=10^{-3}\times I_2$, $\mu_{k,i}=0.3,\tau_{i}=0.001,\forall i=1,2,3,4.$}
}
First, we carry out numerical simulations for Algorithm \ref{ODKF2} (i.e., ESDKF) and Algorithm \ref{ODKF3} with results  given in Fig. \ref{fig:cmp_trigger} and Fig. \ref{fig:Consistence_Trigger}. Fig. \ref{fig:cmp_trigger} shows Algorithm \ref{ODKF3} generally  have better performance than Algorithm \ref{ODKF2}.
Meanwhile, Fig. \ref{fig:Consistence_Trigger}  shows the triggering time instants of observation update for each sensor. Because of the periodic switching of observation matrices, the triggering time instants of all sensors are also periodic. Thus, compared with Algorithm \ref{ODKF2}, Algorithm \ref{ODKF3} can reduce the frequency of observation update with competitive estimation performance.
Fig. \ref{fig:Consistence_Trigger} also gives the behavior of the RMSE and $\sqrt{\tr\left(P_k\right)}$ of Algorithm \ref{ODKF3}, from which one can see the estimation error covariances of the proposed ESDKF keep stable in the given period and the consistency of each sensor remains. 
%Meanwhile, because of the switching of observation matrices and communication topologies, the RMSE and $\sqrt{\tr\left(P_k\right)}$ show  periodic fluctuations of  rise and fall.
%\begin{figure}
%	\centering
%	\includegraphics[scale=0.5]{Trigger.eps}
%	\caption {Update triggering times of Algorithm \ref{ODKF3}}
%	\label{fig:Trigger}
%\end{figure}
%\begin{figure}
%	\centering
%	\includegraphics[scale=0.5]{Consistence.eps}
%	\caption {Consistency of  Algorithm \ref{ODKF3}}
%	\label{fig:Consistence}
%\end{figure}
\subsection{Asymptotically unbiasedness of Algorithms \ref{ODKF2} and \ref{ODKF3}}
{Next, we give a numerical simulation to verify Theorem \ref{thm_bias} and Theorem \ref{thm_bias2}. In this subsection, the uncertain dynamics is assumed to be $	f\left(x_k,k\right)=\frac1{2k}\left(\begin{smallmatrix}
	\sin\left(x_k\left(3\right)\right)\\\sin\left(x_k\left(4\right)\right)
	\end{smallmatrix}\right)+\left(\begin{smallmatrix}
	1\\1
	\end{smallmatrix}\right),$
	and the state-correlated bias $b_{k,i}$ has two situations:
	{\color{black}
	\begin{itemize}
		\item
		Situation 1:
		\begin{equation*}
		b_{k,i}=\frac1k\mathrm{sat}\left(s_{k,i},2\right), B_{k,i}=\frac{4}{k^2}\quad i=1,2,3,4
		\end{equation*}
		where $s_{k,i}=2\sin\left(x_{k}^2(1)+x_{k}^2(2)\right)+b_{0,i}$, and  $b_{0,i}$ is generated uniformly within [-2,2].
		\item
		Situation 2:
		\begin{equation*}
		b_{k,i}=\left\{\begin{aligned}\frac1k\mathrm{sat}\left(s_{k,i},2\right), B_{k,i}=\frac{4}{k^2},\quad i=1,2,\\
		\mathrm{sat}\left(\tilde s_{k,i},40\right),B_{k,i}=1600, \quad i=3,4,\end{aligned}\right .
		\end{equation*}
	where $\tilde{s}_{k,i}=40\sin\left(x_{k}^2(1)+x_{k}^2(2)\right)+b_{0,i}$, $b_{0,1}$ and $b_{0,2}$ are generated uniformly within [-2,2], $b_{0,3}$ and $b_{0,4}$ are generated uniformly within [-40,40].
	\end{itemize}
}
		In Situation 2, the biases of sensor 3 and sensor 4 are big and do not tend to zero.
	The observation matrices are supposed to be $\bar{H}_{k,1}=\left(\begin{smallmatrix}
	1&0&0&0
	\end{smallmatrix}\right),\bar{H}_{k,2}=\left(\begin{smallmatrix}
	0&1&0&0
	\end{smallmatrix}\right),
	\bar{H}_{k,3}=\left(\begin{smallmatrix}
	1&0&0&0
	\end{smallmatrix}\right),
	\bar{H}_{k,4}=\left(\begin{smallmatrix}
	0&1&0&0
	\end{smallmatrix}\right)$.
	And the parameters of Algorithm 1 and Algorithm 2 are set  $X_{i,0}=-\begin{pmatrix}
	10&10&\sqrt{10}&\sqrt{10}&0&0
	\end{pmatrix}^T$, $P_{i,0}=\left(\begin{smallmatrix}
	11P_0&0_{4\times2}\\
	0_{2\times4}&I_{2}
	\end{smallmatrix}\right)$, { $\hat{Q}_k=10^{-3}\times I_2$, $\mu_{k,i}=0.3,{\tau_{i}=0.001},\forall i=1,2,3,4.$}
	%\begin{align*}
	Fig. \ref{fig:Unbiased} gives the mean estimation error of Algorithm \ref{ODKF2} and Algorithm \ref{ODKF3} under Situation 1 and Situation 2. From this figure, one can see, under Situation 1, both Algorithm \ref{ODKF2} and Algorithm \ref{ODKF3} are asymptotically unbiased; however, under Situation 2, only Algorithm \ref{ODKF3} is asymptotically unbiased. Thus Thereom \ref{thm_bias} and Theorem \ref{thm_bias2} are verified.
	\begin{figure}[htp]
		\centering
		\includegraphics[scale=0.5]{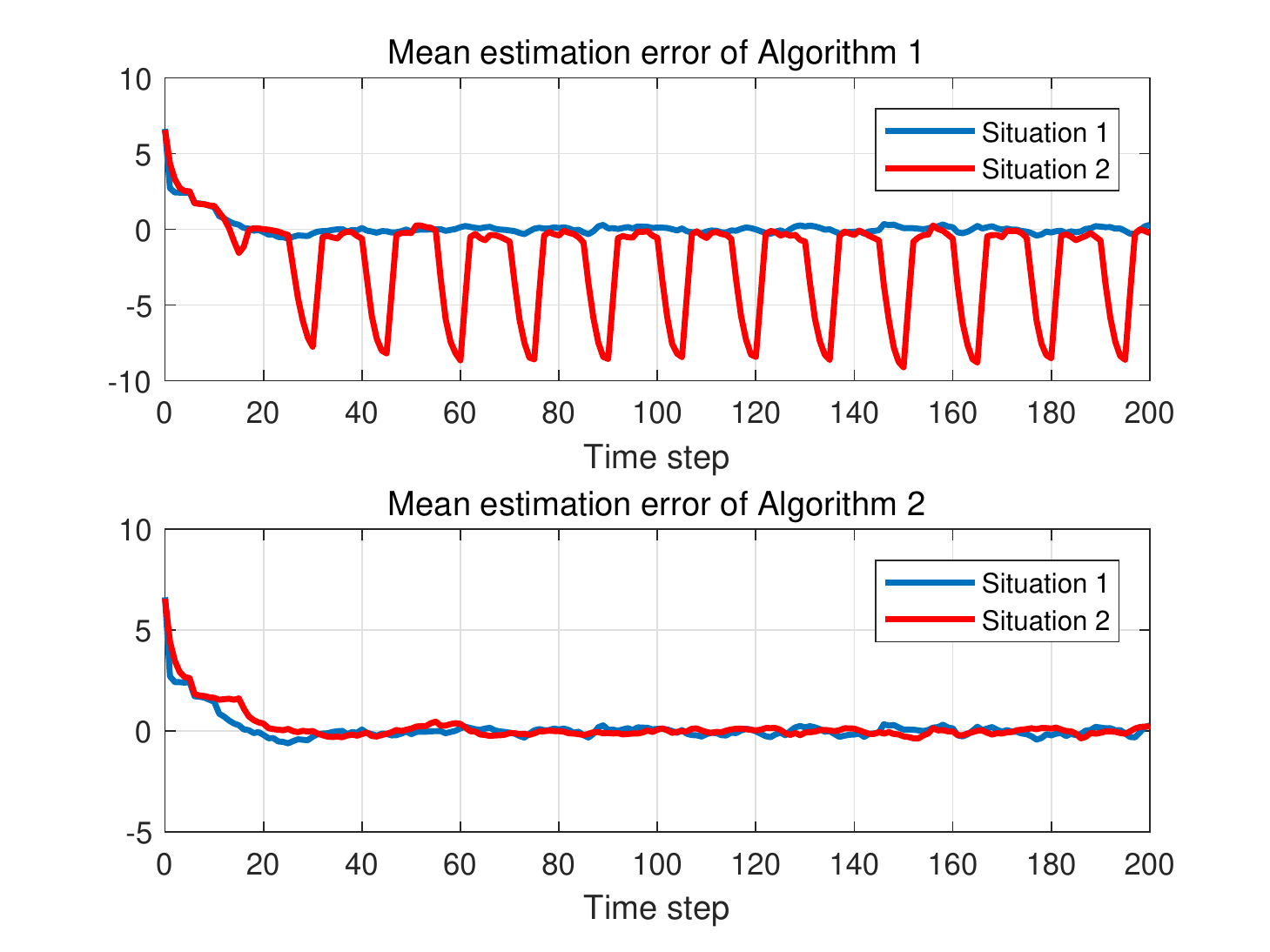}
		\caption {Mean estimation error of Algorithm \ref{ODKF2} and Algorithm \ref{ODKF3}}
		\label{fig:Unbiased}
	\end{figure}
}

\subsection{Comparisons with other algorithms}
\begin{figure}[htp]
	\centering
	\includegraphics[scale=0.3]{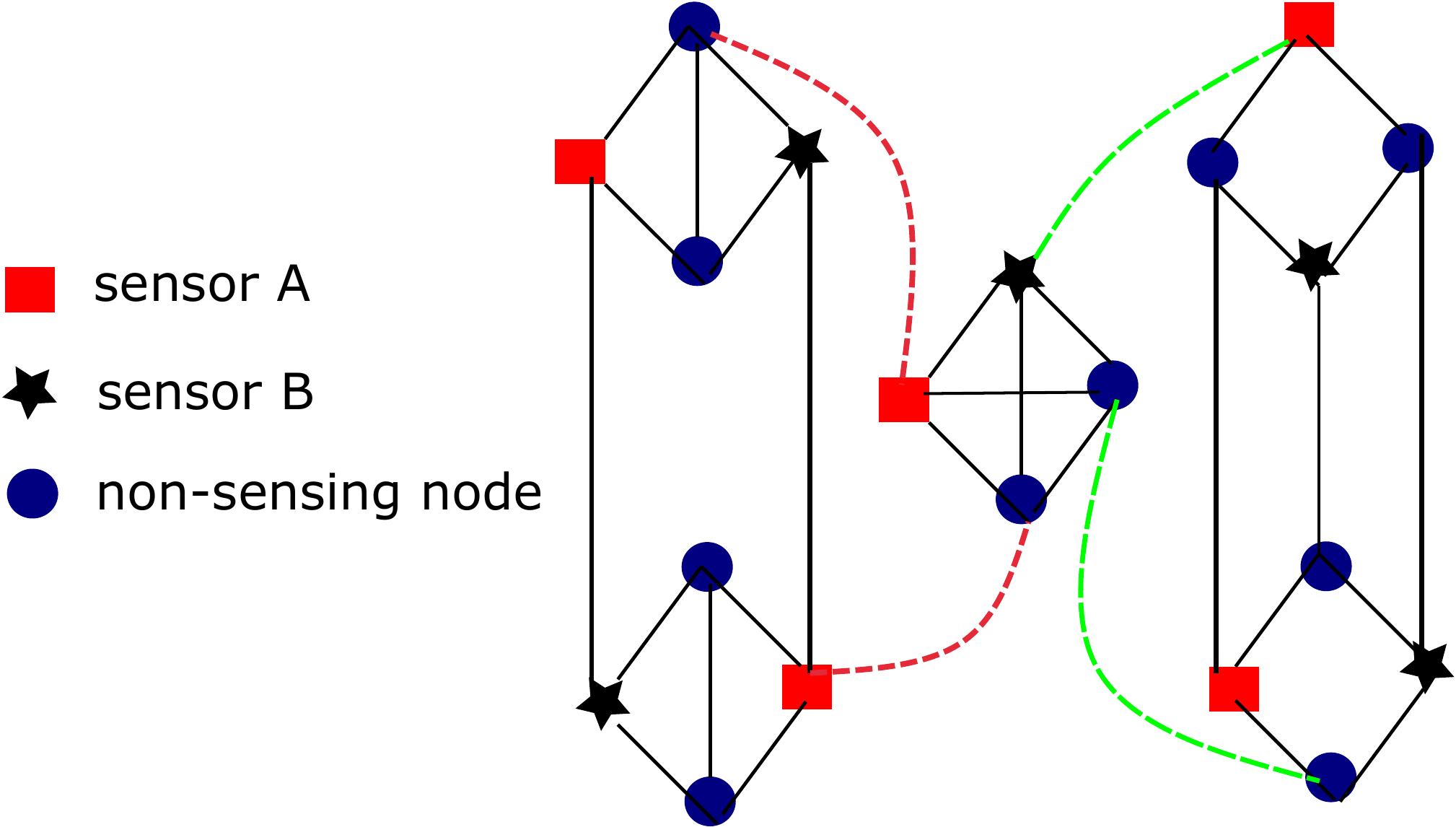}
	\caption {Communication topology of the sensor network}
	\label{fig:Graph}
\end{figure}

{\color{black}To verify that the proposed algorithm can handle singular system matrices, as stated in Theorems \ref{thm_stability} and  \ref{thm_stability2}, in this subsection, let us leave out the physical meaning of system (\ref{eq:sim_mdl}) and assume that the system matrix satisfies $\bar{A}_k=\left(\begin{smallmatrix}
	1&0&T&0\\
	0&1&0&T\\
	0&0&1&0\\
	0&0&0&1
	\end{smallmatrix}\right)$ if $\mathrm{mod}\left(k,10\right)<8$. Otherwise, $\bar{A}_k=\left(\begin{smallmatrix}
	1&0&T&0\\
	0&1&0&T\\
	0&0&1&0\\
	0&0&1&0
	\end{smallmatrix}\right).$
Besides, we consider a network with 20 nodes, where 3 kinds of nodes in this network: sensor A, sensor B and non-sensing node. A non-sensing node has no observation but it is capable to run algorithms and communicate with other nodes.  The observation matrices of sensor A and sensor B are supposed to be $\bar H_A=\left(\begin{smallmatrix}
1&0&0&0
\end{smallmatrix}\right), \bar H_B=\left(\begin{smallmatrix}
0&1&0&0
\end{smallmatrix}\right)$.
The distribution of these 3 kinds of nodes is given in Fig. \ref{fig:Graph}, which is undirected and switching between two graphs. Note that a non-sensing node $i$ can implement algorithms by assuming $\bar H_{k,i}=0$ and $y_{k,i}=0$. In the figure, the  dotted red lines and  blue lines will exist successively for every five time instants. The elements of the adjacency matrix are set to be $a_{i,j}^{\mathcal{G}_s}=\frac1{\left|\mathcal{N}_i^{\mathcal{G}_s}\right|}$.
The setting of uncertain dynamics and bias is the same as  Subsection \ref{subsec:Per}.}

For above system, we carry out  numerical simulations to compare { Algorithm \ref{ODKF3}} with other three algorithms, namely, distributed state estimation with consensus on the posteriors (DSEA-CP) \cite{Battistelli2014Kullback}, centralized Kalman filter (CKF) and centralized extended state based Kalman filter (CESKF) \cite{Bai2017reliable}.
{ For DSEA-CP, the initial estimate is assume to be $\hat{x}_{0,i}=0_{4\times1}, P_{0,i}=P_0,\forall i=1,2,3,4.$ As for CKF, in a data center, the CKF is utilized with the form of standard Kalman filter by employing the collected observations from all sensors. Here, the initial estimate of CKF are $\hat{x}_0=0_{4\times1}$, { $P_0^{\text{CKF}}=P_0$. Similar to CKF, CESKF is utilized with the form of  ESKF \cite{Bai2017reliable} by employing the collected observations from all sensors. As for the parameter of CESKF are set to be $\hat{X}_0=0_{6\times1}$, $P_0^{\text{CESKF}}=\left(\begin{smallmatrix}
		P_0&0_{4\times2}\\
		0_{2\times4}&I_{2}
		\end{smallmatrix}\right)$, $\hat{Q}_k= I_2$.}}
	
\begin{figure}[htp]
	\centering
	\includegraphics[scale=0.5]{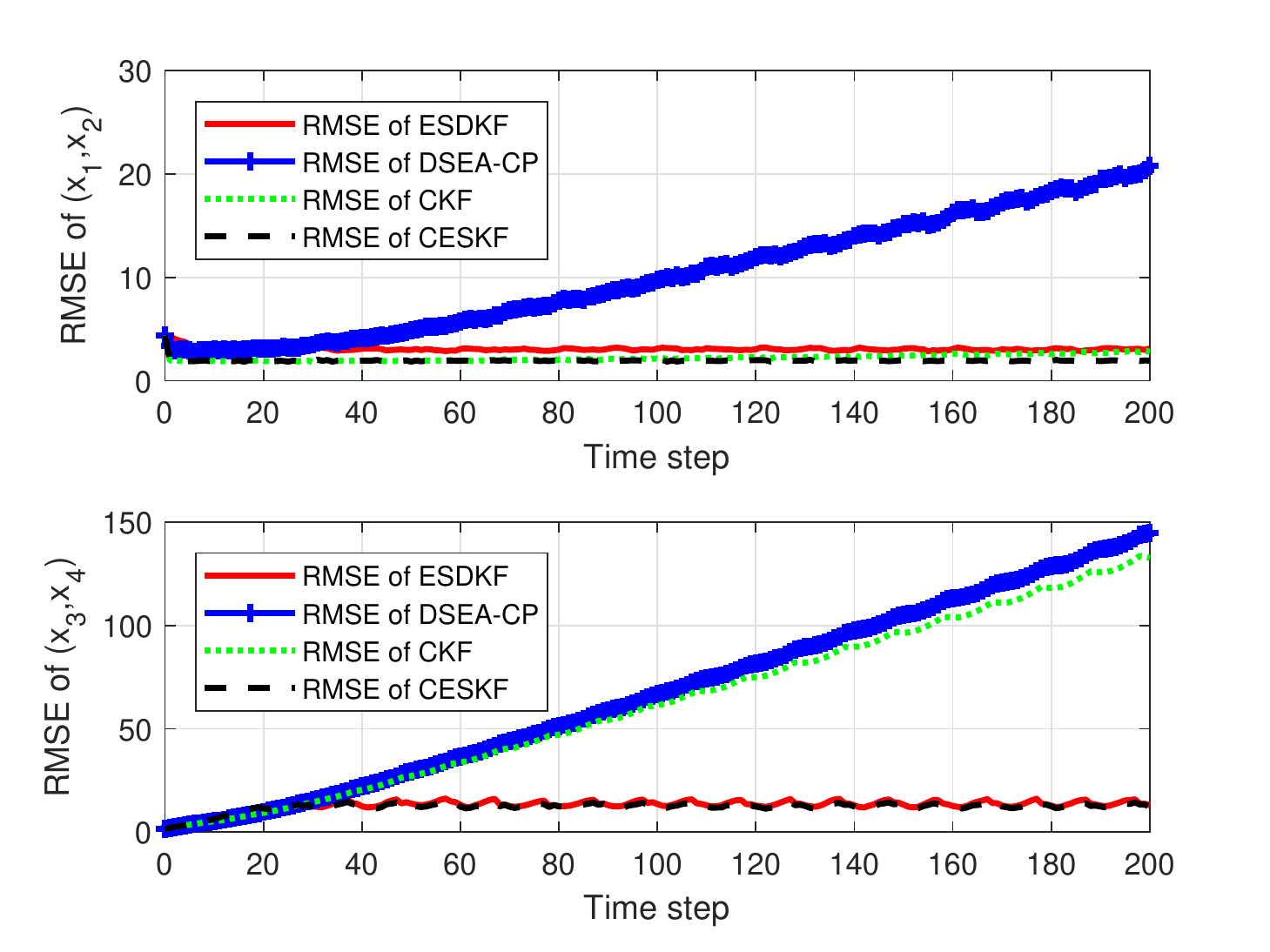}
	\caption {Estimation performance of different algorithms}
	\label{fig:cmp_algorithm}
\end{figure}

The performance comparison result of the above algorithms is shown in Fig. \ref{fig:cmp_algorithm}. { From this figure, one can see that  the RMSE of $x_3$ and $x_4$ for DSEA-CP and CKF become unstable, but the estimation errors of CESKF and Algorithm \ref{ODKF3} still keep stable. { The stability of CESKF and Algorithm \ref{ODKF3} lies in its capability in handling with uncertain nonlinear dynamics. \color{black}Since CESKF is a centralized filter without being affected by the quantized channels and the switching of communication topologies, its estimation error is smaller than ESDKF.}
	For the DSEA-CP, due to the existence of unbounded uncertain dynamics and the switching topologies, both the position estimation error and the velocity estimation error are divergent. As for CKF, since observations (position information) of all sensors are available, the position estimation error is stable. However, the velocity estimation error is divergent   since the existence of unbounded uncertain dynamics.}

Based on the above results, we can see the proposed Algorithm \ref{ODKF2} and Algorithm \ref{ODKF3} are effective distributed filtering algorithms for the considered scenarios.
\section{Conclusion}\label{Conclusion}
In this paper,  we studied a distributed filtering problem  for a class of general uncertain stochastic  systems. By treating the nonlinear uncertain dynamics as  an extended state, we proposed a novel consistent distributed Kalman filter based on quantized sensor communications.
To alleviate the effect of biased observations, the event-triggered observation update based  distributed Kalman filter was presented with  a tighter bound of error covariance than  that of the time-driven one by designing a proper threshold.
Based on mild conditions, the boundedness of the estimation error covariances and the asymptotic unbiasedness of state estimate for both the proposed two distributed filters were proved. 	

\begin{ack}                               % Place acknowledgements
	%This work was supported in part by the NSFC61603380,  the National Key Research and Development Program of China (2016YFB0901902), the National Basic Research Program of China under Grant No. 2014CB845301.
	The authors would like to thank the editors and
	anonymous reviewers for their very constructive comments,
	which greatly improved the quality of this work.
\end{ack}
	
\appendix
\section{Proof of Proposition \ref{lem_iff}}\label{app_iff}
Recall $A_{k}= \begin{pmatrix}
\bar A_{k} & \bar G_{k} \\
0 & I_p
\end{pmatrix}$, and $\sup_{k\in\mathbb{N}}\{\bar G_{k}\bar G_{k}^T\}<\infty$,
then we have
\begin{align}
A_{k}A_{k}^T
&=\begin{pmatrix}\label{pf_iff}
\bar A_{k}\bar A_{k}^T+ \bar G_{k}\bar G_{k}^T& \bar G_{k} \\
\bar G_{k}^T & I_p
\end{pmatrix}.
\end{align}
First, we consider the proof of 1).
We start with the sufficiency of 1). According to (\ref{pf_iff}), we have
$A_{k}A_{k}^T
\leq  2\left( \begin{smallmatrix}
\bar A_{k}\bar A_{k}^T+ \bar G_{k}\bar G_{k}^T& 0 \\
0 & I_p
\end{smallmatrix}\right). $
%\begin{align*}
%A_{k}A_{k}^T
%\leq& 2\begin{pmatrix}
%\bar A_{k}\bar A_{k}^T+ \bar G_{k}\bar G_{k}^T& 0 \\
%0 & I_p
%\end{pmatrix}.
%\end{align*}
Thus, $\sup_k\{\bar A_{k}\bar A_{k}^T\}<\infty$ and $\sup_k\{\bar G_{k}\bar G_{k}^T\}<\infty$ lead to $\sup_k\{ A_{k} A_{k}^T\}<\infty$.
We then consider the necessity of 1).  If $\sup_k\{ A_{k} A_{k}^T\}<\infty$, it follows from (\ref{pf_iff}) that $\sup_k\{\bar A_{k}\bar A_{k}^T+ \bar G_{k}\bar G_{k}^T\}<\infty.$
Due to $\bar G_{k}\bar G_{k}^T\geq 0$, then $\sup_k\{\bar A_{k}\bar A_{k}^T\}<\infty$.
Next, we consider the proof of 2). 	
According to (\ref{pf_iff}), it holds that $A_{k}A_{k}^T=\left( \begin{smallmatrix}
I_n& \bar G_{k} \\
0 & I_p
\end{smallmatrix}\right)\left( \begin{smallmatrix}
\bar A_{k}\bar A_{k}^T& 0 \\
0 & I_p
\end{smallmatrix}\right) \left( \begin{smallmatrix}
I_n& 0 \\
\bar G_{k}^T & I_p
\end{smallmatrix}\right).$
Thus, $\lambda_{min}(\bar A_{k}\bar A_{k}^T)=\min\{\lambda_{min}(\bar A_{k}\bar A_{k}^T),1\}.$
%\begin{align}\label{pf_min}
%\lambda_{min}(\bar A_{k}\bar A_{k}^T)=\min\{\lambda_{min}(\bar A_{k}\bar A_{k}^T),1\}.
%\end{align}
We study the sufficiency of 2).
If $\{\bar A_{k}|\bar A_{k}\in\mathbb{R}^{n\times n},k\in\mathbb{N}\}$ has an L-SS  $\{T_{l},l\in\mathbb{N}\}$ which satisfies
$\lambda_{min}(\bar A_{T_{l}+s}\bar A_{T_{l}+s}^T)$ $>\beta, s\in[0:L).$
Then for the same time sequence $\{T_{l},l\in\mathbb{N}\}$, it holds that $\lambda_{min}(A_{T_{l}+s}A_{T_{l}+s}^T)>\beta^*=\min\{\beta,1\}, s\in[0:L).$
Thus, the time sequence $\{T_{l},l\in\mathbb{N}\}$ is also an L-SS of $\{ A_{k}| A_{k}\in\mathbb{R}^{(n+p)\times (n+p)},k\in\mathbb{N}\}$.
We then consider the necessity of 2). 	If $\{ A_{k}| A_{k}\in\mathbb{R}^{(n+p)\times (n+p)},,k\in\mathbb{N}\}$ has an L-SS $\{\bar T_{l},l\in\mathbb{N}\}$, then $\lambda_{min}( A_{\bar T_{l}+s} A_{\bar T_{l}+s}^T)>\bar\beta, s\in[0:L).$
Recall $\lambda_{min}(\bar A_{k}\bar A_{k}^T)=\min\{\lambda_{min}(\bar A_{k}\bar A_{k}^T),1\}.$,  then we have $\lambda_{min}(\bar A_{\bar T_{l}+s}\bar A_{\bar T_{l}+s}^T)$ $>\bar\beta, s\in[0:L).$
As a result, the time sequence $\{\bar T_{l},l\in\mathbb{N}\}$ is also an L-SS of $\{\bar A_{k}|\bar A_{k}\in\mathbb{R}^{n\times n},,k\in\mathbb{N}\}$.	
{
\section{Proof of Proposition \ref{lem_observable}}\label{pf_observ}
Denote $\Phi_{j,k}=\Phi_{j,j-1}\cdots \Phi_{k+1,k}(j>k),$ 	$\Phi_{k,k}=I_{n+p}$, and $\Phi_{k+1,k}=A_{k}$, then it holds that $\Phi_{j,k}=\begin{pmatrix}
\bar A_{j-1} & \bar G_{j-1} \\
0 & I_p
\end{pmatrix}\times\cdots\times\begin{pmatrix}
\bar A_{k} & \bar G_{k} \\
0 & I_p
\end{pmatrix}
=\begin{pmatrix}
\bar \Phi_{j,k} \quad& \tilde\Phi_{j,k+1} \\
0 & I_p
\end{pmatrix},$
where $\tilde\Phi_{j,k+1}=\sum_{i=k+1}^{j}\bar \Phi_{j,i}\bar G_{i-1}.$
Thus, we obtain $\Phi^T_{j,k} H_{j,i}^T(R_{j,i}+B_{j,i})^{-1} H_{j,i}\Phi_{j,k}
=\begin{pmatrix}
\Theta_{i,j,k}^{1,1} & \Theta_{i,j,k}^{1,2} \\
(\Theta_{i,j,k}^{1,2})^T & \Theta_{i,j,k}^{2,2}
\end{pmatrix},$
%\begin{align*}
%&\Phi^T_{j,k} H_{j,i}^T(R_{j,i}+B_{j,i})^{-1} H_{j,i}\Phi_{j,k}\\
%=&\begin{pmatrix}
%\Theta_{1,1} & \Theta_{1,2} \\
%\Theta_{1,2}^T & \Theta_{2,2}
%\end{pmatrix},
%\end{align*}
where $\Theta_{i,j,k}^{1,1}=\bar\Phi^T_{j,k} \bar H_{j,i}^T(R_{j,i}+B_{j,i})^{-1} \bar H_{j,i}\bar\Phi_{j,k}$, $\Theta_{i,j,k}^{1,2}=\bar\Phi^T_{j,k} \bar H_{j,i}^T(R_{j,i}+B_{j,i})^{-1} \bar H_{j,i}\tilde\Phi_{j,k+1}$ and \\ $\Theta_{i,j,k}^{2,2}=(\bar H_{j,i}\tilde\Phi_{j,k+1})^T(R_{j,i}+B_{j,i})^{-1}(\bar H_{j,i}\tilde\Phi_{j,k+1}).$
%\begin{align*}
%\Theta_{i,j,k}^{1,1}&=\bar\Phi^T_{j,k} \bar H_{j,i}^T(R_{j,i}+B_{j,i})^{-1} \bar H_{j,i}\bar\Phi_{j,k}\\
%\Theta_{i,j,k}^{1,2}&=\bar\Phi^T_{j,k} \bar H_{j,i}^T(R_{j,i}+B_{j,i})^{-1} \bar H_{j,i}\tilde\Phi_{j,k+1}\\
%\Theta_{i,j,k}^{2,2}&=(\bar H_{j,i}\tilde\Phi_{j,k+1})^T(R_{j,i}+B_{j,i})^{-1}(\bar H_{j,i}\tilde\Phi_{j,k+1}).
%\end{align*}
%For any $\alpha>0$,

For $\bar N\in\mathbb{N}^+$, we denote
\begin{align*}
&\sum_{i=1}^{N}\sum_{j=k}^{k+\bar N}\begin{pmatrix}
\Theta_{i,j,k}^{1,1} & \Theta_{i,j,k}^{1,2} \\
(\Theta_{i,j,k}^{1,2})^T & \Theta_{i,j,k}^{2,2}
\end{pmatrix}
:=\begin{pmatrix}
\bar\Theta_{k,\bar N}^{1,1} & \bar\Theta_{k,\bar N}^{1,2} \\
(\bar\Theta_{k,\bar N}^{1,2})^T & \bar\Theta_{k,\bar N}^{2,2}
\end{pmatrix}.
\end{align*}
where
\begin{align}\label{eq_notation}
\begin{split}
\bar\Theta_{k,\bar N}^{1,1}&=\sum_{i=1}^{N}\sum_{j=k}^{k+\bar N}\Theta_{i,j,k}^{1,1}\\
\bar\Theta_{k,\bar N}^{1,2} &=\sum_{i=1}^{N}\sum_{j=k}^{k+\bar N}\Theta_{i,j,k}^{1,2} \\
\bar\Theta_{k,\bar N}^{2,2}&=\sum_{i=1}^{N}\sum_{j=k}^{k+\bar N}\Theta_{i,j,k}^{2,2}
\end{split}
\end{align}
1) Necessity. If   the reformulated system \eqref{system3} is uniformly collectively observable,
then there exist $M, \bar N\in\mathbb{N}^+$ and $\alpha>0$ such that
$\begin{pmatrix}
\bar\Theta_{k,\bar N}^{1,1} &\bar\Theta_{k,\bar N}^{1,2}  \\
(\bar\Theta_{k,\bar N}^{1,2})^T & \bar\Theta_{k,\bar N}^{2,2}-\alpha I_{p}
\end{pmatrix}-\alpha I_{n+p}>0, \forall k\geq M.$
Thus, $\bar\Theta_{k,\bar N}^{1,1}>0$, then in light of Schur Complement \cite{Seber2007A},  for $k\geq \bar M$, $(\bar\Theta_{k,\bar N}^{2,2}-\alpha I_{p})-( \bar\Theta_{k,\bar N}^{1,2})^T(\bar\Theta_{k,\bar N}^{1,1}-\alpha I_{n})^{-1} \bar\Theta_{k,\bar N}^{1,2}>0.$

2) Sufficiency. If there exist  $\bar N,M\in\mathbb{N}^+$  and $\alpha>0$, such that $\bar\Theta_{k,\bar N}^{1,1}-\alpha I_{n} >0$, for $k\geq \bar M$. For $k\geq \bar M$, by $(\bar\Theta_{k,\bar N}^{2,2}-\alpha I_{p})-( \bar\Theta_{k,\bar N}^{1,2})^T(\bar\Theta_{k,\bar N}^{1,1}-\alpha I_{n})^{-1} \bar\Theta_{k,\bar N}^{1,2}>0$ and Schur Complement \cite{Seber2007A}, for $k\geq \bar M$, $\begin{pmatrix}
\bar\Theta_{k,\bar N}^{1,1} & \bar\Theta_{k,\bar N}^{1,2} \\
(\bar\Theta_{k,\bar N}^{1,2})^T & \bar\Theta_{k,\bar N}^{2,2}
\end{pmatrix}>\alpha I_{n+p}$ holds. Thus, the reformulated system \eqref{system3} is uniformly collectively observable.
}

\section{Proof of Lemma \ref{thm_W}}\label{pf_1}
Here we utilize an inductive method for the proof. 	
At the initial moment, under Assumption \ref{ass_noise},
$E\{(\hat X_{0,i}-X_{0})(\hat X_{0,i}-X_{0})^T\} \leq P_{0,i}$.
Suppose  $E\{(\hat X_{k-1,i}-X_{k-1})(\hat X_{k-1,i}-X_{k-1})^T\} \leq P_{k-1,i}.$
Recall
$\bar e_{k,i}=\bar X_{k,i}-X_{k}=A_{k-1}e_{k-1,i}-Du_{k-1}-\omega_{k-1}.$
%\begin{equation*}
%\begin{split}
%\bar e_{k,i}&=\bar X_{k,i}-X_{k}\\
%&=A_{k-1}\hat X_{k-1,i}+D\hat u_{k-1}-A_{k-1}X_{k-1}\\
%&\quad-D u_{k-1}-\omega_{k-1}\\
%&=A_{k-1}(\hat X_{k-1,i}-X_{k-1})+D(\hat u_{k-1}-u_{k-1})-\omega_{k-1},\\
%&=A_{k-1}e_{k-1,i}+D(\hat u_{k-1}-u_{k-1})-\omega_{k-1}.
%\end{split}
%\end{equation*}
Since the estimation error $e_{k-1,i}$ is measurable to $\sigma\{\mathcal{F}_{k-2},v_{k-1,i},i\in\mathcal{V}\}$, and $\omega_{k-1}$ is  independent from $\mathcal{F}_{k-2}$ and $v_{k-1,i},i\in\mathcal{V}$, we have $E\{e_{k-1,i}\omega_{k-1}^T\}=0.$
Similarly,  it holds that $E\{u_{k-1}\omega_{k-1}^T\}=0$. Then,
due to $E\{(\sqrt{\theta_{k,i}} x+\frac{y}{\sqrt{\theta_{k,i}}})(\sqrt{\theta_{k,i}} x+\frac{y}{\sqrt{\theta_{k,i}}})^T\}\geq 0$, we have the inequality $E\{xy^T+yx^T\}\leq E\{\theta_{k,i} xx^T+\frac{1}{\theta_{k,i}}yy^T\}$, $\forall \theta_{k,i}>0$. Then, we have $E\{\bar e_{k,i}\bar e_{k,i}^T\}\leq (1+\theta_{k,i})A_{k-1}P_{k-1,i}A_{k-1}^T+(1+\frac{1}{\theta_{k,i}})\bar Q_{k-1}+\tilde Q_{k-1}.$
%\begin{align}\label{P_bar}
%&E\{\bar e_{k,i}\bar e_{k,i}^T\}\\
%%=&A_{k-1}E\{e_{k-1,i}e_{k-1,i}^T\}A_{k-1}^T\nonumber\\
%%&+A_{k-1}E\{\bar e_{k,i}(\hat u_{k-1}-u_{k-1})^T\}D^T+\tilde Q_{k-1}\nonumber\\
%%&+DE\{(\hat u_{k-1}-u_{k-1})\bar e_{k,i}^T\}A_{k-1}^T \nonumber\\
%%&+DE\{(\hat u_{k-1}-u_{k-1})(\hat u_{k-1}-u_{k-1})^T\}D^T\nonumber\\
%%\leq& (1+\theta_{k,i})A_{k-1}E\{e_{k-1,i}e_{k-1,i}^T\}A_{k-1}^T+\tilde Q_{k-1}\nonumber\\
%%%&+(1+\frac{1}{\theta_{k,i}})DE\{(\hat u_{k-1}-u_{k-1})(\hat u_{k-1}-u_{k-1})^T\}D^T\nonumber\\
%%%\leq& (1+\theta_{k,i})A_{k-1}P_{k-1,i}A_{k-1}^T+\tilde Q_{k-1}\nonumber\\
%%&+(1+\frac{1}{\theta_{k,i}})4p D\cdot diag\{q_{k-1}(1),\dots,q_{k-1}(p)\}D^T\nonumber\\
%\leq& (1+\theta_{k,i})A_{k-1}P_{k-1,i}A_{k-1}^T+(1+\frac{1}{\theta_{k,i}})\bar Q_{k-1}+\tilde Q_{k-1},\nonumber
%\end{align}
%where
%\begin{equation}\label{eq_Q_bar}
%%\begin{split}
%{ \tilde Q_{k}=blockdiag\{Q_{k},0^{p\times p}\}, \bar Q_{k}=D\hat Q_{k}D^T.}
%%\end{split}
%\end{equation}
According to  the definition of $\bar P_{k,i}$, $E\{\bar e_{k,i}\bar e_{k,i}^T\}\leq\bar P_{k,i}$ hold.
At the update stage, there is $\tilde e_{k,i}=\tilde X_{k,i}-X_{k}
=(I-K_{k,i} H_{k,i})\bar e_{k,i}+K_{k,i}b_{k,i}+K_{k,i}v_{k,i}.$
Under Assumption \ref{ass_noise} and the fact that $\bar e_{k,i}$ is measurable to $\mathcal{F}_{k-1}$, it follows that
$E\{\bar e_{k,i}v_{k,i}^T\}=E\{E\{\bar e_{k,i}v_{k,i}^T|\mathcal{F}_{k-1}\}\}
%=E\{E\{(A_{k-1}e_{k-1,i}+w_{k-1})v_{k,i}^T|\mathcal{F}_{k-1}\}\}
=E\{A_{k-1}e_{k-1,i}E\{v_{k,i}^T|\mathcal{F}_{k-1}\}\}
+E\{E\{w_{k-1}v_{k,i}^T|\mathcal{F}_{k-1}\}\}
=E\{w_{k-1}v_{k,i}^T\}=0$.
Similarly,  it holds that $E\{b_{k,i}v_{k,i}^T\}=0$ by noting that $b_{k,i}$ is measurable to $\mathcal{F}_{k-1}$. Then,  $\forall \mu_{k,i}>0$, we have
$E\{\tilde e_{k,i}\tilde e_{k,i}^T\}\leq (1+\mu_{k,i})(I-K_{k,i} H_{k,i})E\{\bar e_{k,i}\bar e_{k,i}^T\}(I-K_{k,i} H_{k,i})^T
+K_{k,i}E\{v_{k,i}v_{k,i}^T\}K_{k,i}^T+\frac{1+\mu_{k,i}}{\mu_{k,i}}K_{k,i}E\{b_{k,i}b_{k,i}^T\}K_{k,i}^T\leq\tilde P_{k,i}$.
%\begin{align}\label{P_k}
%%	\begin{split}
%&E\{\tilde e_{k,i}\tilde e_{k,i}^T\}\nonumber\\
%%=&(I-K_{k,i} H_{k,i})E\{\bar e_{k,i}\bar e_{k,i}^T\}(I-K_{k,i} H_{k,i})^T\nonumber\\
%%&+K_{k,i}E\{v_{k,i}v_{k,i}^T\}K_{k,i}^T+(I-K_{k,i} H_{k,i})E\{\bar e_{k,i}b_{k,i}^T\}\nonumber\\
%%&+E\{b_{k,i}\bar e_{k,i}^T\}(I-K_{k,i} H_{k,i})^T\nonumber\\
%\leq&(1+\mu_{k,i})(I-K_{k,i} H_{k,i})E\{\bar e_{k,i}\bar e_{k,i}^T\}(I-K_{k,i} H_{k,i})^T\nonumber\\
%&+K_{k,i}E\{v_{k,i}v_{k,i}^T\}K_{k,i}^T+\frac{1+\mu_{k,i}}{\mu_{k,i}}K_{k,i}E\{b_{k,i}b_{k,i}^T\}K_{k,i}^T\nonumber\\
%%\leq &(1+\mu_{k,i})(I-K_{k,i} H_{k,i})\bar P_{k,i}(I-K_{k,i} H_{k,i})^T\nonumber\\
%%&+		K_{k,i}\bigg(R_{k,i}+\frac{1+\mu_{k,i}}{\mu_{k,i}}B_{k,i}\bigg)K_{k,i}^T\nonumber\\
%\leq&\tilde P_{k,i}.
%%	\end{split}
%\end{align}

{
Recall the existence of  quantization  operation with respect to $\tilde X_{k,i}+\xi_{k,i}$, where $\xi_{k,i}$ stands for the dithering noise vector. We denote the estimation error $\check e_{k,i}:=\check X_{k,i}-X_{k}=\tilde e_{k,i}+\xi_{k,i}+\vartheta_{k,i}$, where $\vartheta_{k,i}$ is the quantization error vector of $\tilde X_{k,i}+\xi_{k,i}$.
By Assumption \ref{ass_dither},   $\xi_{k,i}+\vartheta_{k,i}$ is independent of $\tilde e_{k,i}$. Then
$E\{\check e_{k,i}\check e_{k,i}^T\}\leq E\{\tilde e_{k,i}\tilde e_{k,i}^T\}+E\{(\xi_{k,i}+\vartheta_{k,i})(\xi_{k,i}+\vartheta_{k,i})^T\}
\leq \tilde P_{k,i}+\Delta_i^2\bar nI_{\bar  n}
=  \check P_{k,i}+P_{k,i}^*+\Delta_i^2\bar nI_{\bar  n},$
%\begin{align*}
%&E\{\check e_{k,i}\check e_{k,i}^T\}\\
%\leq&E\{\tilde e_{k,i}\tilde e_{k,i}^T\}+E\{(\xi_{k,i}+\vartheta_{k,i})(\xi_{k,i}+\vartheta_{k,i})^T\}\\
%\leq&\tilde P_{k,i}+\frac{\Delta^2\bar n}{2}I_{\bar  n}\\
%=& \check P_{k,i}+P_{k,i}^*+\frac{\Delta^2\bar n}{2}I_{\bar  n},
%\end{align*}
where $P_{k,i}^*\in\mathbb{R}^{\bar n\times \bar n}$ is the symmetric quantization error matrix of $\check P_{k,i}$. It holds that
$P_{k,i}^*\leq \lambda_{max}(P_{k,i}^*)I_{\bar n}\leq \|P_{k,i}^* \|_FI_{\bar n}:=\sqrt{\left(\sum_{s=1}^{\bar n}\sum_{l=1}^{\bar n}\left(P_{k,i}^*(s,l)\right)^2\right)}I_{\bar n}\leq \frac{\Delta_i\bar n}{2}I_{\bar n}$, where $\|\cdot \|_F$ is the Frobenius norm, and $P_{k,i}^*(s,l)$ is the $(s,l)$th element of $P_{k,i}$.
Thus, we have $E\{\check e_{k,i}\check e_{k,i}^T\}\leq \check P_{k,i}+\frac{\bar n\Delta_i(2\Delta_i+1)}{2}I_{\bar n}$.
Notice that $e_{k,i}=\hat X_{k,i}-X_{k}=P_{k,i}(\sum_{j\in \mathcal{N}_{i}(k),j\neq i}a_{i,j}(k)\mathcal{\tilde P}_{k,j}^{-1}\check e_{k,j}+a_{i,i}(k)\tilde P_{k,i}^{-1}\tilde e_{k,i})$, where $\mathcal{\tilde P}_{k,j}=\check P_{k,j}+\frac{\bar n\Delta_i(2\Delta_i+1)}{2}I_{\bar n},$ if $i\neq j$, and $j\in\mathcal{N}_{i}$.}
%\begin{equation}\label{e_k01}
%\begin{split}
%e_{k,i}&=\hat X_{k,i}-X_{k}\\
%%&=P_{k,i}\sum_{j\in \mathcal{N}_{i}(k)}a_{i,j}(k)\tilde P_{k,j}^{-1}(\tilde X_{k,j}-X_{k})\\
%&=P_{k,i}\sum_{j\in \mathcal{N}_{i}(k)}a_{i,j}(k)\tilde P_{k,j}^{-1}\tilde e_{k,j},
%\end{split}
%\end{equation}
%where
%\begin{equation}\label{eq_P_fusion}
%\begin{split}
%&P_{k|k,i}=\bigg(\sum_{j\in \mathcal{N}_{i}(k)}a_{i,j}(k)  \tilde P_{k,j}^{-1}\bigg)^{-1}.
%\end{split}
%\end{equation}
According to the consistent estimate of Covariance Intersection \cite{Niehsen2002Information}, there is $E\{e_{k,i}e_{k,i}^T\}\leq P_{k,i}.$
Therefore, the proof is finished.
\section{Proof of Lemma \ref{thm_K}}\label{pf_K}
By Lemma \ref{thm_W}, it holds that $\tilde{P}_{k,i}= (1+\mu_{k,i})\Delta\tilde{P}_{k,i},$		
with
\begin{align}\label{pf_P2}
\Delta\tilde{P}_{k,i}\triangleq &(I-K_{k,i} H_{k,i})\bar P_{k,i}(I-K_{k,i} H_{k,i})^T+	K_{k,i}\tilde{R}_{k,i}K_{k,i}^T,\nonumber\\
%=&\bar P_{k,i}-K_{k,i}H_{k,i}\bar P_{k,i}-\bar P_{k,i}H_{k,i}^TK_{k,i}^T\nonumber\\
%&+K_{k,i}H_{k,i}\bar P_{k,i}H_{k,i}^TK_{k,i}^T+K_{k,i} \tilde{R}_{k,i}K_{k,i}^T\nonumber\\
%=&\bar P_{k,i}-K_{k,i}H_{k,i}\bar P_{k,i}-\bar P_{k,i}H_{k,i}^TK_{k,i}^T\nonumber\\
%&+K_{k,i}(H_{k,i}\bar P_{k,i}H_{k,i}^T+\tilde{R}_{k,i})K_{k,i}^T\nonumber\\
=&(K_{k,i}-K_{k,i}^*)(H_{k,i}\bar P_{k,i}H_{k,i}^T+\tilde{R}_{k,i})(K_{k,i}-K_{k,i}^*)^T\nonumber\\
&+(I-K_{k,i}^*H_{k,i})\bar P_{k,i},
\end{align}
where  $\tilde{R}_{k,i}=\frac{R_{k,i}}{1+\mu_{k,i}}+\frac{B_{k,i}}{\mu_{k,i}} $, $K_{k,i}^*=\bar P_{k,i}H_{k,i}^T(H_{k,i}\bar P_{k,i}H_{k,i}^T+ \tilde{R}_{k,i})^{-1}$.
Thus,  it is seen from  (\ref{pf_P2}) that $\tilde{P}_{k,i}$ is minimized in the sense of positive definiteness (i.e., $\tr(\tilde{P}_{k,i})$ is minimized)  when $K_{k,i}=K_{k,i}^*$.
%As a result,
%\begin{equation*}
%\begin{cases}
%\tilde{P}_{k,i}&=(1+\mu_{k,i})(I-K_{k,i}H_{k,i})\bar P_{k,i}\\
%K_{k,i}&=\bar P_{k,i}H_{k,i}^T(H_{k,i}\bar P_{k,i}H_{k,i}^T+ \tilde{R}_{k,i})^{-1}\\
%\tilde{R}_{k,i}&=\frac{R_{k,i}}{1+\mu_{k,i}}+\frac{B_{k,i}}{\mu_{k,i}}.
%\end{cases}
%\end{equation*}
\section{Proof of Lemma \ref{lem_inver}}\label{app_inver}
For the proof of 1), exploiting the matrix inverse formula on $\tilde P_{ k,i}$ directly yields the  conclusion.	
Next, we consider the proof of 2). First, we prove there exists a constant positive definite matrix $\bar P$, such that $P_{k,i}\geq \bar P$.
Consider $\bar P_{k,i}\geq \frac{1+\theta_{k,i}}{\theta_{k,i}} \bar Q_{k-1}+\tilde Q_{k-1}\geq \frac{1+\theta_{2}}{\theta_{2}} \bar Q_{k-1}+\tilde Q_{k-1}\geq Q^*>0,$
%\begin{align*}
%\bar P_{k,i}&\geq \frac{1+\theta_{k,i}}{\theta_{k,i}} \bar Q_{k-1}+\tilde Q_{k-1}\\
%&\geq \frac{1+\theta_{2}}{\theta_{2}} \bar Q_{k-1}+\tilde Q_{k-1}\geq Q^*>0,
%\end{align*}
where $Q^*$ can be obtained by noting $\inf_k Q_{k}\geq \underline Q>0$ and  Assumption \ref{ass_F}.
According to 1) of Lemma \ref{lem_inver}, then we have $\tilde P_{k,i}^{-1}\leq\frac{Q^*}{1+\mu_{k,i}}+H_{k,i}^T\Delta R_{k,i}^{-1}H_{k,i}\leq Q_*,$
where $Q_*>0$ is obtained by employing the condition 3) of Assumption \ref{ass_noise}.
Recall $P_{k,i}=\bigg(\sum_{j\in \mathcal{N}_{i}(k)}a_{i,j}(k)  \tilde P_{k,j}^{-1}\bigg)^{-1},$
then $P_{k,i}^{-1}\leq Q_*$, which means  $P_{k,i}\geq Q_*^{-1}>0$.
Consider the time sequence $\{T_{l},l\in\mathbb{N}\}$, which is the L-SS of $\{A_{k},k\in\mathbb{N}\}$.
Under Assumption \ref{ass_A}, there exists a scalar $\beta$, such that $A_{T_l+s}P_{T_l+s,i}A_{T_l+s}^T\geq \beta Q_*^{-1}>0.$
Due to sup$_{k}$$Q_{k}\leq \bar Q<\infty$, there is a scalar $\varpi>0$, such that $Q_{T_l+s}\leq \varpi A_{T_l+s}P_{T_l+s,i}A_{T_l+s}^T$.
Then $\bar P_{T_l+s+1,i}= A_{T_l+s}P_{T_l+s,i}A_{T_l+s}^T+Q_{T_l+s}\leq (1+\varpi)A_{T_l+s}P_{T_l+s,i}A_{T_l+s}^T.$
Let $\eta=\frac{1}{1+\varpi}$, then the conclusion 2) of this lemma holds.
\section{Proof of Proposition \ref{prop_theta}}\label{app_propo_theta}
	%	See the proof in Appendix \ref{app_prop_theta}.
	Consider $\tr(\bar P_{k,i})$, then we have $\tr(\bar P_{k,i})=(1+\theta_{k,i}) \tr(A_{k-1}P_{k-1,i} A_{k-1}^T)
	+ \frac{1+\theta_{k,i}}{\theta_{k,i}} \tr(\bar Q_{k-1})+tr(\tilde Q_{k-1}).$
	Hence $\theta_{k,i}^*=\arg\min\limits_{\theta_{k,i}}\tr(\bar P_{k,i})= \arg\min\limits_{\theta_{k,i}}f_k(\theta_{k,i}),$
	where
	$f_k(\theta_{k,i})=\theta_{k,i}\tr(A_{k-1}P_{k-1,i} A_{k-1}^T)+\frac{\tr(\bar Q_{k-1}) }{\theta_{k,i}},$
	which is minimized if $\theta_{k,i}^*\tr(A_{k-1}P_{k-1,i} A_{k-1}^T)=\frac{\tr(\bar Q_{k-1}) }{\theta_{k,i}^*}.$
	As a result, $\theta_{k,i}^*=\sqrt{\frac{\tr(\bar Q_{k-1})}{\tr(A_{k-1}P_{k-1,i} A_{k-1}^T)}}$.		
	Since $P_{k-1,i}>0$, $\bar Q_{k-1}>0$ and $A_{k-1}\neq 0$, we have $\theta_{k,i}^*>0$.	
{
\section{Proof of Lemma \ref{lem_conver}}\label{app_conver}
	By Theorem 1 of \cite{Cat2010Diffusion}, (1) can be proved. 	
For (2), we have $\frac{x_{k}}{\delta^k}=\left(\frac{\rho}{\delta}\right)^kx_0+\sum_{i=0}^{k}\left(\frac{\rho}{\delta}\right)^{k-i}\frac{m_i}{\delta^i}$.
Due to $\rho<\delta<1$, $\left(\frac{\rho}{\delta}\right)^kx_0\rightarrow 0.$ Denote $\bar \rho=\frac{\rho}{\delta}\in(0,1)$ and $\bar m_{i}=\frac{m_i}{\delta^i}=o(1)$, then  construct a sequence $\{\bar x_{k}\}$ satisfying $\bar x_{k+1}=\bar \rho \bar  x_{k}+\bar m_k$ with $\bar  x_{0}=0$. 	By (1), we have $\bar  x_{k}=o(1)$. In light of  $\bar  x_{k}=\sum_{i=0}^{k}\left(\bar \rho \right)^{k-i}\bar m_i$,  $\sum_{i=0}^{k}\left(\frac{\rho}{\delta}\right)^{k-i}\frac{m_i}{\delta^i}=\sum_{i=0}^{k}\left(\bar \rho \right)^{k-i}\bar m_i \rightarrow 0$  as $k\rightarrow\infty$. Hence,  $x_{k}=o(\delta^k)$.
For (3), consider $x_{k}k^{M-1}=\rho^kk^{M-1}x_0+k^{M-1}\sum_{i=0}^{k}\rho ^{k-i} m_i$. Notice $\rho^kk^{M-1}=o(1)$, then consider the convergence of the second term, namely, $k^{M-1}\sum_{i=0}^{k}\rho ^{k-i} m_i$. Due to $m_{k}=o(\frac{1}{k^{M}})$, we have
$k^{M-1}\sum_{i=0}^{k}\rho ^{k-i} m_i=\sum_{i=0}^{k}o(\rho ^{k-i} \frac{k^{M-1}}{i^{M}})
= \sum_{i=0}^{k}o(\frac{k^{M-1}}{i^{M}(k-i)^M })=\sum_{i=0}^{k} o(\frac{1}{k})=o(1)$,
%\begin{align*}
%&k^{M-1}\sum_{i=0}^{k}\rho ^{k-i} m_i\\
%=&\sum_{i=0}^{k}o(\rho ^{k-i} \frac{k^{M-1}}{i^{M}})
%= \sum_{i=0}^{k}o(\frac{k^{M-1}}{i^{M}(k-i)^M })\\
%= &\sum_{i=0}^{k} o(\frac{1}{k})
%=o(1),
%\end{align*}
where the second equality is obtained by $\rho ^{k-i}=o(\frac{1}{(k-i)^M})$ and the third equality is obtained by  $i(k-i)\leq \frac{k}{2}$. Thus, $x_{k}k^{M-1}=o(1)$, which means $x_{k}=o(\frac{1}{k^{M-1}})$.
}

\section{Proofs of Lemmas \ref{lem_trigger2}, \ref{lem_trigger} and Proposition \ref{prop_comp}}\label{app_final}
\textbf{Proof of Lemma \ref{lem_trigger2}.}	Employing the matrix inverse formula on $\tilde  P_{k,i}$ yields $\tilde  P_{k,i}^{-1}=\frac{\bar P_{k,i}^{-1}}{1+\mu_{k,i}}+H_{k,i}^T\Delta R_{k,i}^{-1}H_{k,i}$.
where $\Delta R_{k,i}=R_{k,i}+\frac{1+\mu_{k,i}}{\mu_{k,i}}B_{k,i}$.	
Substituting $\tilde  P_{k,i}^{-1}$ into (\ref{eq_trigger}), the conclusion of this lemma holds.

\textbf{Proof of  Lemma \ref{lem_trigger} .} 
If   (\ref{eq_trigger}) is satisfied,  we have $\tilde P_{k,i}^{-1}= \frac{\bar P_{k,i}^{-1}}{1+\mu_{k,i}}+H_{k,i}^T\Delta R_{k,i}^{-1}H_{k,i}.$
%\begin{align*}
%
%\end{align*}
Thus, conclusion of Lemma \ref{lem_trigger} holds in this case.
If the update event in (\ref{eq_trigger}) is not triggered, then $\tilde P_{k,i}=\bar P_{k,i}$. Besides, according to the scheme, it follows that $\tilde P_{k,i}^{-1} \geq \frac{\bar P_{k,i}^{-1}}{1+\mu_{k,i}}+H_{k,i}^T\Delta R_{k,i}^{-1}H_{k,i}-\tau I_n.$

\textbf{Proof of Proposition \ref{prop_comp}.} In light of Lemma \ref{lem_trigger}, for $\tau=0$, $\tilde P_{k,i}^{-1}\geq \frac{\bar P_{k,i}^{-1}}{1+\mu_{k,i}}+H_{k,i}^T\Delta R_{k,i}^{-1}H_{k,i}=\check  P_{k,i}^{-1},$
%			\begin{align*}
%			
%			\end{align*}
which means $\tilde P_{k,i}\leq \check  P_{k,i}$, where $\check  P_{k,i}$ corresponds to the observation update of  Algorithm \ref{ODKF2}. By using the mathematical induction method, the proof of this proposition can be finished.

\bibliographystyle{ieeetr}

\small
\bibliography{references_filtering2}           % and a bib file to produce the
                                 % bibliography (preferred). The
                                 % correct style is generated by
                                 % Elsevier at the time of printing.

%\begin{thebibliography}{99}     % Otherwise use the
                                 % thebibliography environment.
                                 % Insert the full references here.
                                 % See a recent issue of Automatica
                                 % for the style.
%  \bibitem[Heritage, 1992]{Heritage:92}
%     (1992) {\it The American Heritage.
%     Dictionary of the American Language.}
%     Houghton Mifflin Company.
%  \bibitem[Able, 1956]{Abl:56}
%     B.~C.~Able (1956). Nucleic acid content of macroscope.
%     {\it Nature 2}, 7--9.
%  \bibitem[Able {\em et al.}, 1954]{AbTaRu:54}
%     B.~C. Able, R.~A. Tagg, and M.~Rush (1954).
%     Enzyme-catalyzed cellular transanimations.
%     In A.~F.~Round, editor,
%     {\it Advances in Enzymology Vol. 2} (125--247).
%     New York, Academic Press.
%  \bibitem[R.~Keohane, 1958]{Keo:58}
%     R.~Keohane (1958).
%     {\it Power and Interdependence:
%     World Politics in Transition.}
%     Boston, Little, Brown \& Co.
%  \bibitem[Powers, 1985]{Pow:85}
%     T.~Powers (1985).
%     Is there a way out?
%     {\it Harpers, June 1985}, 35--47.

%\end{thebibliography}

%\appendix

                       % in the appendices.
\end{document}